\documentclass [11pt]{article}

\usepackage{jheppub}

\usepackage[utf8]{inputenc}

\usepackage{amsfonts}

\usepackage{amsmath,amssymb}

\usepackage{graphicx}

\usepackage{simplewick}

\usepackage{array}

\usepackage{mmacells} 

\usepackage{color}

\mmaSet{
  leftmargin=2.3em,
  labelsep=-0.8em,
}

\NewDocumentCommand \mmaMyHat { m }
  {\ensuremath{\hat{\text{#1}}}}

\def\bel{\begin{equation}\begin{aligned}}
\def\eel{\end{aligned}\end{equation}}

\def\bea{\begin{eqnarray}} \def\eea{\end{eqnarray}}
\def\be{\begin{equation}} \def\ee{\end{equation}} 

\def\nn{\nonumber}

\newcommand{\ops}[1]{\mathcal{#1}}

\newcommand{\proj}{\Big\vert_{proj}}

\newcommand{\point}{{\bf p}}
\newcommand{\Point}{{\bf P}}

\newcommand{\CombinedDelta}{\kappa}

\newcommand{\tDual}{\tilde t}

\newcommand{\difOperator}[1]{\mathfrak{#1}}

\newcommand{\difOPE}{\mathcal{B}}

\newcommand{\ffixed}[1]{$f_n(#1,s,\bar s)$}

\DeclareMathOperator{\tr}{tr}

\newcommand{\beq}{\begin{equation}}
\newcommand{\eeq}{\end{equation}}

\newcommand{\algn}[1]{\begin{align}#1\end{align}}

\newcommand{\structgeneral}{
	\left[
	\begin{matrix}
	q_1 & q_2 & q_3 & q_4 \\
	\bar q_1 & \bar q_2 &\bar q_3 &\bar q_4	
	\end{matrix}
	\right]
}
\newcommand{\struct}[8]{
	\left[
	\begin{matrix}
	#1 & #2 & #3 & #4 \\
	#5 & #6 & #7 & #8	
	\end{matrix}
	\right]
}
\newcommand{\structgeneralz}{
	\left[
	\begin{matrix}
	q_1 & q_2 & q_3 & q_4 \\
	\bar q_1 & \bar q_2 &\bar q_3 &\bar q_4\\
	\hline
	z_1 & z_2 & z_3 & z_4	
	\end{matrix}
	\right]
}
\newcommand{\structgeneralconcretez}[4]{
	\left[
	\begin{matrix}
	q_1 & q_2 & q_3 & q_4 \\
	\bar q_1 & \bar q_2 &\bar q_3 &\bar q_4\\
	\hline
	#1 & #2 & #3 & #4
	\end{matrix}
	\right]
}

\newcommand{\conj}[1]{{\overline{#1}}}

\newcommand{\fref}[1]{{\color{gray} \texttt{[#1]}}}

\newcommand{\OO}{\mathcal{O}}
\newcommand{\FF}{\mathcal{F}}
\newcommand{\PP}{\mathcal{P}}
\newcommand{\TT}{\mathcal{T}}

\newcommand{\MM}{\mathcal{M}}
\renewcommand{\SS}{\mathcal{S}}
\newcommand{\<}{\langle}
\renewcommand{\>}{\rangle}

\newcommand{\II}{\mathbb{I}}
\newcommand{\JJ}{\mathbb{J}}
\newcommand{\KK}{\mathbb{K}}
\newcommand{\LL}{\mathbb{L}}

\title{\boldmath General Bootstrap Equations in 4D CFTs}

\author[a]{Gabriel Francisco Cuomo,}
\author[b]{Denis Karateev}
\author[c]{and Petr Kravchuk}

\affiliation[a]{Institute of Physics, EPFL, CH-1015 Lausanne, Switzerland}
\affiliation[b]{SISSA and INFN, Via Bonomea 265, I-34136 Trieste, Italy}
\affiliation[c]{Walter Burke Institute for Theoretical Physics, Caltech, Pasadena, California 91125, USA}

\preprint{CALT-TH 2017-23, SISSA 23/2017/FISI}

\abstract{
We provide a framework for generic 4D conformal bootstrap computations. It is based on the unification of two independent approaches, the covariant (embedding) formalism  and the non-covariant (conformal frame) formalism. We construct their main ingredients (tensor structures and differential operators) and establish a precise connection between them. We supplement the discussion by additional details like classification of tensor structures of $n$-point functions, normalization of 2-point functions and seed conformal blocks, Casimir differential operators and treatment of conserved operators and permutation symmetries. Finally, we implement our framework in a \texttt{Mathematica} package and make it freely available. 
}

\begin{document}

\maketitle

\section{Introduction}
In recent years a lot of progress has been made in understanding Conformal Field Theories (CFTs) in $d\geq 3$ dimensions using the conformal bootstrap approach~\cite{Mack:1969rr,Polyakov:1970xd,Ferrara:1973yt,Polyakov:1974gs,Mack:1975jr} (see~\cite{Simmons-Duffin:2016gjk,Rychkov:2016iqz} for recent introduction). 
In this paper we focus solely on $d=4$. The 4D conformal bootstrap allows to study fixed points of 4D quantum field theories relevant for describing elementary particles and fundamental interactions. It promises to address the QCD conformal window~\cite{Banks:1981nn} and may be useful for constraining the composite Higgs models, see~\cite{Caracciolo:2014cxa} for discussion.

In the conformal bootstrap approach CFTs are described by the local CFT data, which consists of scaling dimensions and Lorentz representations of local primary operators together with structure constants of the operator product expansion (OPE). The observables of the theory are correlation functions which are computed by maximally exploiting the conformal symmetry and the operator product expansion. Remarkably, the CFT data is heavily constrained by the associativity of the OPE, which manifests itself in the form of consistency equations called the crossing or the bootstrap equations.

The bootstrap equations constitute an infinite system of coupled non-linear equations for the CFT data. In a seminal work~\cite{Rattazzi:2008pe} it was shown how constraints on a finite subset of the OPE data can be extracted numerically from these equations.
In 4D the approach of~\cite{Rattazzi:2008pe} was further developed in~\cite{
Rychkov:2009ij,Rattazzi:2010gj,Poland:2010wg,Rattazzi:2010yc,Poland:2011ey,
Caracciolo:2014cxa,Caracciolo:2009bx,Vichi:2011ux,
Paulos:2014vya,Simmons-Duffin:2015qma,Behan:2016dtz,
El-Showk:2016mxr,ElShowk:2012hu,
Gliozzi:2013ysa}.
In 3D a major advance came with the numerical identification of the 3D Ising~\cite{ElShowk:2012ht,El-Showk:2014dwa} and the $O(N)$ models~\cite{Kos:2014bka,Kos:2013tga,Kos:2015mba,Kos:2016ysd}. An analytic approach to the bootstrap equations was proposed in~\cite{Komargodski:2012ek,Fitzpatrick:2012yx} and further developed in~\cite{
Alday:2015eya,Alday:2015ewa,Alday:2015ota,Alday:2016mxe,Alday:2016njk,Alday:2016jfr,
Kaviraj:2015cxa,Kaviraj:2015xsa,Dey:2016zbg,
Simmons-Duffin:2016wlq}. Other approaches include~\cite{Gliozzi:2013ysa,Gliozzi:2014jsa,Mazac:2016qev,Rychkov:2015naa,Gopakumar:2016cpb}. 

Most of these studies, however, focus on correlation functions of scalar operators, and thus only have access to the scaling dimensions of traceless symmetric operators and their OPE coefficients with a pair of scalars. In order to derive constraints on the most general elements of the CFT data, one has to consider more general correlation functions. To the best of our knowledge, the only published numerical studies of a 4-point function of non-scalar operators in non-supersymmetric theories up to date were done in 3D for a 4-point function of Majorana fermions~\cite{Iliesiu:2015qra,Iliesiu:2017nrv} and for a 4-point function of conserved abelian currents~\cite{Dymarsky:JJJJ}.

One reason for the lack of results on 4-point functions of spinning operators is that such correlators are rather hard to deal with. In order to set up the crossing equations for a spinning 4-point function, first, one needs to find a basis of its tensor structures and second, to compute all the relevant conformal blocks.
The difficulty of this task increases with the dimension $d$ due to an increasing complexity of the $d$-dimesnional Lorentz group. For instance, the representations of the 4D Lorentz group are already much richer than the ones in 3D.

The problem of constructing tensor structures has a long history \cite{Ferrara:1972xe,Sotkov:1976xe,mack1977,Osborn:1993cr,Weinberg:2010fx,Giombi:2011rz,Costa:2011mg,Weinberg:2012mz,Costa:2014rya,Iliesiu:2015qra}. In 4D all the 3-point tensor structures were obtained in~\cite{SimmonsDuffin:2012uy} and classified in~\cite{Elkhidir:2014woa} using the covariant embedding formalism approach. Unfortunately, in this approach 4- and higher-point tensor structures are hard to analyze due to a growing number of non-linear relations between the basic building blocks. This problem is alleviated in the conformal frame approach~\cite{Mack:1976pa,Osborn:1993cr,Kravchuk:2016qvl}. In~\cite{Kravchuk:2016qvl} a complete classification of general conformally invariant tensor structures was obtained in a non-covariant form.

The problem of computing the conformal blocks for scalar 4-point functinons was solved by a variety of methods in~\cite{Dolan:2000ut,Dolan:2003hv,Dolan:2011dv,ElShowk:2012ht,Hogervorst:2013sma,SimmonsDuffin:2012uy,Kos:2013tga,Hogervorst:2016hal}. Spinning conformal blocks were considered in~\cite{Costa:2011dw,SimmonsDuffin:2012uy,Costa:2014rya,Rejon-Barrera:2015bpa,Iliesiu:2015qra,Iliesiu:2015akf,Penedones:2015aga,Costa:2016xah,Costa:2016hju,Schomerus:2016epl}. Remarkably, in~\cite{Costa:2011dw} it was found that the Lorentz representations of external operators can be changed by means of differential operators. In 3D, this relates all bosonic conformal blocks to conformal blocks with external scalars. These results were extended to 3D fermions in~\cite{Iliesiu:2015qra,Iliesiu:2015akf} completing in principle the program of computing general conformal blocks in 3D.

Results of~\cite{Costa:2011dw} concerning traceless symmetric operators apply also to 4D, but are not sufficient even for the analysis of an  OPE of traceless symmetric operators since such an OPE also contains non-traceless symmetric operators. The first expression for a 4D spinning conformal block was obtained in~\cite{Rejon-Barrera:2015bpa} for the case of 2 scalars and 2 vectors.
A systematic study of conformal blocks in 4D with operators in arbitrary representations was done in~\cite{Echeverri:2015rwa}, where the results of~\cite{Costa:2011dw} were extended to reduce a general conformal block to a set of simpler conformal blocks called the seed blocks.
In the consequent work~\cite{Echeverri:2016dun} all the seed conformal blocks were computed.

\paragraph{The Goal of the Paper}
The results of~\cite{Elkhidir:2014woa,Echeverri:2015rwa,Echeverri:2016dun,Kravchuk:2016qvl} are in principle sufficient for formulating the bootstrap equations for arbitrary correlators in 4D. Nevertheless, due to a large amount of scattered non-trivial and missing ingredients there is still
a high barrier for performing 4D bootstrap computations. The goal of this paper is to describe all the ingredients needed for setting up the 4D bootstrap equations in a coherent manner using consistent conventions and to implement all these ingredients into a \texttt{Mathematica} package.

In particular, we first unify the results of~\cite{Elkhidir:2014woa,Echeverri:2015rwa,Echeverri:2016dun} with some extra developments and corrections.
We then use the conformal frame approach~\cite{Kravchuk:2016qvl} to solve the problem of constructing a complete basis of 4-point tensor structures in 4D in an extremely simple way.
We provide a precise connection between the embedding and the conformal frame approaches making possible an easy transition between two formalisms at any time.

We implement the formalism in a \texttt{Mathematica} package which allows one to work with 2-, 3- and 4-point functions and to construct arbitrary spin crossing equations in 4D CFTs. The package can be downloaded from \href{https://gitlab.com/bootstrapcollaboration/CFTs4D#cfts4d}{https://gitlab.com/bootstrapcollaboration/CFTs4D}.
Once it is installed one gets an access to a (hopefully) comprehensive documentation and examples. We also refer to the relevant functions from the package throughout the paper as \fref{function}.

\paragraph{Structure of the paper}
In the main body of the paper we describe the basic concepts applicable to the most generic correlators with no additional symmetries or conservation conditions. We comment on how these extra complications can be taken into account, and delegate a more detailed treatment to the appendices.

In section~\ref{sec:Outline} we outline the path to the explicit crossing equations for operators of general spin, abstracting from a specific implementation. In section~\ref{sec:EFA} we describe the implementation of the ideas from section~\ref{sec:Outline} in the embedding formalism. In section~\ref{sec:conformal_frame} we give an alternative implementation in the conformal frame formalism. Section~\ref{sec:example} is devoted to demonstration of the package in an elementary example of a correlator with one vector and three scalar operators. We conclude in section~\ref{sec:conclusions}.

Appendices~\ref{app:conventions4D} and~\ref{sec:DetailsOfTheEmbeddingFormalism} summarize our conventions in 4D Minkowski space and 6D embedding space, as well as cover the action of $\PP$- and $\TT$-symmetries. Appendix~\ref{sec:DetailsOfTheEmbeddingFormalism} also contains details of the embedding formalism. In appendix~\ref{app:twopointnorm} we give details on normalization conventions for 2-point functions and seed conformal blocks.  Appendices~\ref{sec:TensotInvariants} and~\ref{sec:BasisthreePointFunctions} contain details on explicitly covariant tensor structures. In appendix~\ref{app:casimir_differential_operators} we describe all 3 Casimir generators of the four-dimensional conformal group. Appendices~\ref{app:cons} and \ref{app:permutations} cover conservation conditions and permutation symmetries.

\section{Outline of the Framework}
\label{sec:Outline}
The local operators in 4D CFT are labeled by $(\ell,\bar \ell)$ representation of the Lorentz group $SO(1,3)$ and the scaling dimension $\Delta$.\footnote{In this paper we consider only the consequences of the conformal symmetry. In particular, we do not consider global (internal) symmetries because they commute with conformal trasformations and thus can be straightforwardly included. We also do not discuss supersymmetry.} In a CFT one can distinguish a special class of {\it primary} operators, the operators which transform homogeneously under conformal transformations~\cite{Mack:1969rr}. In a unitary CFT any local operator is either a primary or a derivative of a primary, in which case it is called a descendant operator.
A primary operator in representation $(\ell,\bar \ell)$ can be written as\footnote{Our conventions relevant for 3+1 dimensional Minkowski spacetime are summarized in appendix~\ref{app:conventions4D}.}
\beq
	\OO_{\alpha_1\ldots\alpha_\ell}^{\dot\beta_1\ldots\dot\beta_{\bar\ell}}(x),
\eeq
symmetric in spinor indices $\alpha_i$ and $\dot\beta_j$. Because of the symmetry in these indices, we can equivalently represent $\OO$ by a homogeneous polynomial in auxiliary spinors $s^\alpha$ and $\bar s_{\dot\beta}$ of degrees $\ell$ and $\bar\ell$ correspondingly
\beq
	\OO(x,s,\bar s)=s^{\alpha_1}\cdots s^{\alpha_\ell}\bar s_{\dot\beta_1}\cdots\bar s_{\dot\beta_{\bar\ell}}\OO_{\alpha_1\ldots\alpha_\ell}^{\dot\beta_1\ldots\dot\beta_{\bar\ell}}(x).
\eeq
We often call the auxiliary spinors $s$ and $\bar s$ the spinor polarizations.
The indices can be restored at any time by using
\begin{equation}
\OO_{\alpha_1\ldots\alpha_\ell}^{\dot\beta_1\ldots\dot\beta_{\bar\ell}}(x)=
\frac{1}{\ell!\,\bar\ell!}\,\prod_{i=1}^\ell\prod_{j=1}^{\bar\ell}
\frac{\partial}{\partial s^{\alpha_i}}\frac{\partial}{\partial \bar s_{\dot\beta_j}}\;\OO(x,s,\bar s).
\end{equation}
In principle the auxiliary spinors $s$ and $\bar s$ are independent quantities, however without loss of generality we can assume them to be complex conjugates of each other, $s_\alpha=\left(\bar s_{\dot\alpha}\right)^*$. This has the advantage that if $\OO$ with $\ell=\bar\ell$ is a Hermitian operator, e.g. for $\ell=\bar\ell=1$,
\beq
	\OO_{\alpha\dot\beta}(x)=\left(\OO_{\beta\dot\alpha}(x)\right)^\dagger,
\eeq
then so is $\OO(x,s,\bar s)$,
\beq
	\OO(x,s,\bar s)=\left(\OO(x,s,\bar s)\right)^\dagger.
\eeq
More generally for non-Hermitian operators we define
\beq\label{eq:conjugation}
	\conj\OO(x,s,\bar s)\equiv \left(\OO(x,s,\bar s)\right)^\dagger,
\eeq
see~\eqref{eq:conjugation_app} for the index-full version.

Conformal field theories possess an operator product expansion (OPE) with a finite radius of convergence~\cite{Mack:1976pa,mack1977,Pappadopulo:2012jk,Rychkov:2015lca}
\beq\label{eq:OPE}
	\OO_1(x_1,s_1,\bar s_1)\OO_2(x_2,s_2,\bar s_2)=\sum_{\OO}\sum_{a}\lambda^a_{\langle \OO_1\OO_2\conj{\OO}\rangle} \difOPE_a(\partial_{x_2},\partial_{s},\partial_{\bar s},\ldots)\OO(x_2,s,\bar s),
\eeq
where $\difOPE_a$ are differential operators in the indicated variables (depending also on $x_1-x_2,s_j,\bar s_j$, where $j=1,2$), which are fixed by the requirement of conformal invariance of the expansion.
Here $\lambda$'s are the OPE coefficients which are not constrained by the conformal symmetry. In general there can be several independent OPE coefficients for a given triple of primary operators, in which case we label them by an index $a$.

The OPE provides a way of reducing any $n$-point function to $2$-point functions, which have canonical form in a suitable basis of primary operators. Therefore, the set of scaling dimensions and Lorentz representations of local operators, together with the OPE coefficients, completely determines all correlation functions of local operators in conformally flat $\mathbb{R}^{1,3}$. For this reason we call this set of data the CFT data in what follows.\footnote{Besides the correlation functions of local operators one can consider extended operators, such as conformal defects, as well as the correlation functions on various non-trivial manifolds. In order to be able to compute these quantities one has to in general extend the notion of the CFT data.} The goal of the bootstrap approach is to constrain the CFT data by using the associativity of the OPE. In practice this is done by using the associativity inside of a 4-point correlation function, resulting in the crossing equations which can be analyzed numerically and/or analytically. In the remainder of this section we describe in detail the path which leads towards these equations.

\subsection{Correlation Functions of Local Operators}\label{sec:correlation_functions_outline}
We are interested in studying $n$-point correlation functions
\begin{equation}\label{eq:n_point_correlation_function}
f_n(\point_1\ldots\point_n)\equiv\langle 0|\mathcal{O}_{\Delta_1}^{(\ell_1,\bar\ell_1)}(\point_1)
\;\ldots\;
\mathcal{O}_{\Delta_n}^{(\ell_n,\bar\ell_n)}(\point_n)|0\rangle,
\end{equation}
where for convenience we defined a combined notation for dependence of operators on coordinates and auxiliary spinors
\begin{equation}
\point_i\equiv (x_i,s_i,\bar s_i).
\end{equation}
We have labeled the primary operators with their spins and scaling dimensions. In general these labels do not specify the operator uniquely (for example in the presence of global symmetries);
we ignore this subtlety for the sake of notational simplicity.
For our purposes it will be sufficient to assume that all operators are space-like separated (this includes all Euclidean configurations obtained by Wick rotation), and thus the ordering of the operators will be irrelevant up to signs coming from permutations of fermionic operators.

The conformal invariance of the system puts strong constraints on the form of~(\ref{eq:n_point_correlation_function}). By inserting an identity operator ${\bf 1}=U U^\dagger$, where $U$ is the unitary operator implementing a generic conformal transformation, inside this correlator and demanding the vacuum to be invariant $U|0\rangle=0$, one arrives at the constraint
\begin{equation}\label{eq:n_point_correlation_function_constraint}
\langle 0| \big(U^\dagger\mathcal{O}_{\Delta_1}^{(\ell_1,\bar\ell_1)}U\big)
\ldots \big(U^\dagger\mathcal{O}_{\Delta_n}^{(\ell_n,\bar\ell_n)}U\big)|0\rangle=
\langle 0| \mathcal{O}_{\Delta_1}^{(\ell_1,\bar\ell_1)}\ldots \mathcal{O}_{\Delta_n}^{(\ell_n,\bar\ell_n)}|0 \rangle.
\end{equation} 
The algebra of infinitesimal conformal transformations, as well as their action on the primary operators are summarized in our conventions in appendix~\ref{app:conventions4D}.

The general solution to the above constraint has the following form,
\begin{equation}\label{eq:correlation_function_structure}
f_n(x_i,s_i,\bar s_i)=\sum_{I=1}^{N_n}g_n^I(\mathbf{u})\;\mathbb{T}_n^I(x_i,s_i,\bar s_i),
\end{equation}
where $\mathbb{T}_n^I$ are the conformally-invariant tensor structures which are fixed by the conformal symmetry up to a $\mathbf{u}$-dependent change of basis, and $\mathbf{u}$ are cross-ratios which are the scalar conformally-invariant combinations of the coordinates $x_i$. The structures $\mathbb{T}_n^I$ and their number $N_n$ depend non-trivially on the $SO(1,3)$ representations of $\OO_i$, but rather simply on $\Delta_i$, so we can write 
\beq\label{eq:split_4D}
	\mathbb{T}_n^I(x_i,s_i,\bar s_i)=\mathcal{K}_n(x_i)\hat{\mathbb{T}}_n^I(x_i,s_i,\bar s_i),
\eeq
where all $\Delta_i$-dependence is in the ``kinematic'' factor $\mathcal{K}_n$\footnote{This does not uniquely fix the factorization, and we will make a choice based on convenience later.} and all the the $\Delta_i$ enter $\mathcal{K}_n$ through the quantity
\begin{equation}\label{eq:scaling6D}
\CombinedDelta\equiv\Delta+\frac{\ell+\bar\ell}{2}.
\end{equation}
Note that ${\mathbb{T}}$ and $\hat{\mathbb{T}}$ are homogeneous polynomials in the auxiliary spinors, schematically,
\begin{equation}\label{eq:correlator_ss}
\mathbb{T}_n^I,\hat{\mathbb{T}}_n^I\sim\prod_{i=1}^{n} s_i^{\ell_i}\bar s_i^{\bar\ell_i}.
\end{equation}

In the rest of this subsection we give an overview of the structure of $n$-point correlation functions for various $n$, emphasizing the features specific to 4D.

\paragraph{2-point functions}
A 2-point function can be non-zero only if it involves two operators in complex-conjugate representations, $(\ell_1,\bar\ell_1)=(\bar\ell_2,\ell_2)$, and with equal scaling dimensions, $\Delta_1=\Delta_2$. In fact, it is always possible to choose a basis for the primary operators so that the only non-zero 2-point functions are between Hermitian-conjugate pairs of operators. We always assume such a choice.

The general 2-point function \fref{n2CorrelationFunction} then has an extremely simple form given by
\begin{equation}\label{eq:2PointFunction}
\langle \conj\OO_{\Delta}^{(\bar\ell,\ell)}(\point_1)
\OO_{\Delta}^{(\ell,\bar\ell)}(\point_2)\rangle=
c_{\langle\conj\OO\OO\rangle}\,
\underbrace{x_{12}^{-2\,\CombinedDelta_1}}_{=\mathcal{K}_2}
\underbrace{\big[\hat\II^{12}\big]^{\ell}\big[\hat\II^{21}\big]^{\bar\ell}}_{=\hat{\mathbb{T}}_2},
\end{equation}
where $c_{\langle\conj\OO\OO\rangle}$ is a constant. There is a single tensor structure $\hat{\mathbb{T}}_2$, and the building blocks $\hat{\mathbb{I}}^{ij}$ are defined in appendix~\ref{sec:TensotInvariants}. Changing the normalization of $\OO$ one can rescale the coefficient $c_{\langle\conj\OO\OO\rangle}$ by a positive factor. The phase is fixed by the requirement of unitarity, see appendix~\ref{app:twopointnorm}. We can make the following choice
\begin{equation}\label{eq:normalization_2_point_function}
c_{\langle\conj\OO\OO\rangle}=i^{\ell-\bar\ell},\quad
c_{\langle\OO\conj\OO\rangle}=(-)^{\ell-\bar\ell}c_{\langle\conj\OO\OO\rangle}=i^{\bar\ell-\ell},
\end{equation}
where the factor $(-)^{\ell-\bar\ell}$ appears due to the spin statistics theorem.

\paragraph{3-point functions}
A generic form of a 3-point function \fref{n3ListStructures,\\ n3ListStructuresAlternativeTS} is given by\footnote{For notational convenience we use lowercase index $a$ instead of capital index $I$ to label the 3-point tensor structures.}
\begin{equation}\label{eq:3_point_function}
\langle 
\OO_{\Delta_1}^{(\ell_1,\bar\ell_1)}(\point_1)
\OO_{\Delta_2}^{(\ell_2,\bar\ell_2)}(\point_2)
\OO_{\Delta_3}^{(\ell_3,\bar\ell_3)}(\point_3)
\rangle=\mathcal{K}_3\sum_{a=1}^{N_3}
\lambda^a_{\langle\OO_1\OO_2\OO_3\rangle} \;\hat{\mathbb{T}}^a_3,
\end{equation}
where the kinematic factor \fref{n3KinematicFactor} is given by
\begin{equation}\label{eq:kinematic_factor_n=3}
\mathcal{K}_3=
\prod_{i<j} |x_{ij}|^
{
-\CombinedDelta_i-\CombinedDelta_j+\CombinedDelta_k
}.
\end{equation}
The necessary and sufficient condition for the 3-point tensor structures $\hat{\mathbb{T}}^a_3$ to exist is that the 3-point function contains an even number of fermions and the following inequalities hold,
\begin{equation}\label{eq:sufficient_condition_non_zero_3_point_function}
\quad |\ell_i-\bar\ell_i|\leq \ell_j+\bar\ell_j+\ell_k+\bar\ell_k,\quad \text{for all distinct }i,j,k.
\end{equation}
A general discussion on how to construct a basis of tensor structures $\hat{\mathbb{T}}^a_3$ is given in section~\ref{sec:EFA}. For convenience we summarize this construction for 3-point functions in appendix~\ref{sec:BasisthreePointFunctions}.

The fact that the OPE coefficients enter 3-point functions follows simply from using the OPE~\eqref{eq:OPE} and the form of~\eqref{eq:2PointFunction} in the left hand side of \eqref{eq:3_point_function}. It is also clear that one can always choose the bases for $\difOPE_a$ and $\hat{\mathbb{T}}_3^a$ to be compatible.

There is a number of relations the OPE coefficients $\lambda^a_{\<\OO_1\OO_2\OO_3\>}$ have to satisfy. 
The simplest one comes from applying complex conjugation to both sides of~\eqref{eq:3_point_function}.
On the left hand side one has
\begin{equation}
\langle\OO_1\OO_2\OO_3\rangle^* = \langle\conj\OO_3\conj\OO_2\conj\OO_1\rangle.
\end{equation}
Using the properties of tensor structures under conjugation summarized in appendix~\ref{sec:TensotInvariants} one obtains a relation of the form
\begin{equation}
\left(\lambda^a_{\<\OO_1\OO_2\OO_3\>}\right)^*=C^{ab}\,\lambda^b_{\<\conj\OO_3\conj\OO_2\conj\OO_1\>},
\end{equation}
where the matrix $C^{ab}$ is often diagonal with $\pm 1$ entries.
Other constraints arise from the possible $\PP$- and $\TT$-symmetries (see appendix~\ref{app:conventions4D}), conservation equations (see appendix~\ref{app:cons}), and permutation symmetries (see appendix~\ref{app:permutations}). Importantly all these conditions give linear equations for $\lambda$'s, which can be solved in terms of an independent set of real quantities $\hat\lambda$ as
\begin{equation}\label{eq:algebraic_constraints}
\lambda^a_{\langle \OO_1 \OO_2 \OO_3\rangle}=\sum_{\hat a=1}^{\hat N_3} P^{\,a\,\hat{a}}_{\langle \OO_1 \OO_2 \OO_3\rangle}\hat{\lambda}^{\hat{a}}_{\langle \OO_1 \OO_2 \OO_3\rangle},\quad \hat N_3 < N_3.
\end{equation}

It will be important for the calculation of conformal blocks that we can actually construct all the tensor structures $\mathbb{T}^a_3$ in~(\ref{eq:3_point_function}) by considering a simpler 3-point function with two out of three operators having canonical spins $(\ell'_1,\bar\ell'_1)$ and $(\ell'_2,\bar\ell'_2)$, chosen in a way such that the 3-point function has a single tensor structure
\begin{equation}
\langle 
\OO_{\Delta'_1}^{(\ell'_1,\bar\ell'_1)}
\OO_{\Delta'_2}^{(\ell'_2,\bar\ell'_2)}
\OO_{\Delta_3}^{(\ell_3,\bar\ell_3)}
\rangle=
\lambda\; \mathbb{T}_{seed}.
\end{equation}
A simple choice is to set as many spin labels to zero as possible, for example
\begin{equation}\label{eq:choosing_seed_3_point_functions}
\ell_1'=\bar\ell_1'=\ell_2'=0,
\quad
\bar\ell_2'=|\ell_3-\bar\ell_3|.
\end{equation}
As we review in section~\ref{sec:spinning_differential_operators} one can then construct a set of differential operators $\mathbb{D}^a$ 
acting on the coordinates and polarization spinors of the first two operators such that
\begin{equation}\label{eq:tensor_structures_via_differential_operators}
\mathbb{T}^a_3=\mathbb{D}^a\, \mathbb{T}_{seed}.
\end{equation}

We will call the canonical tensor structure $\mathbb{T}_{seed}$ a seed tensor structure in what follows. Our choice of seed structures is described in appendix~\ref{app:twopointnorm}. When the third field is traceless symmetric, one has obviously $\bar\ell_2'=0$, thus relating a pair of generic operators to a pair of scalars~\cite{Costa:2011dw}.

\paragraph{4-point functions and beyond}
In the case $n=4$ one has
\begin{equation}\label{eq:4_point_function}
\langle 
\OO_{\Delta_1}^{(\ell_1,\bar\ell_1)}(\point_1)
\OO_{\Delta_2}^{(\ell_2,\bar\ell_2)}(\point_2)
\OO_{\Delta_3}^{(\ell_3,\bar\ell_3)}(\point_3)
\OO_{\Delta_4}^{(\ell_3,\bar\ell_4)}(\point_4)
\rangle=%\mathcal{K}_4
\sum_{I=1}^{N_4}
g_4^I(u,v)\;\mathbb{T}^I_4,
\end{equation}
where $g_4^I(u,v)$ are not fixed by conformal symmetry and are functions of the 2 conformally invariant cross-ratios \fref{formCrossRatios} 
\beq\label{eq:cross_ratios_uv}
u=\frac{x_{12}^2x_{34}^2}{x_{13}^2x_{24}^2},
\quad
v=\frac{x_{14}^2x_{23}^2}{x_{13}^2x_{24}^2}.
\eeq
In most of the applications it will be more convenient to use another set of variables $(z,\bar z)$ \fref{changeVariables} defined as
\begin{equation}
u=z\bar z,\;\;
v=(1-z)(1-\bar z).
\end{equation}

We classify and construct all the 4-point tensor structures $\mathbb{T}_4$ \fref{n4ListStructures, n4ListStructuresEF} in section~\ref{sec:conformal_frame}. Following the literature we choose the kinematic factor \fref{n4KinematicFactor} of the form\footnote{In section~\ref{sec:conformal_frame} we never separate the kinematic factor which has an extremely simple form $\left( z\bar z\right)^{-\frac{\CombinedDelta_1+\CombinedDelta_2}{2}}$ in the conformal frame.}
\begin{equation}\label{eq:kinematic_factor_n=4}
\mathcal K_4=
\left(\frac{x_{24}}{x_{14}}\right)^{\CombinedDelta_1-\CombinedDelta_2}
\left(\frac{x_{14}}{x_{13}}\right)^{\CombinedDelta_3-\CombinedDelta_4}
\times
\frac{1}{
x_{12}^{\CombinedDelta_1+\CombinedDelta_2}
x_{34}^{\CombinedDelta_3+\CombinedDelta_4}
}.
\end{equation}

The case of $n\geq 5$ point functions is similar to the $n=4$ case with a difference that the number of conformally invariant cross-ratios is $4n-15$. We briefly discuss the classification of tensor structures for higher-point functions in section~\ref{sec:conformal_frame}.

In general 4- and higher-point functions are subject to the same sort of conditions as 3-point functions. Reality conditions and implications of $\PP$- and $\TT$-symmetries are not conceptually different from the 3-point case. However, implications of permutation symmetries and conservation equations are more involved than those for 3-point functions, see~\cite{Dymarsky:2013wla}, due to the existence of non-trivial conformal cross-ratios~\eqref{eq:cross_ratios_uv}. See also appendices~\ref{app:permutations} and~\ref{app:cons} for details.

\subsection{Decomposition in Conformal Partial Waves}\label{sec:decomposition_in_CPW}
Since the OPE data determines all the correlation functions, the functions $g_4^I(u,v)$ entering \eqref{eq:4_point_function} can also be computed. To compute $g_4^I(u,v)$ we use the s-channel OPE, namely the OPE in pairs $\OO_1\OO_2$ and $\OO_3\OO_4$. One way to do this is to insert a complete orthonormal set of states in the correlator
\beq\label{eq:s-channel_OPE}
f_4\underset{s-OPE}{=}
\contraction{\langle}{\OO_1}{}{\OO_2}
\contraction{\langle\OO_1\OO_2}{\OO_3}{}{\OO_4}
\<\OO_1\OO_2\OO_3\OO_4\>=
		\sum_{|\Psi\>}\<\OO_1\OO_2\vert\Psi\>
	\<\Psi\vert\OO_3\OO_4\>.
\eeq
By virtue of the operator-state correspondence, see for example~\cite{Simmons-Duffin:2016gjk,Rychkov:2016iqz}, the states $|\Psi\>$ are in one-to-one correspondence with the local primary operators $\OO$ and their descendants $\partial^n\OO$. This allows us to express the inner products above in terms of the 3-point functions $\<\OO_1\OO_2\OO\>$ and $\<\conj\OO\OO_3\OO_4\>$ with the primary operator $\OO$ and its conjugate $\conj\OO$,
resulting in the following $s$-channel conformal partial wave decomposition
\beq\label{eq:DefinitonCPWs}
	\contraction{\langle}{\OO_1}{}{\OO_2}
\contraction{\langle\OO_1\OO_2}{\OO_3}{}{\OO_4}
\<\OO_1\OO_2\OO_3\OO_4\>=\sum_\OO \sum_{a,b}
	\lambda^a_{\<\OO_1\OO_2\OO\>}W_{\<\OO_1\OO_2\OO\>
	\<\conj\OO\OO_3\OO_4\>}^{ab}\lambda^b_{\<\conj\OO\OO_3\OO_4\>}.
\eeq
The objects $W^{ab}$ are called the Conformal Partial Waves (CPWs). The summation in~\eqref{eq:DefinitonCPWs} is over all primary operators $\OO$ which appear in both 3-point functions $\<\OO_1\OO_2\OO\>$ and $\<\conj\OO\OO_3\OO_4\>$ and
we can write explicitly
\begin{equation}
\sum_{\OO}=
\sum_{|\ell-\bar\ell|=0}^{\infty}\;
\sum_{\ell=0}^{\infty}\;
\sum_{\Delta,i},
\end{equation}
where $i$ labels the possible degeneracy of operators at fixed spin and scaling dimensions (coming, for example, from a global symmetry).
Note that according to properties of 3-point functions~\eqref{eq:sufficient_condition_non_zero_3_point_function}, there is a natural upper cut-off in the first summation
\begin{equation}
\sum_{|\ell-\bar\ell|=0}^{\infty}
=
\sum_{|\ell-\bar\ell|=0}^{|\ell-\bar\ell|_{max}},
\end{equation}
where
\begin{equation}\label{eq:maxl-l}
|\ell-\bar\ell|_{max}=\min(
\ell_1+\bar\ell_1+\ell_2+\bar\ell_2,\;
\ell_3+\bar\ell_3+\ell_4+\bar\ell_4).
\end{equation}
Furthermore, if the operator $\OO$ is bosonic then $|\ell-\bar\ell|$ assumes only even values; if the operator $\OO$ is fermionic $|\ell-\bar\ell|$ assumes only odd values.
The CPWs can be further rewritten in terms of Conformal Blocks (CB) and tensor structures as
\begin{equation}\label{eq:DefinitionOfConformalBlocks}
W_{\<\OO_1\OO_2\OO\>
	\<\conj\OO\OO_3\OO_4\>}^{ab}=\sum_{I=1}^{N_4}
	G_{\<\OO_1\OO_2\OO\>
	\<\conj\OO\OO_3\OO_4\>}^{I,ab}(u,v)\;\mathbb{T}_4^I,
\end{equation}
inducing the conformal block expansion for $g^I_4$
\beq\label{eq:Computed_g_Function}
	g^I_4(u,v)\underset{s-OPE}{=}\sum_\OO\sum_{a,b} \lambda^a_{\<\OO_1\OO_2\OO\>}G_{\<\OO_1\OO_2\OO\>
	\<\conj\OO\OO_3\OO_4\>}^{I,ab}(u,v)\lambda^b_{\<\conj\OO\OO_3\OO_4\>}.
\eeq

\paragraph{Computation of Conformal Partial Waves}
The computation of CPWs is rather difficult. Luckily there is a way of reducing them to simpler objects called the seed CPWs by means of differential operators \cite{Costa:2011dw,Echeverri:2015rwa}. 

For example, the s-channel CPW appearing due to the exchange of a generic operator
\begin{equation}\label{eq:definition_of_p}
O^{(\ell,\bar\ell)}_\Delta,\quad p\equiv|\ell-\bar\ell|
\end{equation}
by using~\eqref{eq:tensor_structures_via_differential_operators} can be written as
\begin{equation}\label{eq:D_on_seeds}
W^{ab}_{\langle \OO_1 \OO_2 \OO\rangle\langle \conj\OO \OO_3 \OO_4\rangle}=
\mathbb{D}^a_{\langle \OO_1 \OO_2 \OO\rangle}
\mathbb{D}^b_{\langle \conj\OO \OO_3 \OO_4\rangle}
W_{\langle \mathcal F_1^{(0,0)} \mathcal F_2^{(p,0)} \OO\rangle\langle \conj\OO \mathcal F_3^{(0,0)} \mathcal F_4^{(0,p)}\rangle}^{seed},
\end{equation}
where $\FF_i$ are the operators with the same 4D scaling dimensions $\Delta_i$ as $\OO_i$, see section~\ref{sec:spinning_differential_operators}.
The seed CPWs are defined as the s-channel contribution of~\eqref{eq:definition_of_p} to the seed 4-point function 
\begin{equation}\label{eq:seed_four_point_function}
\langle
\FF_1^{(0,0)}\,
\FF_2^{(p,0)}\,
\FF_3^{(0,0)}\,
\FF_4^{(0,p)}
\rangle.
\end{equation}
An important property of the seed 4-point function~(\ref{eq:seed_four_point_function}) is that it has only $p+1$ tensor structures. We will distinguish two dual types of seed CPWs, following the convention of~\cite{Echeverri:2016dun},
\begin{align}
W_{seed}^{(p)} &\equiv
W_{\langle \mathcal F_1^{(0,0)} \mathcal F_2^{(p,0)} \OO\rangle\langle \conj\OO \mathcal F_3^{(0,0)}\mathcal F_4^{(0,p)}\rangle}^{seed},
\quad if\; \ell-\bar\ell\leq 0,\\
W_{dual\;seed}^{(p)} &\equiv
W_{\langle \FF_1^{(0,0)} \FF_2^{(p,0)} \OO\rangle\langle \conj\OO \FF_3^{(0,0)} \FF_4^{(0,p)}\rangle}^{seed},
\quad if\; \ell-\bar\ell\geq 0.
\end{align}
The case $W_{seed}^{(0)}=W_{dual\;seed}^{(0)}$ reproduces the classical scalar conformal block found by Dolan and Osborn~\cite{Dolan:2000ut,Dolan:2003hv}. The seed CPWs \fref{seedCPW} can be written in terms of a set of seed Conformal Blocks $H_{e}^{(p)}(z,\bar z)$ and $\overline H_{e}^{(p)}(z,\bar z)$ as\footnote{The factors $(-2)^{p-e}$ are introduced here to match the original work~\cite{Echeverri:2016dun}.}
\begin{align}\label{eq:seedCPW}
W_{seed}^{(p)} &=\mathcal{K}_4\,\sum_{e=0}^p (-2)^{p-e}\,H_{e}^{(p)}(z,\bar z) \big[\hat \II^{42}\big]^e \big[\hat \II^{42}_{31}\big]^{p-e},\\\label{eq:seedCPWdual}
W_{dual\;seed}^{(p)} &=\mathcal{K}_4\,\sum_{e=0}^p (-2)^{p-e}\,\overline H_{e}^{(p)}(z,\bar z) \big[\hat \II^{42}\big]^e \big[\hat \II^{42}_{31}\big]^{p-e},
\end{align}
where the tensor structures are defined in appendix~\ref{sec:TensotInvariants}.

The seed Conformal Blocks $H_{e}^{(p)}(z,\bar z)$ and $\overline H_{e}^{(p)}(z,\bar z)$ were found\footnote{Notice slight change of notation $H_{here}(z,\bar z)\equiv G_{there}(z,\bar z)$. This change is needed to distinguish $H_{here}(z,\bar z)=G_{here}(u(z,\bar z),\,v(z,\bar z)).$} \fref{plugSeedBlocks, plugDualSeedBlocks} analytically in~(5.36) and~(5.37) in~\cite{Echeverri:2016dun} up to an overall normalization factors, denoted there by $c_{0,-p}^p$ and $\bar c_{0,-p}^p$.
Given the choice of seed 3-point tensor structures \eqref{eq:seedleft}-\eqref{eq:seedrightdual} and normalization of 2-point functions~\eqref{eq:normalization_2_point_function}, we can fix these factors as
\begin{equation}\label{eq:normalization_seeds}
c_{0,-p}^p=(-1)^\ell\;{i}^p
\quad\text{and}\quad
\bar c_{0,-p}^p=2^{-p}\;(-1)^\ell\;{i}^p,
\end{equation}
see appendix~\ref{app:twopointnorm} for details.
Other relevant functions are \fref{plugCoefficients,
plugKFunctions, reduceKFunctionDerivatives, plugPolynomialsPQ}.

\paragraph{The Casimir Equation}
A very important property of the CPWs is that they satisfy the conformal Casimir eigenvalue equations \cite{Dolan:2003hv,Dolan:2011dv}\footnote{DK thanks Hugh Osborn for useful discussion on this topic.} which have the form
\begin{equation}\label{eq:casimir_equations}
\Big(\difOperator C_n - E_n\Big)\,
W^{ab}_{\langle \OO_1 \OO_2 \OO\rangle\langle \conj \OO \OO_3 \OO_4\rangle}
=0,
\end{equation}
where $n=2,3,4$ and $\difOperator C_2$, $\difOperator C_3$ and $\difOperator C_4$ are the quadratic, cubic and quartic Casimir differential operators respectively \fref{opCasimir$n$EF, opCasimir24D}. They are defined
in appendix~\ref{app:casimir_differential_operators} together with their eigenvalues \fref{casimirEigenvalue$n$}, where the conformal generators $\difOperator L_{MN}$ given in appendix~\ref{sec:DetailsOfTheEmbeddingFormalism} are taken to act on 2 different points
\begin{equation}
\difOperator L_{MN}=\difOperator L_{i\,MN}+\difOperator L_{j\,MN},
\end{equation}
with $(ij)=(12)$ or $(ij)=(34)$ corresponding to the s-channel CPWs\footnote{Notice that the eigenvalue of $\difOperator C_3$ taken at $(ij)=(34)$ will differ by a minus sign from the eigenvalue of $\difOperator C_3$ taken at $(ij)=(12)$.}.

The $n=2$ Casimir equation was used in~\cite{Echeverri:2016dun} for constructing the seed CPWs. Given that the seed CPWs are already known, in practice the Casimir equations can be used to validate the more general CPWs computed using the prescription above.

\paragraph{Conserved and Identical Operators, $\PP-$ and $\TT-$symmetries}

As noted in section~\ref{sec:correlation_functions_outline}, in general there are various constraints imposed on 3- and 4-point functions, such as reality conditions, permutation symmetries, conservation, and $\PP-$ and $\TT-$ symmetries. Recall that the most general CPW decomposition is given by~\eqref{eq:Computed_g_Function},
\beq\label{eq:Computed_g_Function_Duplicate}
	g^I_4(u,v)\underset{s-OPE}{=}\sum_\OO\sum_{a,b} \lambda^a_{\<\OO_1\OO_2\OO\>}G_{\<\OO_1\OO_2\OO\>
	\<\conj\OO\OO_3\OO_4\>}^{I,ab}(u,v)\lambda^b_{\<\conj\OO\OO_3\OO_4\>}.
\eeq

According to the discussion around \eqref{eq:algebraic_constraints}, the general solution to these constraints relevant for this expansion is
\begin{equation}\label{eq:n=4ConstraintsOPEa}
\lambda^a_{\langle \OO_1 \OO_2 \OO\rangle}=\sum_{\hat{a}}P^{\,a\,\hat{a}}_{\langle \OO_1 \OO_2 \OO\rangle}\hat{\lambda}^{\hat{a}}_{\langle \OO_1 \OO_2 \OO\rangle}
\quad{\text{and}}\quad
\lambda^b_{\langle \conj \OO \OO_3 \OO_4 \rangle}=\sum_{\hat{b}}P^{\,b\,\hat{b}}_{\langle \conj \OO \OO_3 \OO_4 \rangle}\hat{\lambda}^{\hat{b}}_{\langle \conj \OO \OO_3 \OO_4 \rangle}.
\end{equation}
Besides that,
if the pair of operators $\OO_1$ and $\OO_2$ is the same as the pair of operatirs $\OO_3$ and $\OO_4$,
there has to exist relations of the form
\begin{equation}\label{eq:n=4ConstraintsOPEb}
\lambda^b_{\langle \conj \OO \OO_3 \OO_4 \rangle}=\sum_{b}N^{\,b\,c}_{\langle \conj \OO \OO_3 \OO_4 \rangle}\lambda^{c}_{\langle \OO_1 \OO_2 \conj \OO  \rangle}.
\end{equation}

Once the relations~\eqref{eq:n=4ConstraintsOPEa} and~\eqref{eq:n=4ConstraintsOPEb} are inserted in the general expression~\eqref{eq:Computed_g_Function_Duplicate}, the resulting 4-point function will satisfy all the required constraints which preserve the $s$-channel.\footnote{Possible constraints which do not preserve $s$-channel are permutations of the form $(13)$, etc. Such permutations, if present, are equivalent to the crossing equations discussed below.} In particular, the ``reduced'' CPWs corresponding to the coefficients $\hat\lambda$ will also satisfy these constraints automatically. Note that by construction the reduced CPWs are just the linear combinations of the generic CPWs.

\subsection{The Bootstrap Equations}
\label{sec:BootstrapEquations}
The conformal bootstrap equations are the equations which must be satisfied by the consistent CFT data. They arise as follows. The s-channel OPE~(\ref{eq:s-channel_OPE}) is not the only option to compute 4-point functions, there are in fact two other possibilities. One can use the t-channel OPE expansion
\begin{multline}\label{eq:t-channel_OPE}
f_4\underset{t-OPE}{=}
\contraction{\langle}{\OO}{_1(\point_1) \OO_2(\point_2) \OO_3(\point_3)}{\OO}
\contraction[2ex]{\langle \OO_1(\point_1)}{\OO}{_2(\point_2)}{\OO}
\langle \OO_1(\point_1) \OO_2(\point_2) \OO_3(\point_3) \OO_4(\point_4) \rangle=\\
\contraction{\pm\langle}{\OO}{_1(\point_1)}{\OO}
\contraction{\pm\<\OO_1(\point_1) \OO_2(\point_2)}{\OO}{_3(\point_3)}{\OO}
\pm\<\OO_3(\point_1) \OO_2(\point_2) \OO_1(\point_3) \OO_4(\point_4)\>\Big|_{\point_1\leftrightarrow \point_3}=
\contraction{\pm\langle}{\OO}{_1(\point_1)}{\OO}
\contraction{\pm\<\OO_1(\point_1) \OO_2(\point_2)}{\OO}{_3(\point_3)}{\OO}
\pm\<\OO_1(\point_1) \OO_4(\point_2) \OO_3(\point_3) \OO_2(\point_4)\>\Big|_{\point_2\leftrightarrow \point_4}
\end{multline}
or the u-channel OPE expansion
\begin{multline}\label{eq:u-channel_OPE}
f_4\underset{u-OPE}{=}
\contraction{\langle}{\OO}{_1(\point_1) \OO_2(\point_2)}{\OO}
\contraction[2ex]{\langle \OO_1(\point_1)}{\OO}{_2(\point_2) \OO_3(\point_3)}{\OO}
\langle \OO_1(\point_1) \OO_2(\point_2) \OO_3(\point_3) \OO_4(\point_4) \rangle=\\
\contraction{\pm\langle}{\OO}{_1(\point_1)}{\OO}
\contraction{\pm\<\OO_1(\point_1) \OO_2(\point_2)}{\OO}{_3(\point_3)}{\OO}
\pm\<\OO_4(\point_1) \OO_2(\point_2) \OO_3(\point_3) \OO_1(\point_4)\>\Big|_{\point_1\leftrightarrow \point_4}=
\contraction{\pm\langle}{\OO}{_1(\point_1)}{\OO}
\contraction{\pm\<\OO_1(\point_1) \OO_2(\point_2)}{\OO}{_3(\point_3)}{\OO}
\pm\<\OO_1(\point_1) \OO_3(\point_2) \OO_2(\point_3) \OO_4(\point_4)\>\Big|_{\point_2\leftrightarrow \point_3}.
\end{multline}
In the above relations we permuted operators in the second and third equalities to get back the s-channel configuration.  Minus signs are inserted for odd permutation of fermion operators.

In a consistent CFT the function $f_4$ is unique and does not depend on the channel used to computation it, leading to the requirement that the expressions~(\ref{eq:s-channel_OPE}), (\ref{eq:t-channel_OPE}) and (\ref{eq:u-channel_OPE}) must be equal. These equalities are the bootstrap equations. 
To be concrete we write the $s$-$t$ consistency equation using~(\ref{eq:DefinitonCPWs}) and~(\ref{eq:t-channel_OPE})
\begin{align}
f_4 &\underset{s-OPE}{=}\sum_\OO 
	\lambda^a_{\<\OO_1\OO_2\OO\>}W_{\<\OO_1\OO_2\OO\>
	\<\conj\OO\OO_3\OO_4\>}^{ab}\lambda^b_{\<\conj\OO\OO_3\OO_4\>},\\
f_4 &\underset{t-OPE}{=}\pm
	\sum_\OO 
	\lambda^a_{\<\OO_3\OO_2\OO\>}W_{\<\OO_3\OO_2\OO\>
	\<\conj\OO\OO_1\OO_4\>}^{ab}\lambda^b_{\<\conj\OO\OO_1\OO_4\>}\Bigg|_{\point_1\leftrightarrow \point_3}.
\end{align}
In this example the tensor structures $\hat{\mathbb{T}}_n^I$ transform under permutation of points $\point_i\leftrightarrow \point_j$ as
\begin{equation}\label{eq:PermutationsInTensorStructures}
\hat{\mathbb{T}}_{\<\OO_3\OO_2\OO_1\OO_4\>}^I
\Big|_{\point_1\leftrightarrow \point_3}=M_{\point_1\leftrightarrow \point_3}^{IJ}\;\hat{\mathbb{T}}_{\<\OO_1\OO_2\OO_3\OO_4\>}^J,
\end{equation}
since they form a basis.
Further decomposing these expressions using the basis of tensor structures one can compute the unknown $g_4^I(z,\bar z)$ 
\begin{align}\label{eq:BootstrapEq1}
	g^I_4(z,\bar z) &\underset{s-OPE}{=}\sum_\OO\sum_{a,b} \lambda^a_{\<\OO_1\OO_2\OO\>}G_{\<\OO_1\OO_2\OO\>
	\<\conj\OO\OO_3\OO_4\>}^{I,ab}(z,\bar z)\lambda^b_{\<\conj\OO\OO_3\OO_4\>},\\\label{eq:BootstrapEq2}
	g^I_4(z,\bar z) &\underset{t-OPE}{=}\pm\,M_{\point_1\leftrightarrow\point_3}^{IJ}\,\sum_\OO\sum_{a,b} \lambda^a_{\<\OO_3\OO_2\OO\>}G_{\<\OO_3\OO_2\OO\>
	\<\conj\OO\OO_1\OO_4\>}^{J,ab}(1-z,1-\bar z)\lambda^b_{\<\conj\OO\OO_1\OO_4\>}.
\end{align}
Equating~\eqref{eq:BootstrapEq1} and~\eqref{eq:BootstrapEq2} we get $N_4$ independent equations. In a presence of additional constraints discussed in appendices~\ref{app:conventions4D},~\ref{app:cons} and~\ref{app:permutations}, not all the $N_4$ equations are independent, and one should chose only those equations which correspond to the independent degrees of freedom. In the conventional numerical approach to conformal bootstrap, when Taylor expanding the crossing equations around $z=\bar z=1/2$, one should also be careful to understand which Taylor coefficients are truly independent. Among other things, this depends on the analyticity properties of tensor structures $\mathbb{T}_4$, see appendix~A of \cite{Kravchuk:2016qvl} for a discussion.

\section{Embedding Formalism}
\label{sec:EFA}
This section is meant to be a summary and a review of the embedding formalism (EF)~\cite{Dirac:1936fq,Weinberg:2010fx,Costa:2011mg,SimmonsDuffin:2012uy} approach to 4D correlators. The discussion is based on the works~\cite{Elkhidir:2014woa,Echeverri:2015rwa} with some developments and corrections.

The key observation is that the 4D conformal group is isomorphic to $SO(4,2)$, the linear Lorentz group in 6D. It is then convenient to \emph{embed} the 4D space into the 6D space where the group acts linearly, lifting the 4D operators to 6D operators.  In particular, the linearity of the action of the conformal group in 6D allows one to easily build conformally invariant objects. However, non-trivial relations between these exist, posing problems for constructing the basis of tensor structures already in the case of 4-point functions.
This motivates the introduction of a different formalism described in section~\ref{sec:conformal_frame}.

The details of the 6D EF, its connection to the usual 4D formalism, and the relevant conventions are reviewed in  appendix~\ref{sec:DetailsOfTheEmbeddingFormalism}. In this section we discuss only the construction of $n$-point tensor structures and the spinning differential operators.
Our presentation focuses on the EF as a practical realization of the framework discussed in section~\ref{sec:Outline}.\footnote{Note that most of the results discussed in section~\ref{sec:Outline}, like the explicit construction of 2- and 3-point tensor strucutures~\cite{Costa:2011mg,SimmonsDuffin:2012uy,Elkhidir:2014woa} and the existence of the spinning differential operators~\cite{Costa:2011dw,Echeverri:2015rwa} were originally obtained within the EF.}

\paragraph{Embedding}
Let us first review the very basics of the EF. We label the points in the 6D space by $X^M=\{X^\mu,\,X^+,\,X^-\}$, with the metric given by
\beq
	X^2=X^\mu X_\mu+X^+X^-.
\eeq
The 4D space is then identified with the $X^+=1$ section of the lightcone $X^2=0$, and the coordinates on this section are chosen to be $x^\mu=X^\mu$.

A generic 4D operator $\OO_{\alpha_1\ldots\alpha_\ell}^{\dot\beta_1\ldots\dot\beta_{\bar\ell}}(x)$ in spin-$(\ell,\bar\ell)$ representation can be uplifted according to~\eqref{eq:operator_embedding} to a 6D operator $O^{a_1...a_\ell}_{b_1...b_{\bar\ell}}(X)$ defined on the lightcone $X^2=0$ and totally symmetric in its both sets of indices. We can define an index-free operator $O(X,S,\overline{S})$ using the 6D polarizations $S_a$ and $\overline{S}^{b}$ by
\begin{equation}
\label{eq:embeddingFormulaMainText}
O(X,S,\overline{S})\equiv O^{a_1...a_\ell}_{b_1...b_{\bar\ell}}(X)S_{a_1}\ldots S_{a_\ell}\overline{S}^{b_1}\ldots\overline{S}^{b_{\bar\ell}}.
\end{equation}
The 6D operators are homogeneous in $X$ and the 6D polarizations,
\begin{equation}\label{eq:scaling_6D}
O(X,S,\overline{S})\sim X^{-\CombinedDelta}\,S^\ell\,\overline S^{\bar\ell},\quad
\CombinedDelta= \Delta+\frac{\ell+\bar\ell}{2}.
\end{equation}
It is sometimes useful to assign the 4D scaling dimensions to the basic 6D objects as
\begin{equation}\label{eq:4D_scaling_dimesnion_6D_objects}
\Delta[X]=-1
\quad\text{and}\quad
\Delta[S]=\Delta[\overline S]=-\frac{1}{2}.
\end{equation}

According to~\eqref{eq:gauge_redundancy} there is a lot of freedom in choosing the lift $O(X,S,\overline S)$. We can express this freedom by saying that the operators differing by gauge terms proportional to $S \overline{\mathbf{X}}$, $\overline S \mathbf{X}$ or $\overline S S$ are equivalent. Note that $O(X,S,\overline S)$ is a priori defined only on the lightcone $X^2=0$, but it is convenient to extend it arbitrarily to all values of $X$. This gives an additional redundancy that the operators differing by terms proportional to $X^2$ are equivalent.

The 4D field can be recovered via a projection operation defined in appendix \ref{sec:DetailsOfTheEmbeddingFormalism},
\begin{equation}
\OO(x,s,\bar s)=%(X^{+})^{\CombinedDelta_\OO}
O(X,S,\overline{S})\bigg|_{proj},
\end{equation}
which essentially substitutes $X,\,S,\,\overline{S}$ with some expressions depending on $x,\,s,\, \bar s$ only. All the gauge terms proportional to $S \overline{\mathbf{X}},\; \overline S \mathbf{X},\; \overline S S$ or $X^2$ vanish under this operation.

Sometimes it is convenient to work with index-full form $O^{a_1...a_\ell}_{b_1...b_{\bar\ell}}(X)$ and to fix part of the gauge freedom by requiring it to be traceless. We can restore the traceless form from the index-free expression $O(X,S,\overline{S})$ by
\begin{equation}\label{eq:extractingIndices6D}
O^{a_1...a_\ell}_{b_1...b_{\bar\ell}}(X)=\frac{2}{\ell!\,\bar{\ell}!\,(2+\ell+\bar{\ell})!}\,
\left(\prod_{i=1}^\ell\partial^{a_i}\right)\left(\prod_{j=1}^{\bar\ell}\partial_{b_j}\right)
O(X,S,\overline{S}),
\end{equation}
where\footnote{These operators are constructed to map terms proportional to $\overline S S$ to other terms proportional to $\overline S S$.
In the equivalence class of uplifts, given an operator $O(X,S,\overline S)$ one can find another operator $O^\prime(X,S,\overline S)=O(X,S,\overline S)+(\overline S S)(\ldots)_O$ which differs from $O$  by terms proportional to $\overline S S$ and encodes a traceless operator $O^{a_1...a_\ell}_{b_1...b_{\bar\ell}}(X)$. Since after taking the maximal number of derivatives the $\overline S S$ terms can only map to zero, we can safely replace $O$ by $O'$.
The action on $O^\prime(X,S,\overline S)$ is proportional to the action of $\frac{\partial}{\partial S_a}$ and $\frac{\partial}{\partial \overline S^a}$ and thus provides an inverse operation to~\eqref{eq:embeddingFormulaMainText}.}
\begin{equation}\label{eq:Todorov6D}
\partial^a\equiv\left (S\cdot\frac{\partial}{\partial S}+\overline S\cdot\frac{\partial}{\partial \overline S}+3\right)\frac{\partial}{\partial S_a}-\overline S^a\left(\frac{\partial}{\partial S\cdot \partial \overline S}\right),
\end{equation}
\begin{equation}\label{eq:Todorov6Ddual}
\partial_b\equiv\left (S\cdot\frac{\partial}{\partial S}+\overline S\cdot\frac{\partial}{\partial \overline S}+3\right)\frac{\partial}{\partial \overline S^b}-S_b\left(\frac{\partial}{\partial S\cdot \partial \overline S}\right).
\end{equation}

\paragraph{Correlation functions}
A correlation function of 6D operators on the light cone must be SO(4,2) invariant and obey the homogeneity property~(\ref{eq:scaling_6D}). Consequently, it has the following generic form
\begin{equation}\label{eq:generic_correlator_6D}
\langle O_{\Delta_1}^{(\ell_1,\bar\ell_1)}(\Point_1)\ldots O_{\Delta_n}^{(\ell_n,\bar\ell_n)}(\Point_n)\rangle
=\sum_{I=1}^{N_n}g_I(\mathbf{U})T^I(X,S,\overline S),
\end{equation}
where $T^I(X,S,\overline S)$ are the 6D homogeneous $SU(2,2)$ invariant tensor structures and $g_I(\mathbf{U})$ are functions of 6D cross-ratios, i.e. homogeneous with degree zero SO(4,2) invariant functions of coordinates on the projective light cone. We also defined a short-hand notation
\begin{equation}
\Point\equiv(X,S,\overline{S}).
\end{equation}
Tensor structures split in a scaling-dependent and in a spin-dependent parts as
\begin{equation}
T^I(X,S,\overline S)=K_n\hat T^I(X,S,\overline S),\quad
T^I,\hat T_n^I\sim\prod_{i=1}^{n} S_i^{\ell_i}\overline S_i^{\bar\ell_i}.
\end{equation}
The object $K_n$ is the 6D kinematic factor and $\hat T^I$ are the $SO(4,2)$ invariants of degree zero in each coordinate. The main invariant building block is the scalar product\footnote{Notice a difference in the definition of $X_{ij}$ compared to~\cite{Elkhidir:2014woa,Echeverri:2015rwa,Echeverri:2016dun}: $X_{ij}^\text{here}=-2X_{ij}^\text{there}$.}
\begin{equation}\label{eq:the_6D_scalar_product}
X_{ij}\equiv - 2\,(X_i\cdot X_j),
\end{equation} 
The 6D kinematic factors \fref{n3KinematicFactor, n4KinematicFactor} are given by
\begin{equation}
K_2\equiv X_{12}^{-\frac{\CombinedDelta_1}{2}},\quad
K_3\equiv
\prod_{i<j} X_{ij}^{-\frac{\CombinedDelta_i+\CombinedDelta_j-\CombinedDelta_k}{2}},
\end{equation}
and
\begin{equation}
K_4\equiv
\left(\frac{X_{24}}{X_{14}}\right)^\frac{\CombinedDelta_1-\CombinedDelta_2}{2}
\left(\frac{X_{14}}{X_{13}}\right)^\frac{\CombinedDelta_3-\CombinedDelta_4}{2}
\times
\frac{1}{
X_{12}^\frac{\CombinedDelta_1+\CombinedDelta_2}{2}
X_{34}^\frac{\CombinedDelta_3+\CombinedDelta_4}{2}
}.
\end{equation}
We also define the 6D cross-ratios by taking products of $X_{ij}$ factors. For $n=4$ only two cross ratios can be formed
\begin{equation}\label{eq:cross_invariant_ratios_6D}
U\equiv\frac{X_{12}^2X_{34}^2}{X_{13}^2X_{24}^2},
\;\;\;
V\equiv\frac{X_{14}^2X_{23}^2}{X_{13}^2X_{24}^2}.
\end{equation}
With these definitions, under projection we recover the usual 4D expressions:
\begin{equation}\label{eq:projection_dot}
X_{ij}\proj=x_{ij}^2,\quad
K_n\proj=\mathcal{K}_n,\quad
U\proj=u,\quad
V\proj=v.
\end{equation}
Finally, given a correlator in the embedding space one can recover the 4D correlator
\begin{equation}
\langle \OO_{\Delta_1}^{(\ell_1,\bar\ell_1)}(\point_1)\ldots \OO_{\Delta_n}^{(\ell_n,\bar\ell_n)}(\point_n)\rangle=
\langle O_{\Delta_1}^{(\ell_1,\bar\ell_1)}(\Point_1)\ldots O_{\Delta_n}^{(\ell_n,\bar\ell_n)}(\Point_n)\rangle\proj,
\end{equation}
with the projections of the 6D invariants entering the 6D correlator given in the formula~\eqref{eq:projection_dot} and appendix~\ref{sec:TensotInvariants}.

\subsection{Construction of Tensor Structures}\label{sec:constructing_tensor_structures_EF}
Let us discuss the construction of tensor structures $\hat{T}_n^I(X,S,\overline{S})$. In index-free notation, this is equivalent to finding all $SU(2,2)$ invariant homogeneous polynomials in $S,\,\overline S$.
All $SU(2,2)$ invariants are built fully contracting the indices of the following objects:
\begin{equation}\label{eq:ingredients}
\delta^a_b,\,\epsilon_{abcd},\,\overline\epsilon^{abcd},\,\mathbf{X}_{i\,ab},\,\overline{\mathbf{X}}^{ab}_j,\,S_{k\,a},\,\overline S^a_l.
\end{equation}
With the exception of taking traces over the coordinates $\tr[\mathbf{X}_i\overline{\mathbf{X}}_j\ldots \mathbf{X}_k\overline{\mathbf{X}}_l]$,
\footnote{All such traces can be reduced to the scalar product $X_{ij}=-\text{Tr}[\mathbf{X}_i\overline{\mathbf{X}}_j]/2$.} all other tensor structures are built out of simpler invariants of degree two or four in $S$ and $\overline S$.

\paragraph{List of non-normalized invariants}
\label{sec:invariants_list}
By taking into account eq. \eqref{eq:XXbar} and the relations \eqref{eq:light_cone_condition} and \eqref{eq:gauge_choice}, it is possible to identify a set of invariants with the properties discussed above. These can be conveniently divided in five classes. The number of possible invariants increases with the number of points $n$. Below we provide a complete list of them for $n\leq 5$ and indicate their transformation property under the 4D parity. In what follows the indices $i,j,k,l,\ldots$ are assumed to label different points. 

\subparagraph{Class I} constructed from $\overline{S}_i$ and $S_j$ belonging to two different operators.
\begin{equation}\label{eq:classIinvariants}
\begin{array}{>{\displaystyle}l>{\displaystyle}l>{\displaystyle}l>{\displaystyle}l>{\displaystyle}l>{\displaystyle}l}

n\geq 2: & I^{ij}  &\equiv  &(\overline{S}_i S_j)  &  \stackrel{\mathcal P}{\longrightarrow}  & -I^{ji},\\

n\geq 4: &  I^{ij}_{kl}   &\equiv   &(\overline{S}_i \mathbf{X}_k \overline{\mathbf{X}}_l S_j)  &\stackrel{\mathcal P}{\longrightarrow} &-I^{ji}_{lk},\\

n\geq 6: & \ldots &   & \ldots &  & \ldots
\end{array}
\end{equation}

\subparagraph{Class II} constructed from $\overline{S}_i$ and $S_i$ belonging to the same operator.
\begin{equation}\label{eq:classIIinvariants}
\begin{array}{>{\displaystyle}l>{\displaystyle}l>{\displaystyle}l>{\displaystyle}l>{\displaystyle}l>{\displaystyle}l}

n\geq 3: & J^i_{jk}  &\equiv  &(\overline{S}_i \mathbf{X}_j \overline{\mathbf{X}}_k S_i)  &  \stackrel{\mathcal P}{\longrightarrow}  &  -J^i_{kj}=J^i_{jk},\\

n\geq 5: &  J^i_{jklm}    &\equiv   &(\overline{S}_i \mathbf{X}_j \overline{\mathbf{X}}_k \mathbf{X}_l \overline{\mathbf{X}}_m S_i)  &\stackrel{\mathcal P}{\longrightarrow} &-J^i_{mlkj},\\

n\geq 7: & \ldots &  & \ldots &  & \ldots
\end{array}
\end{equation}

\subparagraph{Class III} constructed from $S_i$ and $S_j$ belonging to two different operators.
\begin{equation}\label{eq:classIIIinvariants}
\begin{array}{>{\displaystyle}l>{\displaystyle}l>{\displaystyle}l>{\displaystyle}l>{\displaystyle}l>{\displaystyle}l>{\displaystyle}l>{\displaystyle}l}

n\geq 3: & K_k^{ij}  &\equiv  &(S_i\overline{\mathbf{X}}_k S_j)  
   &  \stackrel{\mathcal P}{\longleftrightarrow}  
   &  \overline{K}_k^{ij} & \equiv & (\overline{S}_i\mathbf{X}_k\overline{S}_j),\\
   
n\geq 5: & K_{klm}^{ij} & \equiv  & (S_i\overline{\mathbf{X}}_k\mathbf{X}_l\overline{\mathbf{X}}_m S_j) 
   & \stackrel{\mathcal P}{\longleftrightarrow}
   & \overline{K}_{klm}^{ij} & \equiv & (\overline{S}_i\mathbf{X}_k\overline{\mathbf{X}}_l\mathbf{X}_m\overline{S}_j),\\

n\geq 7: & \ldots &   & \ldots &  & \ldots &  & \ldots
\end{array}
\end{equation}

\subparagraph{Class IV}constructed from $S_i$ and $S_i$ belonging to the same operator.
\begin{equation}
\begin{array}{>{\displaystyle}l>{\displaystyle}l>{\displaystyle}l>{\displaystyle}l>{\displaystyle}l>{\displaystyle}l>{\displaystyle}l>{\displaystyle}l}

n\geq 4: & L^i_{jkl}  &\equiv  &(S_i \overline{\mathbf{X}}_j \mathbf{X}_k \overline{\mathbf{X}}_l S_i) 
   &  \stackrel{\mathcal P}{\longleftrightarrow}  
   &  \overline{L}^i_{jkl} & \equiv &  (\overline{S}_i\mathbf{X}_j\overline{\mathbf{X}}_k\mathbf{X}_l \overline{S}_i),\\
   
n\geq 6: & \ldots &  & \ldots &  & \ldots &  & \ldots
\end{array}
\end{equation}

\subparagraph{Class V} constructed from four $S$ or four $\overline{S}$ belonging to different operators.
\begin{equation}
n \geq4:\;\; M^{ijkl}\equiv\epsilon(S_i S_j S_k S_l)\stackrel{\mathcal P}{\longleftrightarrow}
\overline M^{ijkl}\equiv\epsilon(\overline{S}_i \overline{S}_j \overline{S}_k \overline{S}_l).
\end{equation}

\paragraph{Basic linear relations}
Simple properties \fref{applyEFProperties} arise due to the relation~\eqref{eq:XXbar}.
For instance
\begin{equation}
J^i_{jk}=-J^i_{kj},\quad
K_k^{ij}=-K_k^{ji},\quad
\overline K_k^{ij}=-\overline K_k^{ji}
\end{equation}
for $n\geq 3$. Consequently not all these invariants are independent and it is convenient to work only with a subset of them, for instance
$J^i_{j<k},\;K_k^{i<j},\;\overline K_k^{i<j}$.
For $n\geq 4$ other properties must be taken into account:
\begin{equation}
I^{ij}_{kl}+I^{ij}_{lk}=- X_{kl} I^{ij},\quad L^i_{jkl}=L^i_{[jkl]},\quad M^{ijkl}=M^{[ijkl]},
\quad \overline M^{ijkl}=\overline M^{[ijkl]}.
\end{equation}
These can be used in analogous manner to work only with a subset of invariants, for instance  $I^{i<j}_{k<l}$, $I^{i>j}_{k>l}$, $L^i_{j<k<l}$, $M^{1234}$ and $\overline M^{1234}$.
Another important linear relation is
\begin{equation}\label{eq:LinearRelation}
J^i_{[jk} X_{l]m}=0
\end{equation}
where $m$ is allowed to be equal to $i$.

\paragraph{Non-linear relations} Unfortunately, even after taking into account all the linear relations above, many non-linear relations between products of invariant are present, see equations~\eqref{eq:Jacobi_rel_1} - \eqref{eq:Jacobi_rel_4} for $n\geq 3$ relations \fref{applyJacobiRelations} and appendix A in~\cite{Echeverri:2015rwa} for some $n\geq 4$ relations.\footnote{Mind the difference in notation, see footnote~\ref{foot:difference_in_notation} for details.} We expect that they all arise from~(\ref{eq:jacobi_identities}).\footnote{In principle the Schouten identities might also contribute, see the footnote at page 26 of \cite{SimmonsDuffin:2012uy}; we found however that the Schouten identities, when contracted, give relations equivalent to \eqref{eq:jacobi_identities} for $n\leq 4$.} 
As an example consider the following set of relations
\begin{align}
M^{ijkl} &=-2\,X_{ij}^{-1}\big( K_i^{jk}\,K_j^{il}-K_i^{jl}\,K_j^{ik} \big),\\
\overline M^{ijkl} &=-2\,X_{ij}^{-1}\big( \overline K_i^{jk}\,\overline K_j^{il}-\overline K_i^{jl}\,\overline K_j^{ik} \big).
\end{align}
They show that $M^{ijkl}$ and $\overline M^{ijkl}$ can be rewritten in terms of other invariants; hence class V objects are never used.
All the relations obtained by fully contracting~\eqref{eq:ingredients} with \eqref{eq:jacobi_identities} in all possible ways, involve at most products of two invariants in class $I-IV$. In fact, we will see in section~\ref{sec:CF_to_EF} that all non-linear relations have a quadratic nature. However, these quadratic relations can be combined together to form relations involving products of three or more invariants.\footnote{In other words, we have a graded ring of invariants and an ideal $I$ of relations between them. The goal is to find a basis of independent invariants of a given degree modulo $I$. In principle, $I$ is generated by a quadratic basis, but it is not trivial to reduce invariants modulo this basis. One would like to find a better basis, e.g. a Gr\"obner basis, which then will contain higher-order relations.} See appendix~\ref{sec:BasisthreePointFunctions} for an example of such phenomena in the $n=3$ case.

\paragraph{Normalization of invariants}
The $\hat{T}_n^I(X,S,\overline{S})$ are required to be of degree zero in all coordinates. It
is then convenient to introduce the following normalization factors
\begin{equation}
N_{ij}\equiv X_{ij}^{-1},\;\;
N^{ij}_{k}\equiv \sqrt{\frac{X_{ij}}{X_{ik}X_{kj}}},\;\;N_{ijk}\equiv\frac{1}{\sqrt{X_{ij}X_{jk}X_{ki}}}.
\end{equation}
Using these factors \fref{normalizeInvariants, denormalizeInvariants} it is possible to define normalized type I and type II tensor structures 
\begin{equation}\label{eq:normalization_1}
\hat{I}^{ij}\equiv I^{ij},\;\;
\hat{I}^{ij}_{kl}\equiv N_{kl}I^{ij}_{kl},\;\;
\hat{J}^{i}_{jk}\equiv N_{jk}J^{i}_{jk},\;\;
\hat{J}^{i}_{jklm}\equiv N_{jk} N_{lm}J^{i}_{jklm},
\end{equation}
and normalized type III and type IV tensor structures
\begin{equation}\label{eq:normalization_2}
\hat{K}^{ij}_k\equiv N^{ij}_k K^{ij}_k,\;\;
\hat{K}^{ij}_{klm}\equiv N_{klm} K^{ij}_{klm},\;\;
\hat{L}^i_{jkl} \equiv N_{jkl} L^i_{jkl},
\end{equation}
with the analogous expressions for parity conjugated invariants $\hat{\overline{K}}_k^{ij}$, $\hat{\overline{K}}_{klm}^{ij}$ and $\hat{\overline{L}}^i_{jkl}$. In appendix \ref{sec:TensotInvariants} we provide an explicit 4D form of these invariants after projection.
Notice the slight change of notation from previous works\footnote{\label{foot:difference_in_notation}The correspondence with the notation of~\cite{Elkhidir:2014woa,Echeverri:2015rwa,Echeverri:2016dun} is as follows: $
\hat{I}^{ij}\sim I_{ij},\;\;
-2\, \hat{I}^{ij}_{kl}\sim\hat{J}_{ij,\,kl},\;\;
-2 \,\hat{J}^{i}_{jk}\sim J_{i,\,jk},\;\;
\sqrt{-2}\, \hat{K}^{ij}_k\sim K_{k,\,ij},
\sqrt{-2} \,\hat{\overline K}^{ij}_k\sim\overline K_{k,\,ij},\;\;
\sqrt{-8}\,\hat{L}^i_{jkl} \sim K_{i,jkl},\;\;
\sqrt{-8}\,\hat{\overline{L}}^i_{jkl}\sim \overline K_{i,jkl}
$, where the expressions in the l.h.s. represent our notation and the expressions in the r.h.s. represent their notation.}.

\paragraph{Basis of tensor structures}
\label{sec:basis_6D}
Given an $n$-point function, one can construct a set of tensor structures \fref{n3ListStructures, n3ListStructuresAlternativeTS, n4ListStructuresEF} by taking products of basic invariants as
\begin{equation}\label{eq:set_of_tensor_structures}
\hat T_n^I
=\Big\{
\prod_{i,j,\ldots}
\underbrace{
\big[\hat I^{ij}\big]^{\#}
}_{n\geq2}
\underbrace{
\big[\hat J^{i}_{jk}\big]^{\#}
\big[\hat K^{jk}_{i}\big]^{\#}
\big[\hat{\overline K}^{jk}_{i}\big]^{\#}
}_{n\geq 3}
\underbrace{
\big[\hat I^{ij}_{kl}\big]^{\#}
\big[\hat L^i_{jkl}\big]^{\#}
\big[\hat{\overline L}^i_{jkl}\big]^{\#}
}_{n\geq 4}
\underbrace{
\big[\hat J^{i}_{jklm}\big]^{\#}
\big[\hat K^{jk}_{ilm}\big]^{\#}
\big[\hat{\overline K}^{jk}_{ilm}\big]^{\#}
}_{n\geq 5}
\ldots
\Big\}.
\end{equation}
The subscripts stress that for a given number of points $n$ not all the invariants are defined. 
The non-negative exponents $\#$ are determined by requiring $\hat T_n^I$ to be of degree $(\ell_i,\bar{\ell}_i)$ in $(S_i,\overline{S}_i)$. Generally, not all tensor structures obtained in this way are independent, due to the properties and relations discussed above. The number of relations to take into account increase rapidly with $n$. For $n\leq 3$ the problem of constructing a basis of independent tensor structures has been succesfully solved in \cite{SimmonsDuffin:2012uy,Elkhidir:2014woa}; we review the construction for $n=3$ in appendix \ref{sec:BasisthreePointFunctions}. 
However the increasing number of relations makes this approach inefficient to study general correlators for $n\geq 4$, mainly because many relations which are cubic or higher order in invariants can be written. In section~\ref{sec:conformal_frame} an alternative method of identifying all the independent structures is provided. Using this method we will also prove in section~\ref{sec:EF_to_CF} that any $n$-point function tensor structure is constructed out of $n\leq 5$ invariants, namely the invariants involving five or less points in the formula~(\ref{eq:set_of_tensor_structures}).

\subsection{Spinning Differential Operators}
\label{sec:spinning_differential_operators}
Let us now discuss the EF realization of the \emph{spinning differential operators} used in~\eqref{eq:tensor_structures_via_differential_operators} which allow to relate 3-point tensor structures of correlators with different spins\footnote{This relation is of course purely kinematic, it holds only at the level of tensor structures and does not hold at the level of the full correlator.}
\begin{equation}\label{eq:kinematic_connection_of_operators}
\langle O^{(\ell_i,\bar\ell_i)}_{\Delta_{\mathcal{O}_i}}
O^{(\ell_j,\bar\ell_j)}_{\Delta_{\mathcal{O}_j}}
O^{(\ell,\bar\ell)}_{\Delta_{\mathcal{O}}}
\rangle
\sim\mathbf{D}_{ij}\,
\langle O^{(\ell_i',\bar\ell_i')}_{\Delta'_{\mathcal{O}_i}}
O^{(\ell_j',\bar\ell_j')}_{\Delta'_{\mathcal{O}_j}}
O^{(\ell,\bar\ell)}_{\Delta_{\mathcal{O}}}
\rangle.
\end{equation}
The operators\footnote{\label{foot:differential_basis}We distinguish the operators $\mathbf{D}$ here and the operators $\mathbb{D}$ described in section~\ref{sec:correlation_functions_outline} because acting on the seed tensor structures they generate different bases. The basis spanned by $\mathbf{D}$ is often called the differential basis.} $\mathbf{D}_{ij}$ are written as a product of basic differential operators which were found in~\cite{Echeverri:2015rwa}
\begin{equation}\label{eq:spinning_differential_operator_full}
\mathbf{D}_{ij}=\Big\{
\prod_{i,j=1,2}
\nabla_{ij}^{\#}
I_{ij}^{\#}
\bar d_{ij}^{\#}
d_{ij}^{\#}
D_{ij}^{\#}
\widetilde D_{ij}^{\#}
\Big\}.
\end{equation}
The exponents are determined by matching the spins on both sides of \eqref{eq:kinematic_connection_of_operators}. The basic spinning differential operators are constructed to be insensitive to pure gauge modifications and different extensions of fields outside of the light cone as stressed in \eqref{eq:derivative_consistency}. The action of these operators in 4D can be deduced by using the projection rules given in~\eqref{eq:derivatives_projection}.

We provide here the list of basic differential operators\footnote{Notice a change in the normaliztion of the basic spinning differential operators compared to~\cite{Echeverri:2015rwa}. } entering~\eqref{eq:spinning_differential_operator_full} arranging them in two sets according to the value of $\Delta \ell=|\ell_i+\ell_j-\bar{\ell}_i-\bar{\ell}_j|=0,2$. For $\Delta \ell=0$ we have
\begin{equation}\label{eq:spinning_differential_operators_A}
\begin{array}{>{\displaystyle}l>{\displaystyle}l>{\displaystyle}l}
D_{ij}&
\equiv\frac{1}{2}\overline S_i \Sigma^M \overline\Sigma^N S_i\Big(X_{jM}\frac{\partial}{\partial X^N_i}-X_{jN}\frac{\partial}{\partial X^M_i}\Big)
&\sim \overline S_i S_i, \\
\widetilde D_{ij}&
\equiv\overline S_i \mathbf{X}_j \overline\Sigma^N S_i\frac{\partial}{\partial X^N_j}+2I^{ij}\,S_{ia}\frac{\partial}{\partial S_{ja}}-2I^{ji}\,\overline S^a_i\frac{\partial}{\partial\overline S^a_j}
&\sim \overline S_i S_i,\\
I^{ij} &\equiv \overline S_i S_j&\sim \overline{S}_i S_j,\\
\nabla_{ij} &\equiv  \big[{\mathbf X}_i \overline{\mathbf X}_j]_a^b\,\frac{\partial^2}{\partial S_{i\,a}\;\partial\overline{S}_j^b}&\sim S_i^{-1}\overline S_j^{-1}.
\end{array}
\end{equation}
For $\Delta \ell =2$ we have
\begin{equation}\label{eq:spinning_differential_operators_B}
\begin{array}{>{\displaystyle}l>{\displaystyle}l>{\displaystyle}l}
 d_{ij} &\equiv S_j \overline X_{i} \frac{\partial}{\partial\overline S_i} &
\sim \overline S_i^{-1} S_j \\
\overline d_{ij}&\equiv \overline S_j X_{i} \frac{\partial}{\partial S_i} &
\sim S_i^{-1}\overline S_j.
\end{array}
\end{equation}
Note that for any differential operator $\mathbf{D}_{ij}$ we necessarily have $\Delta\ell$ even, since it has to preserve the total Fermi/Bose statistics of the pair of local operators.

The basic spinning differential operators described above carry the 4D scaling dimension according to~\eqref{eq:4D_scaling_dimesnion_6D_objects}, thus it is convenient to introduce an operator $\Xi$
which formally shifts the 4D dimensions of external operators in a way that effectively makes the 4D scaling dimensions of $\mathbf{D}_{ij}$ vanish. The action of $\Xi$ on basic spinning differential operators is defined as
\begin{equation}\label{eq:shifts_1}
\Xi[D_{ij}]f_n= (D_{ij}f_n)\Big|_{\Delta_j\rightarrow\Delta_j+1},\quad
\Xi[\widetilde D_{ij}]f_n= (\widetilde D_{ij}f_n)\Big|_{\Delta_i\rightarrow\Delta_i+1}
\end{equation}
and
\begin{equation}\label{eq:shifts_2}
\Xi[op]f_n = (op\, f_n)
\Big|_{\Delta_i\rightarrow\Delta_i+1/2}\Big|_{\Delta_j\rightarrow\Delta_j+1/2},
\end{equation}
where $op$ denotes any of the remaining spinning differential operators.\footnote{The shift in the last formula can alternatively be implemented with multiplication by a factor $ X_{ij}^{-1/2}$.}
These formal shifts of course make sense only if the scaling dimensions appear as variables in $f_n$. The use of the dimension-shifting operator $\Xi$ allows to keep the same scaling dimensions in the seed CPWs and the CPW related by~\eqref{eq:D_on_seeds}.

The relevant functions in the package are \fref{opDEF, opDtEF, opdEF, opdbEF, opIEF, opNEF} and $\Xi$.

\section{Conformal Frame}
\label{sec:conformal_frame}

For sufficiently complicated correlation functions one finds a lot of degeneracies in the embedding space construction of tensor structures. There exists an alternative construction \cite{Osborn:1993cr,Kravchuk:2016qvl} which  provides better control under degeneracies. More precisely, it reduces the problem of constructing tensor structures to the well studied problem of finding invariant tensors of orthogonal groups of small rank.

Our aim is to describe the correlation function $f_n(x,s,\bar s)$ whose generic form is given in~\eqref{eq:correlation_function_structure}.
The conformal symmetry relates the values of \ffixed{x} at different values of $x$. There is a classical argument, usually applied to 4-point correlation functions, saying that it is sufficient to know only the value \ffixed{x_{CF}} for some standard choices of $x_{CF}$ such that all the other values of $x$ can be obtained from some $x_{CF}$ by a conformal transformation. This conformal transformation then allows one to compute \ffixed{x} from \ffixed{x_{CF}}. The standard configurations $x_{CF}$ are chosen in such a way that there are no conformal transformations relating two different standard configurations, so that the values \ffixed{x_{CF}} can be specified independently. Following \cite{Kravchuk:2016qvl}, we call the set of standard configurations $x_{CF}$ the conformal frame (CF).

The usefulness of this construction lies in the fact that the values \ffixed{x_{CF}} have to satisfy only a few constraints. In particular, these values have to be invariant only under the conformal transformations which do not change $x_{CF}$~\cite{Kravchuk:2016qvl}. 
Such conformal transformations form a group which we call the ``little group''. The little group is $SO(d+2-n)$ for $n$-point functions in $d$ dimensions.\footnote{For $n\geq 3$ and generic $x$. The little group is trivial for $n\geq d+2$.} For example, for 4-point functions in 4D it is $SO(2)\simeq U(1)$. One can already see a considerable simplification offered by this construction for 4-point functions in 4D, since the invariants of $SO(2)$ are extremely easy to classify.

We use the following choice for the conformal frame configurations $x_{CF}$ for $n\geq 3$,
\algn{
\label{eq:conformalframedefn_1}
	x_1^\mu&=(0,0,0,0),\\
\label{eq:conformalframedefn_2}
	x_2^\mu&=((\bar z-z)/2,0,0,(z+\bar z)/2),\\
\label{eq:conformalframedefn_3}
	x_3^\mu&=(0,0,0,1),\\
\label{eq:conformalframedefn_4}
	x_4^\mu&=(0,0,0,L),\\
\label{eq:conformalframedefn_5}
	x_5^\mu&=(x_5^0,x_5^1,0,x_5^3),
}
where if $n=3$ we can set $z=\bar z=1/2$ and if we have more than $5$ operators, the unspecified positions $x_{\geq 6}$ are completely unconstrained.

Here $L$ is a fixed number, and we always take the limit $L\to+\infty$ to place the corresponding operator ``at inifinity''. In this limit one should use the rescaled operator $\OO_4$ 
\begin{equation}\label{eq:rescaled_O}
\OO_4\to \OO_4\;L^{2\Delta_4}
\end{equation}
inside all correlators to get a finite and non-zero result.

The variables $z,\,\bar z,\,x_5^0,\, x_5^1,\,x_5^3$ and the 4-vectors $x_6,\,x_7,\,\ldots$ are the coordinates on the conformal frame and thus are essentially the conformal cross-ratios. Note that we have $2$ conformal cross-ratios for $4$ points, and $4\,n-15$ for $n$ points with $n\geq 5$. Notice also that for 4-point functions the analytic continuation with $z=\bar z^*$ corresponds to Euclidean kinematics. It is easy to check that there are no conformal generators which take the conformal frame configuration \eqref{eq:conformalframedefn_1} - \eqref{eq:conformalframedefn_5} to another nearby conformal frame configuration.

\subsection{Construction of Tensor Structures}
\subsubsection{Three-point Functions}
As shown in appendix~\ref{sec:BasisthreePointFunctions}, an independent basis for general 3-point tensor structures is relatively easy to construct in EF, and there is no direct need for the conformal frame construction. Nonetheless, in this section we employ the CF to  construct 3-point tensor structures in order to illustrate how the formalism works in a familiar case.\footnote{The CF construction of 3-point functions is not implemented in the package.}

The little group algebra $\mathfrak{so}(1,2)$ which fixes the points $x_1,x_2,x_3$ is defined by the following generators
\beq
\label{eq:so3stab}
	M^{01},\quad M^{02},\quad M^{12},
\eeq
see appendix~\ref{app:conventions4D} for details. According to our conventions, the corresponding generators acting on polarizations $s_\alpha$ are
\beq\label{eq:su2taction}
	\mathcal{S}^{01}=-\frac{1}{2}\sigma^1,\quad
	\mathcal{S}^{02}=-\frac{1}{2}\sigma^2,\quad
	\mathcal{S}^{12}=\frac{i}{2}\sigma^3,
\eeq
and the generators acting on $\bar s^{\dot \alpha}$ are
\beq
	\overline{\mathcal{S}}^{01}=\frac{1}{2}\sigma^1,\quad
	\overline{\mathcal{S}}^{02}=\frac{1}{2}\sigma^2,\quad
	\overline{\mathcal{S}}^{12}=\frac{i}{2}\sigma^3.
\eeq
It is easy to see that if we introduce $t_\alpha\equiv s_\alpha$ and $\tDual_\alpha\equiv \sigma^3_{\alpha\dot\beta}\bar s^{\dot\beta}$, then $t$ and $\tDual$ transform in the same representation of $\mathfrak{so}(1,2)$.

General 3-point structures are put in one-to-one correspondence with the $\mathfrak{so}(1,2)\simeq \mathfrak{su}(2)$ conformal frame invariants built out of $t_i$ and $\tDual_i$,  $i=1,2,3$. 
This gives an explicit implementation of the rule \cite{Mack:1976pa,Osborn:1993cr,Kravchuk:2016qvl} which states that 3-point structures correspond to the invariants of $SO(d-1)=SO(3)$ group
\beq\label{eq:3ptgrouptheoryrule}
	\left((\ell_1,\bar\ell_1)\otimes(\ell_2,\bar\ell_2)\otimes(\ell_3,\bar\ell_3)\right)^{SO(3)}=\left(\ell_1\otimes \bar\ell_1 \otimes \ell_2 \otimes \bar\ell_2 \otimes\ell_3 \otimes \bar\ell_3\right)^{SO(3)}.
\eeq 
Using this rule, we can immediately build independent bases of 3-point structures, for example by first computing the tensor product decompositions 
\beq
\ell_i\otimes\bar\ell_i=\bigoplus_{j_i=|\ell_i-\bar\ell_i|}^{\ell_i+\bar\ell_i}j_i,\qquad (j_i+\ell_i+\bar\ell_i\text{ even})
\eeq
and then for every set of $j_i$ constructing the unique singlet in $j_1\otimes j_2\otimes j_3$ when it exists.

A more direct way, which does not however automatically avoid degeneracies, is to use the basic building blocks for $SO(3)$ invariants, which are the contractions of the form $t_i^\alpha t_{j\,\alpha}$, $t_i^\alpha\tDual_{j\,\alpha}$ and $\tDual_i^\alpha\tDual_{j\,\alpha}$. It is then straightforward  to establish the correspondence with the embedding formalism invariants
\begin{equation}\label{eq:efcfdict}
I^{ij}\propto \tDual_i t_j,\quad
J^i_{jk}\propto \tDual_i t_i,\quad
K^{ij}_k\propto t_i t_j,\quad
\overline K^{ij}_k\propto \tDual_i \tDual_j,
\end{equation}
where it is understood that $i,j,k$ are all distinct. Up to the coefficients, this dictionary is fixed completely by matching the degrees of $s$ and $\bar s$ on each side.

Correspondingly, as in the embedding space formalism, we have relations between these building blocks, which now come from the Schouten identity\footnote{Which itself follows from contracting $\epsilon^{\beta\gamma}$ with the identity $A_{[\alpha}B_\beta C_{\gamma]}=0$ valid for two-component spinors.}
\beq\label{eq:ShoutenIdentity}
	(AB) C_\alpha+(BC)A_\alpha+(CA) B_\alpha=0.
\eeq
For example we can take $A=t_i$, $B=t_k$, $C=\tDual_j$ and contract~\eqref{eq:ShoutenIdentity} with $\tDual_k$ to find
\beq
(t_i t_k)(\tDual_j\tDual_k)+
(t_k\tDual_j)(t_i\tDual_k)+
(\tDual_jt_i)(t_k\tDual_k)=0,
\eeq
which corresponds via the dictionary~\eqref{eq:efcfdict} to an identity of the form
\beq
	\# K^{ik}_j\overline K^{jk}_i + \# I^{jk}I^{ki}+\# I^{ji}J^{k}_{ij}=0.
\eeq
This gives precisely the structure of the relation~\eqref{eq:Jacobi_rel_1}. We thus effectively reproduce the EF construction.

Finally, let us briefly comment on the action of $\PP$ in the 3-point conformal frame. The parity transformation of operators~\eqref{eq:ParityIndexFreeNotation} induces the following transformation of polarizations
\beq\label{eq:sparity}
	s_\alpha\to i \bar s^{\dot\alpha},\quad \bar s^{\dot \beta}\to i s_\beta\quad\Longrightarrow\quad
	t\to i \sigma^3\tDual,\quad\tDual\to i\sigma^3 t.
\eeq
The full parity transformation does not however preserve the conformal frame since it reflects all three spatial axes and thus moves the points $x_2$ and $x_3$. We can reproduce the correct parity action in the conformal frame by supplementing the full parity transformation with $i\pi$ boost in the $03$ plane given by $e^{-i\pi S^{03}}=i\sigma_3$ on $t$ and by $\sigma^3 e^{-i\pi S^{03}}\sigma^3 = -i\sigma_3$ on $\tDual$. This leads to
\beq\label{eq:3ptparity}
	t\to \tDual,\quad \tDual\to -t.
\eeq

Note that according to \eqref{eq:3ptparity} the transformations properties of~\eqref{eq:efcfdict} under parity match precisely the ones found in~\eqref{eq:classIinvariants} - \eqref{eq:classIIIinvariants}.

\subsubsection{Four-point Functions}
\label{sec:fourpointCFstructs}
In the $n=4$ case the little group algebra $\mathfrak{so}(2)\simeq\mathfrak{u}(1)$
which fixes the points $x_1,x_2,x_3,x_4$ is given by the generator 
\beq
	M^{12}.
\eeq
Note that the algebra $\mathfrak{so}(2)$ is a subalgebra of the 3-point little group algebra $\mathfrak{so}(1,2)$ discussed above. According to \eqref{eq:su2taction}, its action on both $t$ and $\tDual$ is given by
\beq	
	\mathcal{S}^{12}=\frac{i}{2}\sigma^3.
\eeq
This generator acts diagonally on $t$ and $\tDual$, so that we can decompose
\beq\label{eq:components_of_s_sb}
	s_\alpha\equiv\begin{pmatrix}
		\xi\\\eta
	\end{pmatrix},
	\quad 
	\bar s_{\dot\beta}\equiv\begin{pmatrix}
		\bar\xi\\\bar\eta
	\end{pmatrix}
	\quad \Longrightarrow\quad
	t\equiv s_\alpha=\begin{pmatrix}
		\xi\\\eta
	\end{pmatrix},
	\quad 	
	\tDual\equiv\sigma^3_{\alpha\dot\beta}\bar s^{\dot\beta}=
	\begin{pmatrix}
		\bar\eta\\
		\bar\xi
	\end{pmatrix}.
\eeq
Note that our convention $\bar s_{\dot\alpha}=(s_\alpha)^*$ implies that $\bar\xi=\xi^*$ and $\bar\eta=\eta^*$.
Appropriately defining the $\mathfrak{u}(1)$ charge $Q$ we can say that 
\begin{equation}
Q[\xi]=Q[\bar\eta]=+1
\quad{\bf and}\quad
Q[\eta]=Q[\bar\xi]=-1.
\end{equation}
Tensor structures of 4-point functions are just the products of $\xi,\bar\xi,\eta,\bar\eta$ of total charge $Q=0$. These are given by \fref{CF4pt,n4ListStructures}
\beq
\begin{matrix}
\structgeneral\equiv\prod_{i=1}^4 \xi_i^{-q_i+\ell_i/2}\eta_i^{q_i+\ell_i/2}\bar \xi_i^{-\bar q_1+\bar \ell_i/2}\bar \eta_i^{\bar q_i+\bar \ell_i/2},\\ q_i\in \{-\ell_i/2,\ldots,\ell_i/2\},\,\bar q_i\in \{-\bar\ell_i/2,\ldots,\bar\ell_i/2\},
\end{matrix}
\label{eq:fourpointcfdef}
\eeq
subject to 
\beq
\sum_{i=1}^4(q_i-\bar q_i)=0.
\eeq
It is clear from the construction that these 4-point structures are all independent, i.e. there are no relations between them. It is in contrast with the embedding space formalism, where there are a lot of relations between various 4 point building blocks.

As a simple example, consider a 4-point function of a $(1,0)$ fermion at position $1$, a $(0,1)$ fermion at position $2$ and two scalars at position $3$ and $4$. The allowed 4-point tensor structures are then
\beq
	\struct{+\tfrac{1}{2}}{0}{0}{0}{0}{+\tfrac{1}{2}}{0}{0}\quad\text{and}\quad\struct{-\tfrac{1}{2}}{0}{0}{0}{0}{-\tfrac{1}{2}}{0}{0}.
\eeq 

To compute the action of space parity, we need to supplement the full spatial parity \eqref{eq:sparity} with a $\pi$ rotation in, say, the $13$ plane in order to make sure that parity preserves the 4-point conformal frame \eqref{eq:conformalframedefn_1} - \eqref{eq:conformalframedefn_4}. In this case the combined transformation is simply a reflection in the $2$'nd coordinate direction. It is easy to compute that this gives the action 
\begin{equation}\label{eq:4ptparity}
\xi\to -i\bar\xi,\quad
\bar \xi\to i\xi,\quad
\eta\to -i\bar\eta,\quad
\bar \eta\to i\eta.
\end{equation}
Note that this does not commute with the action of $\mathfrak{u}(1)$ since the choice of the $13$ plane was  arbitrary -- we could have also chosen the $23$ plane, and $\mathfrak{u}(1)$ rotates between these two choices. It is only important that this reflection reverses the charges of $\mathfrak{u}(1)$ and thus maps invariants into invariants.

From \eqref{eq:4ptparity} we find that the parity acts as
\beq
\PP\structgeneral=i^{-\sum_i \ell_i-\bar\ell_i}\struct{\bar q_1}{\bar q_2}{\bar q_3}{\bar q_4}{q_1}{q_2}{q_3}{q_4}.\label{eq:cf4parity}
\eeq
From the definition~\eqref{eq:fourpointcfdef} we also immediately find the complex conjugation rule
\beq\label{eq:cf4conj}
	\structgeneral^*=\struct{\bar q_1}{\bar q_2}{\bar q_3}{\bar q_4}{q_1}{q_2}{q_3}{q_4}.
\eeq
According to~\eqref{eq:time_reversal_simple_rule}, by combining these two transformations we find the action of time reversal
\beq\label{eq:cf4time}
	\TT\structgeneral=i^{\sum_i \ell_i-\bar\ell_i}\structgeneral.
\eeq

\subsubsection{Five-point Functions and Higher}
In the $n\geq 5$ case there are no conformal generators which fix the conformal frame. It means that all $\xi,\bar\xi,\eta,\bar\eta$ are invariant by themselves.\footnote{\label{ftn:spincenter}More precisely, there is still the $\mathbb{Z}_2$ kernel of the projection $Spin(1,3)\to SO(1,3)$, which gives the selection rule that the full correlator should be bosonic (in this sense $\xi,\bar\xi,\eta,\bar\eta$ are not individually invariant).} This allows us to construct the $n$-point tensor structures
\beq
	\left[\begin{matrix}
	q_1&q_2&\ldots &q_n\\
	\bar q_1&\bar q_2&\ldots &\bar q_n
	\end{matrix}\right]\equiv\prod_{i=1}^n \xi_i^{-q_i+\ell_i/2}\eta_i^{q_i+\ell_i/2}\bar \xi_i^{-\bar q_1+\bar \ell_i/2}\bar \eta_i^{\bar q_i+\bar \ell_i/2},
\eeq
with the only restriction
\beq
	q_i\in\{-\ell_i/2,\ldots	\ell_i/2\},\quad
	\bar q_i\in\{-\bar\ell_i/2,\ldots	\bar\ell_i/2\}.
\eeq

\subsection{Relation with the EF}
In practical applications, 3- and 4-point functions are the most important objects. It is possible to treat 3-point functions in the CF or the EF. Since the latter is explicitly covariant, it is often more convenient. On the other hand, 4-point functions are treated most easily in the conformal frame approach. This creates a somewhat unfortunate situation when we have two formalisms for closely related objects. To remedy this, let us discuss how to go back and forth between the EF and the CF.

\paragraph{Embedding formalism to conformal frame}
\label{sec:EF_to_CF}
It is relatively straightforward to find the map \fref{toConformalFrame} from the embedding formalism tensor structures to the conformal frame ones. First one needs to project the 6D elements to the 4D ones and then to substitute the appropriate values of coordinates according to the choice of the conformal frame.

For 6D coordinates according to~\eqref{eq:projection_coordinates} and the definition of the conformal frame \eqref{eq:conformalframedefn_1} - \eqref{eq:conformalframedefn_4} one has
\algn{\begin{split}
X_1&=(0,0,0,0,1,0),\\
X_2&=((\bar z-z)/2,0,0,(z+\bar z)/2,1,-z\bar z),\\
X_3&=(0,0,0,1,1,-1),\\
X_4&=(0,0,0,L,1,-L^2),
\end{split}
}
and for the 6D polarizations according to~\eqref{eq:projection_auxiliary_vectors} one has
\beq
	(S_i)_a=\begin{pmatrix}
	(s_i)_\alpha\\
	-x_i^\mu\bar\sigma_\mu^{\dot\alpha\beta}(s_i)_\beta
	\end{pmatrix},\quad
	(\overline S_i)^a=\begin{pmatrix}
	(\bar s_i)_{\dot\beta}\bar\sigma^{\dot\beta\alpha}_\mu x_i^\mu\\
	(\bar s_i)_{\dot\alpha}
	\end{pmatrix}.
\eeq
In the last expression it is understood that all the coordinates $x$ belong to the conformal frame $x_{CF}$ \eqref{eq:conformalframedefn_1} - \eqref{eq:conformalframedefn_4}.

The final step is to perform the rescaling~\eqref{eq:rescaled_O} and to take the limit $L\to+\infty$. There is a very neat way to do it by recalling that 6D operators $O$ according to~\eqref{eq:scaling_6D} are homogeneous in 6D coordinates and 6D polarizations, thus
\beq
O(S_4,\overline S_4,X_4)L^{2\Delta_4}=O(S_4,\overline S_4,X_4)L^{2\CombinedDelta_{\OO}-\ell_4-\bar \ell_4}=O(S_4/L,\overline S_4/L,X_4/L^2).
\eeq
It is then clear that the final step is equivalent to the following substitution of the 6D coordinates at the 4th position
\begin{equation}
X_4\to \lim_{L\to +\infty}X_4/L^2=(0,0,0,0,0,-1)
\end{equation}
and for the 6D polarizations
\beq
	(S_4)_a\to \lim_{L\to +\infty}(S_4)_a/L=\begin{pmatrix}
	0\\
	-\bar\sigma_3^{\dot\alpha\beta}(s_4)_\beta
	\end{pmatrix},\quad
	(\overline S_4)^a\to \lim_{L\to +\infty}(\overline S_4)^a/L=\begin{pmatrix}
	(\bar s_4)_{\dot\beta}\bar\sigma^{\dot\beta\alpha}_3 \\
	0
	\end{pmatrix}.
\eeq

\paragraph{Conformal frame to embedding formalism}
\label{sec:CF_to_EF}
As discussed in section~\ref{sec:fourpointCFstructs},
4-point tensor structures are given by products of $\xi_i,\bar\xi_i,\eta_i,\bar\eta_i$ with vanishing total $U(1)$ charge. It is easy to convince oneself that any such product can be represented (not uniquely) by a product of $U(1)$-invariant bilinears
\begin{equation}\label{eq:pairs_4}
\bar\xi_i\xi_j,\quad
\bar\eta_i\eta_j,\quad
\xi_i\eta_j,\quad
\bar\xi_i\bar\eta_j,
\end{equation}
where $i,j=1\ldots4$. 
For $n\geq5$-point a general tensor structure is still represented by a product of bilinears, see footnote~\ref{ftn:spincenter}, but since there is no $U(1)$-invariance condition, the following set of bilinears should also be taken into account
\begin{equation}\label{eq:pairs_n}
\xi_i\xi_j,\quad
\eta_i\eta_j,\quad
\bar\xi_i\bar\xi_j,\quad
\bar\eta_i\bar\eta_j,\quad
\bar\eta_i\xi_j,\quad
\bar\xi_i\eta_j,
\end{equation}
where $i,j=1\ldots n$.

These bilinears themselves are tensor structures with low spin. Noticing that the EF invariants are also naturally bilinears in polarizations we can write a corresponding set of EF invariants with the same spin signatures. Translating these invariants to conformal frame via the procedure described above \fref{toConformalFrame}, one can then invert the result and express the bilinears~\eqref{eq:pairs_4} and~\eqref{eq:pairs_n} in terms of covariant expressions. We could call this procedure covariantization \fref{toEmbeddingFormalism}. The basis of EF structures is over-complete so the inversion procedure is ambiguous and one is free to choose one out of many options.

Since there is a finite number of bilinears~\eqref{eq:pairs_4} and~\eqref{eq:pairs_n} there will be a finite number of covariant tensor structures they can be expressed in terms of after the covariantization procedure. It is then very easy to see that one needs only the class of $n=4$ tensor structures to cover all the bilinears~\eqref{eq:pairs_4} and the class of $n=5$ tensor structures to cover all the bilinears~\eqref{eq:pairs_n}.

The ambiguity of the inversion procedure mentioned above is related to the linear relations between EF structures. Non-linear relations between EF structures arise due to the tautologies such as
\beq
	(\bar\xi_i\xi_j)(\bar\eta_k\eta_l)=(\bar\xi_i\bar\eta_k)(\xi_j\eta_l).
\eeq
This observation in principle allows to classify all relations between $n\geq 4$ EF invariants.

\subparagraph{Example.}
By going to the conformal frame we get
\begin{equation}\label{eq:Example_EF_To_CF}
\hat J^1_{23}=\frac{z}{z-1}\,\bar\xi_1\xi_1-\frac{\bar z}{\bar z-1}\,\bar\eta_1\eta_1,\quad
\hat J^1_{24}=-z\,\bar\xi_1\xi_1+\bar z\,\bar\eta_1\eta_1,\quad
\hat J^1_{34}=-\bar\xi_1\xi_1+\bar\eta_1\eta_1.
\end{equation}
Inverting these relation one gets
\begin{equation}\label{eq:Example_EF_To_CF_Inverted}
\bar\xi_1\xi_1=-\frac{z-1}{z\,(z-\bar z)}\,\Big( (\bar z-1)\, \hat J^{1}_{23}+\hat J^{1}_{24} \Big),\quad
\bar\eta_1\eta_1=-\frac{\bar z-1}{\bar z\,(z-\bar z)}\,\Big( (z-1)\,\hat J^{1}_{23}+\hat J^{1}_{24} \Big).
\end{equation}
We see right away that the invariants $J^1_{23}$, $J^1_{24}$ and $J^1_{34}$ must be dependent. One can easily get a relation between them by plugging~\eqref{eq:Example_EF_To_CF_Inverted} to the third expression~\eqref{eq:Example_EF_To_CF}. The obtained relation will match perfectly the linear relation~\eqref{eq:LinearRelation}.

Note that there is a factor $1/(z-\bar z)$ in \eqref{eq:Example_EF_To_CF_Inverted}, which suggests that the structure $\bar\xi_1\xi_1$ blows up at $z=\bar z$. This is not the case simply by the definition of $\xi$ and $\bar \xi$; instead, it is the combination of structures on the right hand side which develops a zero giving a finite value at $z=\bar z$. However, this value will depend on the way the limit is taken. This is related to the enhancement of the little group from $U(1)=SO(2)$ to $SO(1,2)$ at $z=\bar z$. At $z=\bar z$ it is no longer true that $\bar\xi_1\xi_1$ is a little group invariant. This enhancement implies certain boundary conditions for the functions which multiply the conformal frame invariants. See appendix~A of \cite{Kravchuk:2016qvl} for a detailed discussion of this point.

\subsection{Differentiation in the Conformal Frame}
\label{sec:differentiation_CF}
Now we would like to understand how to implement the action of the embedding formalism differential operators such as (\ref{eq:spinning_differential_operators_A}) and (\ref{eq:spinning_differential_operators_B}) directly in the conformal frame. We need to make two steps.  First, to understand the form of these differential operators in 4D space. This is done by using the projection of 6D differential operators to 4D  given in appendix~\ref{sec:DetailsOfTheEmbeddingFormalism}.
Second, to understand how to act with 4D differential operators directly in the conformal frame. We focus on this step in the remainder of this section. For simplicity, we restrict the discussion to the most important case of four points.

A correlation function in the conformal frame is obtained by restricting its coordinates $x$ to the conformal frame configurations $x_{CF}$. The action of the derivatives $\partial/\partial s$ and $\partial/\partial \bar s$ in polarizations on this correlation function is straightforward, since nothing happens to polarizations during this restriction. The only non-trivial part is the coordinate derivatives $\partial/\partial x_i$: in the conformal frame a correlator only depends on the variables $z$ and $\bar z$ which describe two degrees of freedom of the second operator and it is not immediately obvious how to take say the $\partial/\partial x_1$ derivatives.

The resolution is to recall that 4-point functions according to~\eqref{eq:n_point_correlation_function_constraint} are invariant under generic conformal transformation spanned by $15$ conformal generators $L_{MN}$. By using~\eqref{eq:lorentz_generators} one can see that it is equivalent to 15 differential equations
\beq
	(\difOperator L_{1\,MN}+\difOperator L_{2\,MN}+\difOperator L_{3\,MN}+\difOperator L_{4\,MN})\,f_4(x_i,s_i,\bar s_i)=0.
	\label{eq:confinvariance}
\eeq
The differential operators $\difOperator L_{i\,MN}$ defined in~\eqref{eq:lorentz_generators_6D_differential} together with~\eqref{eq:derivatives_projection} and~\eqref{eq:derivatives_projection_2} are given by linear combinations of derivatives $\partial/\partial x_i$, $\partial/\partial s_i$ and $\partial/\partial \bar s_i$. 
Out of 15 differential equations~\eqref{eq:confinvariance} one equation (for $L_{12}$) expresses the little group invariance under rotations in the $12$ plane and thus when restricted to the 4-point conformal frame~\eqref{eq:conformalframedefn_1} - \eqref{eq:conformalframedefn_4} does not contain derivatives $\partial/\partial x_i$. The remaining $14$ equations
allow to express the $14$ unknown derivatives $\partial/\partial x_i^\mu$ restricted to the conformal frame
in terms of $\partial/\partial x_2^0$, $\partial/\partial x_2^3$, $\partial/\partial s_i$ and $\partial/\partial \bar s_i$. Higher-order derivatives can be obtained in a similar way by differentiating \eqref{eq:confinvariance}.

Computation of general derivatives can be cumbersome, but in practice it is easily automated with \texttt{Mathematica}. We provide a conformal frame implementation of the differential operators \eqref{eq:spinning_differential_operators_A} - \eqref{eq:spinning_differential_operators_B}  \fref{opD4D, opDt4D, opd4D, opdb4D, opI4D, opN4D} 
as well as of the quadratic Casimir operator \fref{opCasimir24D} acting on 4-point functions. As a simple example (although it does not require differentiation in $x$), we display here the action of $\nabla_{12}$ on a generic conformal frame structure
\algn{
	\nabla_{12}\structgeneral g(z,\bar z)=&-\frac{(\ell_1+2q_1)(\bar\ell_2+2\bar q_2)}{4}\struct{q_1-\tfrac{1}{2}}{q_2}{q_3}{q_4}{\bar q_1}{\bar q_2-\tfrac{1}{2}}{\bar q_3}{\bar q_4}zg(z,\bar z)\nn\\
	&+\frac{(\ell_1-2q_1)(\bar\ell_2-2\bar q_2)}{4}\struct{q_1+\tfrac{1}{2}}{q_2}{q_3}{q_4}{\bar q_1}{\bar q_2+\tfrac{1}{2}}{\bar q_3}{\bar q_4}\bar zg(z,\bar z).
}
Other operators, e.g.~\eqref{eq:spinning_differential_operators_A}, give rise to more complicated expressions which however can still be efficiently applied to the seed CPWs.

\section{Package Demonstration}
\label{sec:example}
In this section we demonstrate the \texttt{CFTs4D} package on a simple example of a 4-point function with one vector and three scalar operators. The content of this section is intended to give a flavor of how the package works. This example should not be treated as a part of the package documentation, which is instead available through \texttt{Mathematica} help system together with more detailed and involved examples.

Consider the 4-point function
\beq\label{eq:object_to_compute}
	\<V_\mu(x_1)\phi(x_2)\phi(x_3)\phi(x_4)\>,
\eeq
where it is assumed for simplicity that the scalars are identical. We start by building a basis of tensor structures for the 3-point function $\<\OO^{(\ell,\bar\ell)}(x_0)\phi(x_3)\phi(x_4)\>$, where according to~\eqref{eq:sufficient_condition_non_zero_3_point_function} the operators $\OO$ can only be traceless symmetric with $\ell=\bar\ell$. We load the package and ask for normalized tensor structures
\begin{mmaCell}{Input}
  <<"CFTs4D`"
\end{mmaCell}

\begin{mmaCell}[moredefined={l, threePoint34, n3ListStructures}]{Input}
  $Assumptions=\{l>10\} ;
  threePoint34=n3ListStructures[\{\{0,0\},\{0,0\},\{l,l\}\}]
\end{mmaCell}

\begin{mmaCell}{Output}
  \{\mmaSup{\mmaSubSup{\mmaMyHat{J}}{\{1,2\}}{\{3\}}}{\(l\)}\}
\end{mmaCell}

\noindent This is the classical result that only one structure appears in such 3-point function.
Note that \texttt{n3ListStructures} always labels positions of the operators in 3-point functions as $1,2,3$. We have also made an explicit assumption that $\ell$ is large enough thus permitting the code to build the most generic structures. We go on to construct a list of normalized tensor structures for $\<V_\mu(x_1)\phi(x_2)\OO^{(\ell,\ell)}(x_3)\>$
\begin{mmaCell}[moredefined={threePoint12, n3ListStructures, l}]{Input}
  threePoint12=n3ListStructures[\{\{1,1\},\{0,0\},\{l,l\}\}]
\end{mmaCell}

\begin{mmaCell}{Output}
  \{\mmaSubSup{\mmaMyHat{J}}{\{2,3\}}{\{1\}} \mmaSup{\mmaSubSup{\mmaMyHat{J}}{\{1,2\}}{\{3\}}}{l},\mmaSup{\mmaMyHat{I}}{\{1,3\}} \mmaSup{\mmaMyHat{I}}{\{3,1\}} \mmaSup{\mmaSubSup{\mmaMyHat{J}}{\{1,2\}}{\{3\}}}{-1+l}\}
\end{mmaCell}

\noindent with $\ell=\bar\ell$ being the only case of interest for computing~\eqref{eq:object_to_compute}.

There are two structures available. We now look for the spinning differential operators which generate these structures. Since we are only interested in the exchange of traceless symmetric operators, our seed 3-point function is of the form $\<\FF^{(0,0)}\FF^{(0,0)}\OO^{(\ell,\ell)}\>$ with the following normalized seed tensor structures

\begin{mmaCell}[moredefined={seedStructure12, n3ListStructures, l}]{Input}
  seedStructure12=n3ListStructures[\{\{0,0\},\{0,0\},\{l,l\}\}][[1]]
\end{mmaCell}

\begin{mmaCell}{Output}
  \mmaSup{\mmaSubSup{\mmaMyHat{J}}{\{1,2\}}{\{3\}}}{l}
\end{mmaCell}

\noindent From \eqref{eq:spinning_differential_operator_full} and \eqref{eq:spinning_differential_operators_A} it is clear that the simplest differential operators raising the spin of the first field as $\Delta \ell_1= \Delta \bar{\ell}_1=1$ are given by
\begin{equation}\label{eq:example_3_point_Diff1}
\mathbf{D}^I=\Big\{ D_{12}, \;\widetilde{D}_{12} \Big\},
\end{equation}
so we write

\begin{mmaCell}[moredefined={diffOperators12, opDEF, opDtEF}]{Input}
  diffOperators12=\{\mmaDef{\(\pmb{\Xi}\)}[opDEF][1,2],\mmaDef{\(\pmb{\Xi}\)}[opDtEF][1,2]\}
\end{mmaCell}

\begin{mmaCell}{Output}
  \{\(\Xi\)[opDEF][1,2],\(\Xi\)[opDtEF][1,2]\}
\end{mmaCell}

\noindent We have surrounded the differential operators with $\Xi$ in order to shift all the scaling dimensions appropriately when applying them as explained in~\eqref{eq:shifts_1} and below.
Before proceeding further with the differential operators one needs to compute the kinematic factors

\begin{mmaCell}[moredefined={kinematicSeed12, n3KinematicFactor, l, kinematicStructure12}]{Input}
  kinematicSeed12=n3KinematicFactor[\{\mmaDef{\(\pmb{\Delta}\)}[1],\mmaDef{\(\pmb{\Delta}\)}[2],\mmaDef{\(\pmb{\Delta}\)}[3]\},\{\{0,0\},\{0,0\},\{l,l\}\}];
  kinematicStructure12=n3KinematicFactor[\{\mmaDef{\(\pmb{\Delta}\)}[1],\mmaDef{\(\pmb{\Delta}\)}[2],\mmaDef{\(\pmb{\Delta}\)}[3]\},\{\{1,1\},\{0,0\},\{l,l\}\}];
\end{mmaCell}

\noindent and combine them with the normalized tensor structures. Finally one applies the differential operators\footnote{The package applies differential operators only to non-normalized tensor structures, the function 
\texttt{denormalizeInvariants} is used to pull out the normalization factor explicitly according to~\eqref{eq:normalization_1} and~\eqref{eq:normalization_2}.}

\begin{mmaCell}[moredefined={seedStructure12N, kinematicSeed12, denormalizeInvariants, seedStructure12, diffStructures12N, diffOperators12},morepattern={\#}]{Input}
  seedStructure12N=kinematicSeed12*denormalizeInvariants[seedStructure12];
  diffStructures12N=#[seedStructure12N]&/@diffOperators12;
\end{mmaCell}

\noindent The result \texttt{diffStructures12N} of this calculation is a complicated expression. We simplify it by using the built-in functions and strip of the kinematic factor
\begin{mmaCell}[moredefined={diffStructures12N, applyEFProperties, applyJacobiRelations, diffStructures12, kinematicStructure12, normalizeInvariants, invSbSnorm, invSSnorm, invSbSbnorm},morepattern={\#}]{Input}
  diffStructures12N=diffStructures12N//applyEFProperties//applyJacobiRelations;
  diffStructures12=diffStructures12N/kinematicStructure12//normalizeInvariants;
  diffStructures12=diffStructures12//Simplify
\end{mmaCell}

\begin{mmaCell}{Output}
  \{-2 l \mmaSup{\mmaMyHat{I}}{\{1,3\}} \mmaSup{\mmaMyHat{I}}{\{3,1\}} \mmaSup{\mmaSubSup{\mmaMyHat{J}}{\{1,2\}}{\{3\}}}{-1+l}+\mmaSubSup{\mmaMyHat{J}}{\{2,3\}}{\{1\}} \mmaSup{\mmaSubSup{\mmaMyHat{J}}{\{1,2\}}{\{3\}}}{l} (-1-l+\(\Delta\)[1]-\(\Delta\)[2]+\(\Delta\)[3]),\\ 2 l \mmaSup{\mmaMyHat{I}}{\{1,3\}} \mmaSup{\mmaMyHat{I}}{\{3,1\}} \mmaSup{\mmaSubSup{\mmaMyHat{J}}{\{1,2\}}{\{3\}}}{-1+l}+\mmaSubSup{\mmaMyHat{J}}{\{2,3\}}{\{1\}} \mmaSup{\mmaSubSup{\mmaMyHat{J}}{\{1,2\}}{\{3\}}}{l} (-1+l-\(\Delta\)[1]+\(\Delta\)[2]+\(\Delta\)[3])\}
\end{mmaCell}

\noindent We get the differential basis of tensor structures for $\<V_\mu(x_1)\phi(x_2)\OO^{(\ell,\ell)}(x_3)\>$ which can be converted to the conventional basis~\eqref{eq:tensor_structures_via_differential_operators} via the matrix 
\begin{equation}
T^a =M^{ab} \mathbf{D}^b\, T_{seed}\equiv\mathbb{D}^a\, T_{seed}.
\end{equation}

\begin{mmaCell}[moredefined={inverseM, threePoint12, diffStructures12, M},morepattern={\#}]{Input}
  inverseM=Coefficient[#,threePoint12]&/@diffStructures12/.\{\mmaDef{\(\pmb{\Delta}\)}[3]\(\pmb{\to}\)\mmaDef{\(\pmb{\Delta}\)}\};
  M=Inverse[inverseM]//Factor
\end{mmaCell}

\begin{mmaCell}{Output}
  \{\{\mmaFrac{1}{2 (-1+\(\Delta\))},\mmaFrac{1}{2 (-1+\(\Delta\))}\},\{-\mmaFrac{-1+l+\(\Delta\)-\(\Delta\)[1]+\(\Delta\)[2]}{4 l (-1+\(\Delta\))},-\mmaFrac{1+l-\(\Delta\)-\(\Delta\)[1]+\(\Delta\)[2]}{4 l (-1+\(\Delta\))}\}\}
\end{mmaCell}

\noindent In other words, we have
\begin{equation}\label{eq:example_3_point_Matrix1}
M^{ab}=\frac{1}{2(\Delta-1)}
  \begin{pmatrix}
    1 & 1 \\
     \frac{\Delta_1-\Delta_2-\Delta-\ell+1}{4\ell} & \frac{\Delta_1-\Delta_2+\Delta-\ell-1}{4\ell}
  \end{pmatrix}.
\end{equation}

We proceed to compute the conformal partial waves. We start with the seed CPW corresponding to $p=0$,
\begin{mmaCell}[moredefined={seedEF, seedCPW}]{Input}
  seedEF=seedCPW[0];
\end{mmaCell}
This gives the standard EF expression for the scalar CPW (with a rather lengthy kinematic factor). At the level of 4-point functions it is more convenient to apply differential operators directly in the CF (even though we could have continued working in the EF), so we convert \texttt{seedEF} to the conformal frame expression
\begin{mmaCell}[moredefined={seedCF, seedEF, toConformalFrame, changeVariables}]{Input}
  seedCF=seedEF//toConformalFrame//changeVariables//Simplify
\end{mmaCell}
\begin{mmaCell}{Output}
  \mmaSup{(z \mmaOver{z}{_})}{\mmaFrac{1}{2} (-\(\Delta\)[1]-\(\Delta\)[2])} H[\{0,0\},\{\(\Delta\)[1],\(\Delta\)[2],\(\Delta\)[3],\(\Delta\)[4]\},\{0,0\}][z,\mmaOver{z}{_}]
\end{mmaCell}
Here \texttt{H[...]} represents the scalar conformal block. In a spinning cases this expression would explicitly contain the components of polarizations $\xi_i,\eta_i$, etc. as in the right hand-side of~\eqref{eq:fourpointcfdef}. It is however more convenient to convert this expression to the more abstract form where all the tensor structures are represented by the objects \texttt{CF4pt[...]} in a spirit of the left hand-side of~\eqref{eq:fourpointcfdef}
\begin{mmaCell}[moredefined={seedCF, collapseCFStructs},morepattern={\#}]{Input}
  seedCF=seedCF//collapseCFStructs[\{\mmaDef{\(\pmb{\Delta}\)}[1],\mmaDef{\(\pmb{\Delta}\)}[2],\mmaDef{\(\pmb{\Delta}\)}[3],\mmaDef{\(\pmb{\Delta}\)}[4]\}]
\end{mmaCell}
\begin{mmaCell}{Output}
  CF4pt[\{\(\Delta\)[1],\(\Delta\)[2],\(\Delta\)[3],\(\Delta\)[4]\},\{0,0,0,0\},\{0,0,0,0\},\{0,0,0,0\},\{0,0,0,0\},\\\mmaSup{(z \mmaOver{z}{_})}{\mmaFrac{1}{2} (-\(\Delta\)[1]-\(\Delta\)[2])} H[\{0,0\},\{\(\Delta\)[1],\(\Delta\)[2],\(\Delta\)[3],\(\Delta\)[4]\},\{0,0\}][z,\mmaOver{z}{_}]]
\end{mmaCell}
Here the structure \texttt{CF4pt[...]} contains explicitly the four external scaling dimensions $\Delta_i$, the four spins $\ell_i$, the four spins $\bar\ell_i$, and the parameters $q_i$ and $\bar q_i$, followed by the coefficient corresponding to this structure.

The advantage of working with abstract structures is that one can precompute a relatively simple rule of how a differential operator acts on the most generic structure and then apply it to any structure very quickly. In our case, we write
\begin{mmaCell}[moredefined={structure12Rules, operatorRule, opD4D, opDt4D}]{Input}
  structure12Rules=\{operatorRule[\mmaDef{\(\pmb{\Xi}\)}[opD4D]][1,2],\\    operatorRule[\mmaDef{\(\pmb{\Xi}\)}[opDt4D]][1,2]\};
\end{mmaCell}
which computes such rules for \texttt{opD4D} and \texttt{opDt4D}. Note that here we use the 4D operator instead of their 6D analogues. We now compute the action of these differential operators combining it with the rotation $M^{ab}$
\begin{mmaCell}[moredefined={CPWs, M, seedCF, structure12Rules, CF4pt, simplifyInCF4pt},morepattern={\#}]{Input}
  CPWs = M.(seedCF/.structure12Rules)//Expand[#,CF4pt]&//simplifyInCF4pt;
\end{mmaCell}

The expressions inside \texttt{CPWs} are relatively simple combinations of derivatives of the functions \texttt{H[...]}, which are however still too bulky to be displayed here. We can check that we get the right type of 4-point tensor structures. For instance one of two structures has the following form expected from~\eqref{eq:fourpointcfdef} for $\ell_1=\bar\ell_1=1$
\begin{mmaCell}[moredefined={CPWs}]{Input}
  CPWs[[1,1,1;;-2]]
\end{mmaCell}
\begin{mmaCell}{Output}
  CF4pt[\{\(\Delta\)[1],\(\Delta\)[2],\(\Delta\)[3],\(\Delta\)[4]\},\{1,0,0,0\},\{1,0,0,0\},\{-\mmaFrac{1}{2},0,0,0\},\{-\mmaFrac{1}{2},0,0,0\}]
\end{mmaCell}

We will now check that the quadratic Casimir equation is satisfied by the CPWs computed above. First we derive the ``replacement'' rule for the Casimir operator analogously to \texttt{opD4D}
\begin{mmaCell}[moredefined={ruleCasimir, operatorRule, opCasimir4D}]{Input}
  ruleCasimir = operatorRule[opCasimir4D][1,2];
\end{mmaCell}
We then obtain the Casimir equations
\begin{mmaCell}[moredefined={casimirEquation, CPWs, ruleCasimir, casimirEigenvalue2, CF4pt, mapInCF4pt, specialSimplifyH},morepattern={\#}]{Input}
  casimirEquation=((CPWs/.ruleCasimir)-casimirEigenvalue2[0]CPWs)//...;
\end{mmaCell}
\begin{mmaCell}[moredefined={casimirEquationFull, casimirEquation, plugSeedBlocks, plugCoefficients, plugPrefactor, plugKFunction}]{Input}
  casimirEquationFull=casimirEquation//plugSeedBlocks[1]//...;
\end{mmaCell}
In the above code excerpts $\ldots$ indicate some technical steps which can be found in the package documentation. The result is that \texttt{casimirEquationFull} contains the Casimir equation given in terms of the hypergeometric functions. We can now evaluate the equation numerically at some random point to convince ourselves that it is indeed satisfied
\begin{mmaCell}[moredefined={casimirEquationFull, l, expandCFStructs},morepattern={i_, i}]{Input}
  casimirEquationFull/.\{\mmaDef{\(\pmb{\Delta}\)}[i_]:>10+i,\mmaDef{\(\pmb{\Delta}\)}\(\pmb{\to}\)13,l\(\pmb{\to}\)2,...\}//expandCFStructs//Chop
\end{mmaCell}

\begin{mmaCell}{Output}
  \{0,0\}
\end{mmaCell}
Here $\ldots$ stand for a substitution of random high-precision numerical values for $z$ and $\bar z$.

\section{Conclusions}
\label{sec:conclusions}
In this paper we have described a framework for performing computations in 4D CFTs by unifying two different approaches, the covariant embedding formalism and the non-covariant conformal frame formalism. This framework allows to work with general 2-, 3- and 4-point functions and thus to construct the 4D bootstrap equations for the operators in arbitrary spin representation, ready for further numerical or analytical analysis.

In the embedding formalism we have explained the recipe for constructing tensor structures of $n$-point functions in the 6D embedding space. We have also summarized the so called spinning differential operators relating generic CPWs to the seed CPWs. The conformally covariant expressions in 4D are easily obtained from the 6D expressions by using the so called projection operation. For the objects like kinematic factors and 2-, 3-, and 4-point tensor structures we have performed the projection operation explicitly.

The construction of a \emph{basis} of tensor structures in the embedding formalism requires however the knowledge of a complete set of non-linear relations between products of the basic conformal invariants. Starting from $n=4$ it is rather difficult to find such a set of relations and thus the embedding formalism turns out to be practically inefficient for $n\geq 4$. This problem is solved using the conformal frame approach.

In the conformal frame we have provided a complete basis for $(n\geq 3)$-point tensor structures in a remarkably simple form. 
For instance in the $n=4$ case the tensor structures are simply monomials in polarization spinors with vanishing total charge under the $U(1)$ little group. In the $n<4$ cases the little group is larger and constructing its singlets becomes harder whereas the embedding formalism is easily manageable. Since the embedding formalism is also explicitly covariant it becomes preferable for working with 2- and 3-point functions.

With practical applications in mind, we have found the action of various differential operators on 4-point functions in the conformal frame formalism. We have also shown how to apply permutations in the conformal frame. These results allow one to work with the 4-point functions (and, consequently, the crossing equations)  entirely within the conformal frame formalism.

We have established a connection between the tensor structures constructed in the embedding and the conformal frame formalisms. The embedding formalism to conformal frame transition is straightforward and amounts to performing the 4D projection of the 6D structures and setting all the coordinates to the conformal frame. The conformal frame to the embedding formalism transition is slightly more complicated since it is not uniquely defined due to redundancies among the allowed 6D structures. After ``translating'' all the basic 6D structures to the conformal frame one inverts these relations by choosing only the independent 6D structures.

Finally, we have implemented our framework as a Mathematica package freely available at~\href{https://gitlab.com/bootstrapcollaboration/CFTs4D#cfts4d}{https://gitlab.com/bootstrapcollaboration/CFTs4D}. It can perform any manipulations with 2-, 3- and 4-point functions in both formalism switching between them when needed. A detailed documentation is incorporated in the package with many explicit examples.

In the appendices we made our best effort to establish consistent conventions; we have provided a proper normalization of 2-point functions and the seed conformal blocks and summarized all the Casimir differential operators available in 4D. We have also given some extra details on permutation symmetries and conserved operators.

It is our hope that this paper will aid the development of conformal bootstrap methods in 4D and will facilitate their application to spinning correlation functions, such as 4-point functions involving fermionic operators, global symmetry currents and stress-energy tensors.

\section*{Acknowledgments}
We thank Alejandra Castro, Tolya Dymarsky, Emtinan Elkhidir, Gabriele Ferretti, Diego Hofman, Hugh Osborn, Jo\~ao Penedones, Riccardo Rattazzi, Fernando Rej\'on-Barrera, Slava Rychkov, Volker Schomerus, David Simmons-Duffin, Marco Serone and Alessandro Vichi for useful discussions. We particularly thank Marco Serone for his valuable comments on the draft and Emtinan Elkhidir for collaboration on the initial stages of this work.
DK and PK are grateful to the organizers of Boostrap 2016 workshop and the Galileo Galilei Institute for Theoretical Physics where the main ideas of this project were born. PK would like to thank the Institute for Advanced Study, where part of this work was completed, for hospitality.
This work is supported in part by the DOE grant DE-SC0011632 (PK).

\appendix

\section{Details of the 4D Formalism}
\label{app:conventions4D}
We work in the signature $-+++$ and denote the diagonal 4D Minkowski metric by $h_{\mu\nu}$. We mostly follow the conventions of Wess and Bagger~\cite{Wess:1992cp}.

The representations of the connected Lorentz group in 4D are labeled by a pair of non-negative integers $(\ell,\bar\ell)$. These representations can be constructed as the highest-weight irreducible components in a tensor product of the two basic spinor representations $(1,0)$ and $(0,1)$. 

We denote the objects in the left-handed spinor representation $(1,0)$ as $\psi_\alpha$, $\alpha=1,2$, and the objects in its dual representation as $\psi^\alpha$. The original and the dual representations are equivalent via the identification
\beq\label{eq:leftduality}
	\psi_\alpha=\epsilon_{\alpha\beta}\psi^\beta,\quad \psi^\alpha=\epsilon^{\alpha\beta}\psi_\beta,
\eeq
where
\beq
	\epsilon^{12}=-\epsilon^{21}=\epsilon_{21}=-\epsilon_{12}=+1.
\eeq
Because of the equivalence between $(1,0)$ and its dual representation, we will not be careful to distinguish them in the text, the distinction in formulas will be clear from the location of indices.

The right-handed spinor representation $(0,1)$ is the complex conjugate of the left-handed spinor representation, and the objects transforming in $(0,1)$ representation will be denoted as $\chi_{\dot\alpha}$. Here the dot should not be considered as part of the index, but rather as an indication that this index transforms in $(0,1)$ and not in $(1,0)$ representation.
For example, the definition of $(0,1)$ representation is essentially
\beq
	\psi^\dagger_{\dot\alpha}=(\psi_\alpha)^\dagger.
\eeq
The dual of $(0,1)$ is equivalent to $(0,1)$ via the conjugation of \eqref{eq:leftduality}
\beq
	\chi_{\dot\alpha}=\epsilon_{\dot\alpha\dot\beta}\chi^{\dot\beta},\quad \chi^{\dot\alpha}=\epsilon^{\dot\alpha\dot\beta}\chi_{\dot\beta},
\eeq
where $\epsilon_{\dot\alpha\dot\beta}\equiv\epsilon_{\alpha\beta},\,\epsilon^{\dot\alpha\dot\beta}\equiv\epsilon^{\alpha\beta}$. We use the contraction conventions
\beq
	\psi_1\psi_2=\psi_1^{\alpha}\psi_2{}_{\alpha},\quad \chi_1\chi_2=\chi_1{}_{\dot\alpha}\chi_2^{\dot\alpha}.
\eeq

The tensor product $(1,0)\otimes(0,1)=(1,1)$ is equivalent to the vector representation, and the equivalence is established by the 4D sigma matrices $\sigma^\mu_{\alpha\dot\beta}$ and $\bar\sigma^{\mu\dot\alpha\beta}$, which we define as
\beq
	 \sigma^0=\begin{pmatrix}
	 -1 & 0\\
	 0 & -1
	 \end{pmatrix},\quad
	 \sigma^1=\begin{pmatrix}
	 0 & 1\\
	 1 & 0
	 \end{pmatrix},\quad
	 \sigma^2=\begin{pmatrix}
	 0 & -i\\
	 i & 0
	 \end{pmatrix},\quad
	 \sigma^3=\begin{pmatrix}
	 1 & 0\\
	 0 & -1
	 \end{pmatrix},\quad
\eeq
and $\bar\sigma^0=\sigma^0,\;\bar\sigma^1=-\sigma^1,\;\bar\sigma^2=-\sigma^2,\;\bar\sigma^3=-\sigma^3$. 
For a convenient summary of relations involving sigma-matrices see for example~\cite{Dreiner:2008tw}.\footnote{One should download and compile the version with mostly plus metric. Notice also a factor of ${i}$ difference between their $\sigma^{\mu\nu}$ and $\bar\sigma^{\mu\nu}$ and ours $\SS^{\mu\nu}$ and $\bar\SS^{\mu\nu}$.}

For primary operators we adopt the convention to write them out with dotted indices upstairs and the undotted indices downstairs
\beq
	\OO^{\dot\alpha_1\ldots\dot\alpha_{\bar\ell}}_{\beta_1\ldots\beta_\ell}.
\eeq
In this notation the index-full version of~\eqref{eq:conjugation} is
\beq\label{eq:conjugation_app}
	\conj{\OO}^{\dot\beta_1\ldots\dot\beta_{\ell}}_{\alpha_1\ldots\alpha_{\bar\ell}}\equiv
    (-1)^{\bar\ell-\ell}\;
	\epsilon_{\alpha_1\alpha_1'}\ldots\epsilon_{\alpha_{\bar\ell}\alpha'_{\bar\ell}}\epsilon^{\dot\beta_1\dot\beta'_1}\cdots\epsilon^{\dot\beta_\ell\dot\beta'_\ell}\OO^\dagger{}^{\alpha'_1\ldots\alpha'_{\bar\ell}}_{\dot\beta'_1\ldots\dot\beta'_\ell}.
\eeq

\paragraph{Action of Conformal Generators}
We denote the conformal generators by $P,K,D,M$. We choose to work with anti-Hermitian generators (related to the Hermitian ones by a factor of $i$)
\beq
	D^\dagger=-D,\quad P^\dagger=-P,\quad K^\dagger=-K,\quad M^\dagger=-M,
\eeq
which allow us to avoid many factors of $i$ in the formulas below (note that even though $D$ is anti-Hermitian, its adjoint action has real eigenvalues).
These generators satisfy the following algebra
\beq
[D,D] = 0,\quad
[D,P_{\mu}] =P_\mu,\quad
[D,K_{\mu}] =-K_\mu,
\eeq
\beq
[P_\mu,P_\nu] =0,\quad
[K_\mu,K_\nu] =0,\quad
[K_\mu,P_{\nu}] = 2 h_{\mu\nu}D - 2 M_{\mu\nu},
\eeq
\beq
[M_{\mu\nu},D] =0,\quad
[M_{\mu\nu},P_{\rho}] =h_{\nu\rho}P_\mu-h_{\mu\rho}P_\nu,\quad
[M_{\mu\nu},K_{\rho}] = h_{\nu\rho}K_\mu-h_{\mu\rho}K_\nu,
\eeq
\beq\label{eq:algebraLorentzGenerators4D}
[M_{\mu\nu},M_{\rho\sigma}]=h_{\nu\rho}M_{\mu\sigma}-h_{\mu\rho}M_{\nu\sigma}-h_{\nu\sigma}M_{\mu\rho}+h_{\mu\sigma}M_{\nu\rho}.
\eeq
The action of the conformal generators on primary fields is given by
\algn{\label{eq:DilatationAction}
[D,\OO(x,s,\bar s)]&=(x^\mu\partial_\mu+\Delta)\,\OO(x,s,\bar s),\\
[P_\mu,\OO(x,s,\bar s)]&=\partial_\mu\,\OO(x,s,\bar s),\\
[K_\mu,\OO(x,s,\bar s)]&=(2x_\mu x^\sigma-x^2\delta^\sigma_\mu)\partial_\sigma\OO(x,s,\bar s)+2(\Delta\, x_\mu-x^\sigma \MM_{\mu\sigma}) \OO(x,s,\bar s),\\
\label{eq:LorentzAction}
[M_{\mu\nu},\OO(x,s,\bar s)]&=(x_\nu\partial_\mu-x_\mu\partial_\nu)\OO(x,s,\bar s)+\MM_{\mu\nu}\OO(x,s,\bar s),
}
where the spin generators are
\beq
	\MM_{\mu\nu}\OO(x,s,\bar s)=\left(-s^\alpha(\SS_{\mu\nu})_\alpha{}^\beta\frac{\partial}{\partial s^\beta}-\bar s_{\dot\alpha}(\bar\SS_{\mu\nu})^{\dot\alpha}{}_{\dot\beta}\frac{\partial}{\partial \bar s_{\dot\beta}}\right)\OO(x,s,\bar s).
\eeq
We have defined here the generators of the left- and right-handed spinor representations
\beq\label{eq:LorentzGeneratorSpinor}
	(\SS_{\mu\nu})_\alpha{}^\beta=-\frac{1}{4}(\sigma_\mu\bar\sigma_\nu-\sigma_\nu\bar\sigma_\mu)_{\alpha}{}^\beta,
\eeq
\beq\label{eq:LorentzGeneratorSpinorConjugated}
	(\bar \SS_{\mu\nu})^{\dot\alpha}{}_{\dot\beta}=-\frac{1}{4}(\bar\sigma_\mu\sigma_\nu-\bar\sigma_\nu\sigma_\mu)^{\dot\alpha}{}_{\dot\beta},
\eeq
which satisfy the same commutation relations as $M_{\mu\nu}$.
Notice that as usual the differential operators in the right hand side of~\eqref{eq:DilatationAction}-\eqref{eq:LorentzAction} have the commutation relations opposite to those of the Hilbert space operators in the left hand side. This is because if the Hilbert space operators $A$ and $B$ act on fields by differential operators $\mathfrak{A}$ and $\mathfrak{B}$, then their product $AB$ acts by $\mathfrak{BA}$.

\paragraph{Action of Space Parity}
If a theory preserves parity, there exists a unitary operator $\PP$ with the following commutation rule with Lorentz generators
\beq
	\PP M_{0i}\PP^{-1}=-M_{0i},\quad \PP M_{ij}\PP^{-1} = M_{ij},
\eeq
where $i,j=1,2,3$. Applying this to \eqref{eq:LorentzAction} at $x=0$, we see that  
\beq
	[M_{\mu\nu},\PP\OO_\alpha(0)\PP^{-1}]=(\bar \SS_{\mu\nu})^{\dot\alpha}{}_{\dot\beta}\PP\OO_\beta(0)\PP^{-1}.
\eeq
This implies that we can define an operator $\widetilde\OO$ as
\beq\label{eq:ParitytransformationOnFermion}
	\widetilde\OO^{\dot\alpha}(x)\equiv-i\PP\OO_\alpha(\PP x)\PP^{-1}
\eeq
which transform as a primary operator in the representation $(0,1)$. We also have $\PP x^0=x^0,\,\PP x^k=-x^k,\,\,k=1,2,3.$  More generally, it is easy to check that we can consistently define
\beq
	\widetilde\OO^{\dot\alpha_1\ldots\dot\alpha_\ell}_{\beta_1\ldots\beta_{\bar\ell}}(x)\equiv(-i)^{\ell+\bar\ell}\PP\OO^{\dot\beta_1\ldots\dot\beta_{\bar\ell}}_{\alpha_1\ldots\alpha_{\ell}}(\PP x)\PP^{-1}.
\eeq
The factor of $i$ was introduced to reproduce the standard parity action on traceless symmetric operators in the $\widetilde\OO=\OO$ case.

The above definition provides the most generic action of parity on the operators $\OO$ which can be slightly rewritten as
\algn{
\label{eq:ParityTransformationGeneric}
	\PP\OO^{\dot\beta_1\ldots\dot\beta_{\bar\ell}}_{\alpha_1\ldots\alpha_{\ell}}(x)\PP^{-1}&=i^{\ell+\bar\ell}\widetilde{\OO}^{\dot\alpha_1\ldots\dot\alpha_{\ell}}_{\beta_1\ldots\beta_{\bar\ell}}(\PP x),
}
or equivalently in index-free notation
\beq\label{eq:ParityIndexFreeNotation}
	\PP\OO(x,s,\bar s)\PP^{-1}=\widetilde{\OO}(\PP x,\PP s,\PP \bar s),\quad (\PP \bar s)_{\dot\alpha}=i s^{\alpha},\,\quad (\PP s)^{\alpha}=i \bar s_{\dot \alpha}.
\eeq
Notice that if $\OO$ transforms in the $(\ell,\bar\ell)$ representation then the operator $\widetilde\OO$ transforms in $(\bar\ell,\ell)$ and may or may not be related to the operator $\conj{\OO}$ defined in~\eqref{eq:conjugation} or to $\OO$ itself if $\ell=\bar\ell$. This depends on a specific theory. What is important for us is that in a theory which preserves $\PP$ there is a relation between correlators involving $\OO_i$ and $\widetilde\OO_i$
\algn{
	\<0|\OO_1(x_1,s_1,\bar s_1)\cdots\OO_n&(x_n,s_n,\bar s_n)|0\>=\nn\\
	=&\<0|\PP\OO_1(x_1,s_1,\bar s_1)\PP^{-1}\cdots\PP\OO_n(x_n,s_1,\bar s_1)\PP^{-1}|0\>\nn\\
	=&\<0|\widetilde{\OO}_1(\PP x_1,\PP s_1,\PP \bar s_1)\cdots\widetilde{\OO}_n(\PP x_n,\PP s_n,\PP \bar s_n)|0\>.
}
Written in terms of tensor structures this equality reads as
\beq
	\sum_I \mathbb{T}_n^I g_n^I=\sum_I (\PP\widetilde{\mathbb{T}}_n^I) \widetilde g_n^I,
\eeq
where $\PP\widetilde{\mathbb{T}}_n^I$ is given by $\widetilde{\mathbb{T}}_n^I$ with $x\to\PP x$, $s\to \PP s$, $\bar s\to \PP\bar s$ and $\widetilde{\mathbb{T}}_n^I$ are the tensor structures appropriate to the correlators with the operators $\widetilde{\OO}_i$.\footnote{If there are any parity-odd cross-ratios (i.e. $n\geq 6$) then $\widetilde g$ should have these with reversed signs.} We provide the rules for the action of $\PP$ on various tensor structures in equations~\eqref{eq:InvariantsParityA},~\eqref{eq:InvariantsParityB} and~\eqref{eq:cf4parity}~\fref{applyPParity}.

\paragraph{Action of Time Reversal}
If a theory has time reversal symmetry, there exists an anti-unitary operator $\TT$ with the following commutation rule with Lorentz generators
\beq
	\TT M_{0i}\TT^{-1}=-M_{0i},\quad \TT M_{ij}\TT^{-1} = M_{ij},
\eeq
where $i,j=1,2,3$. Applying it to \eqref{eq:LorentzAction} at $x=0$, we see that  
\beq
	[M_{\mu\nu},\TT\OO_\alpha(0)\TT^{-1}]= \left[(\bar \SS_{\mu\nu})^{\dot\alpha}{}_{\dot\beta}\right]^*\TT\OO_\beta(0)\TT^{-1}.
\eeq
This implies that $\TT\OO_\beta(0)\TT^{-1}$ transforms as $\psi^\beta$ and we can define the operator $\widehat\OO$ as
\algn{
\label{eq:TimeReversalTransformationOnFermion}
	\widehat\OO_{\alpha}(x)&\equiv-i\epsilon_{\alpha\beta}\TT\OO_\beta(\TT x)\TT^{-1},
}
where $\TT x^0=-x^0,\,\TT x^k=x^k,\,\,k=1,2,3$.
One can similarly define
\algn{
	\widehat\OO^{\dot\alpha}(x)&\equiv i\epsilon^{\dot\alpha\dot\beta}\TT\OO^{\dot\beta}(\TT x)\TT^{-1}
}
and extend the above definitions to arbitrary representations in an obvious way.
For traceless symmetric operators in the $\widehat{\OO}=\OO$ case, this reproduces the standard time reversal action. In index-free notation we can write\footnote{Note that $\TT s$ and $\TT\bar s$ are not complex conjugates of each other even if $s$ and $\bar s$ are, so to avoid confusion here we do not assume that $s$ and $\bar s$ are complex-conjugate. There is always a second complex conjugation (see below), so this is only intermediate.}
\beq\label{eq:TimeReversalIndexFreeNotation}
	\TT\OO(x,s,\bar s)\TT^{-1}=\widehat\OO(\TT x,\TT s,\TT \bar s),\quad (\TT s)^\alpha= is_{\dot \alpha}^*,\, (\TT\bar s)_{\dot\alpha}= -i(\bar s^*)^{\alpha}.
\eeq
Again, $\widehat{\OO}$ may or may not be related to $\OO$ depending on a theory. The only important point is that there is a relation between correlators with $\OO_i$ and $\widehat{\OO}_i$ in a theory preserving the time reversal symmetry
\algn{
	\<0|\OO_1(x_1,s_1,\bar s_1)\cdots\OO_n&(x_n,s_n,\bar s_n)|0\>=\nn\\
	=&\left[\<0|\TT\OO_1(x_1,s_1,\bar s_1)\TT^{-1}\cdots\TT\OO_n(x_n,s_1,\bar s_1)\TT^{-1}|0\>\right]^*\nn\\
	=&\left[\<0|\widehat\OO_1(\TT x_1,\TT s_1,\TT \bar s_1)\cdots\widehat\OO_n(\TT x_n,\TT s_n,\TT \bar s_n)|0\>\right]^*,
}
where the conjugation happens because of the anti-unitarity of $\TT$.\footnote{As an extreme example $\TT i\TT^{-1}=-i$, so we have $i=\<0|i|0\>=[\<0|\TT i\TT^{-1}|0\>]^*\neq\<0|\TT i\TT^{-1}|0\>$.} Written in terms of tensor structures this equality reads as
\beq
	\sum_I \mathbb{T}_n^I g_n^I=\sum_I (\TT\widehat {\mathbb{T}}_n^I) (\widehat g_n^I)^*,
\eeq
where $\TT\widehat {\mathbb{T}}_n^I$ is given by $(\widehat {\mathbb{T}}_n^I)^*$ with the replacements $x\to \TT x$, $s\to\TT s$, $\bar s\to \TT\bar s$ made before the conjugation and $\widehat {\mathbb{T}}_n^I$ are the structures appropriate for the operators $\widehat{\OO}_i$.

Computing $\TT\widehat {\mathbb{T}}_n^I$ is easy, since we can construct $\TT$ conjugation from $\PP$ and the rotation $e^{i\pi M^{03}+\pi M^{12}}$. The latter rotation sends $s\to s$, $\bar s\to -\bar s$, which takes $\TT s$ and $\TT\bar s$ to $\PP s$ and $\PP \bar s$. The end result is
\begin{equation}\label{eq:time_reversal_simple_rule}
\TT\widehat {\mathbb{T}}_n^I=\left(\PP\widehat {\mathbb{T}}_n^I\right)^*.
\end{equation}
We list the rules for the action of $\TT$ on tensor structures in equations \eqref{eq:InvariantsTimeReversalA},~\eqref{eq:InvariantsTimeReversalB} and~\eqref{eq:cf4time} \fref{applyTParity}.

\section{Details of the 6D Formalism}
\label{sec:DetailsOfTheEmbeddingFormalism}
In this appendix we describe our conventions for the 6D embedding space. We mostly follow~\cite{SimmonsDuffin:2012uy,Elkhidir:2014woa}.

We work in the signature $\{-++++-\}$, and we denote the 6D metric by $h_{MN}$.
We often use the lightcone coordinates
\begin{equation}
X^\pm \equiv X^4\pm X^5,
\end{equation}
and write the components of 6D vectors as
\begin{equation}
X^M=\{X^\mu,\,X^+,\,X^-\}.
\end{equation}
The metric in lightcone coordinates has the components 
\beq
h_{+-}=h_{-+}=\frac{1}{2},\quad h^{+-}=h^{-+}=2.
\eeq

The 6D Lorentz group $Spin(2,4)$ is isomorphic to the $SU(2,2)$ group. The latter can be defined as the group of 4 by 4 matrices $U$ which act on 4-component complex vectors $V_a$ and preserve the sesquilinear form
\beq
	\<V,W\> = g^{\bar a b} (V_a)^* W_b, \quad \<UV,UW\> = \<V,W\>.
\eeq 
Here the metric tensor $g^{\bar a b}$ is a Hermitian matrix with eigenvalues $\{+1,+1,-1,-1\}$, which we choose to be
\beq
	g^{\bar ab}\equiv g^{ b \bar a}\equiv\begin{pmatrix}
		0 & 0 & i & 0\\
		0 & 0 & 0 & i\\
		-i & 0 & 0 & 0\\
		0 & -i & 0 & 0
	\end{pmatrix}_{ab}.
\eeq
The bar over the index $\bar a$ indicates that this index transforms in a complex conjugate representation. In other words, we say that $V_a$ transforms in the fundamental representation while
\beq
	V^*_{\bar a} \equiv (V_a)^*
\eeq
transforms in the complex conjugate of the fundamental representation (that is, by matrices $U^*$). The metric $g^{\bar a b}$ establishes an isomorphism between the complex conjugate representation and the dual representation
\beq
	\overline V^a\equiv g^{a 
	\bar b}\overline V_{\bar b}.
\eeq
We say that $\overline V^a$ transforms in the anti-fundamental representation (that is, the anti-fundamental representation is the dual of the fundamental representation). The inverse isomorphism is established by the tensor
\beq
	g_{\bar a b}\equiv g_{b \bar a}\equiv -g^{\bar a b}.
\eeq
We have the relations
\beq
	g_{a\bar b}g^{\bar b c}=g^{c\bar b}g_{\bar b a}=\delta_a^c,\quad (g^{a\bar b})^*=g^{\bar a b}.
\eeq

The isomorphism between $Spin(2,4)$ and $SU(2,2)$ can be established by identifying the vector representation of $Spin(2,4)$ with the exterior square of the fundamental or anti-fundamental representations of $SU(2,2)$.\footnote{The fundamental and anti-fundamental representations themselves are the two spinor representations of $Spin(2,4)$.} This equivalence is provided by the invariant tensors $\Sigma^M_{ab}$ and $\overline \Sigma^{M\, ab}$ defined by
\begin{equation}
\Sigma_{ab}^\mu=\begin{pmatrix}
0 & -(\sigma^\mu\epsilon)_\alpha^{\;\;\dot\beta} \\
(\bar\sigma^\mu\epsilon)^{\dot\alpha}_{\;\;\beta} & 0
\end{pmatrix},\quad
\Sigma_{ab}^+=\begin{pmatrix}
0 & 0 \\
0 & 2\,\epsilon^{\dot\alpha\dot\beta}
\end{pmatrix},\quad
\Sigma_{ab}^-=\begin{pmatrix}
-2\,\epsilon^{\dot\alpha\dot\beta} & 0 \\
0 & 0
\end{pmatrix},
\end{equation}
and
\begin{equation}
\overline\Sigma^{\mu\,ab}=\begin{pmatrix}
0 & -(\epsilon\sigma^\mu)^\alpha_{\;\;\dot\beta} \\
(\epsilon\bar\sigma^\mu)_{\dot\alpha}^{\;\;\beta} & 0
\end{pmatrix},\quad
\overline\Sigma^{+\,ab}=\begin{pmatrix}
-2\,\epsilon^{\alpha\beta} & 0 \\
0 & 0
\end{pmatrix},\quad
\overline\Sigma^{-\,ab}=\begin{pmatrix}
0 & 0 \\
0 & 2\,\epsilon^{\dot\alpha\dot\beta}
\end{pmatrix}.
\end{equation}
These tensors have the following simple conjugation properties,
\begin{equation}\label{eq:sigmareality}
\left(\Sigma_{ab}^M\right)^* =g_{\bar a a'}g_{\bar b b'}\overline\Sigma^{M\,a'b'}
\quad\quad
\big(\overline \Sigma^{M\, ab}\big)^* =g^{\bar a a'}g^{\bar b b'}\Sigma^{M}_{a'b'}.
\end{equation}
The above sigma-matrices satisfy many useful relations, for an incomplete list of them see appendix A in~\cite{Elkhidir:2014woa}. 
%-------------------------------------------------------------------------
%Useful relations for evaluating traces
%\begin{align}
%Tr[\Sigma^A\overline\Sigma^B]=Tr[\overline\Sigma^A\Sigma^B]&=4\,\eta^{AB},\\
%Tr[\Sigma^A\overline\Sigma^B\Sigma^C\overline\Sigma^D]=
%Tr[\overline\Sigma^A\Sigma^B\overline\Sigma^C\Sigma^D]&=4\,\big(\eta^{AB}\eta^{CD}-\eta^{AC}\eta^{BD}+\eta^{AD}\eta^{BC} \big).
%\end{align}
%-------------------------------------------------------------------------
Using the sigma matrices we define the coordinate matrices
\begin{equation}\label{eq:coordinates}
\mathbf{X}_{ab}\equiv X_M\Sigma^M_{ab}=-\mathbf{X}_{ba},
\quad\;\;
\overline{\mathbf{X}}^{ab}\equiv X_M\overline{\Sigma}^{M\,ab}=-\overline{\mathbf{X}}^{ba},
\end{equation}
which satisfy the algebra
\begin{equation}\label{eq:XXbar}
\hspace{1mm}_a(\mathbf{X}_i\overline{\mathbf{X}}_j)^b
+
\hspace{1mm}_a(\mathbf{X}_j\overline{\mathbf{X}}_i)^b
=2\,(X_i\cdot X_j)\delta_a^b.
\end{equation}

We can now identify the $SU(2,2)$ generators corresponding to the standard 6D Lorentz generators
\begin{equation}\label{eq:generators_SU(2,2)}
\Sigma^{MN} \equiv \frac{1}{4}\, (\Sigma^M\overline \Sigma^N- \Sigma^N\overline \Sigma^M),\quad
\overline\Sigma^{MN} \equiv \frac{1}{4}\, (\overline\Sigma^M \Sigma^N- \overline\Sigma^N \Sigma^M),
\end{equation}
satisfying the commutation relations
\algn{
	[\Sigma^{MN},\Sigma^{PQ}]&=h^{NP}\Sigma^{MQ}-h^{MP}\Sigma^{NQ}-h^{NQ}\Sigma^{MP}+h^{MQ}\Sigma^{NP},\\
	[\overline\Sigma^{MN},\overline\Sigma^{PQ}]&=h^{NP}\overline\Sigma^{MQ}-h^{MP}\overline\Sigma^{NQ}-h^{NQ}\overline\Sigma^{MP}+h^{MQ}\overline\Sigma^{NP},
}
thus establishing the isomorphism $Spin(2,4)\simeq SU(2,2)$ at Lie algebra level.

By comparing the expressions for $\Sigma_{\mu\nu}$ and $\overline{\Sigma}^{\mu\nu}$ with $\SS^{\mu\nu}$ and $\overline{\SS}^{\mu\nu}$, we find that under the Lorentz $Spin(1,3)$ subgroup of $Spin(2,4)$ the fundamental and anti-fundamental representations of $SU(2,2)$ decompose as
\beq
	V_a=\begin{pmatrix}
	V_\alpha\\ V^{\dot\alpha}
	\end{pmatrix},\quad
	\overline W^a=\begin{pmatrix}
	\overline W^\alpha\\ \overline W_{\dot\alpha}
	\end{pmatrix}.
\eeq
In other words, we write $V_\alpha$ or $V^{\dot\alpha}$ to refer to first two or second two components of $V_a$, and analogously for $\overline{W}^a$.

\paragraph{Conformal algebra in 6D notation}
We can identify explicitly the conformal generators with the 6D Lorentz algebra
\beq\label{eq:lorentz_definition}
	M_{\mu\nu}=L_{\mu\nu},\quad D=L_{45},\quad P_\mu=L_{5\mu}-L_{4\mu},\quad K_\mu=-L_{4\mu}-L_{5\mu}.
\eeq
With these conventions, the generators $L_{MN}$ satisfy the algebra
\beq
	[L_{MN},L_{PQ}]=h_{NP}L_{MQ}-h_{MP}L_{NQ}-h_{NQ}L_{MP}+h_{MQ}L_{NP}.
\eeq
These generators act on the 6D primary operators as
\begin{equation}\label{eq:lorentz_generators}
[L_{MN},O(X,S,\overline S)]=\difOperator L_{MN} O(X,S,\overline S),
\end{equation}
where the differential 6D generator is defined as
\begin{equation}\label{eq:lorentz_generators_6D_differential}
\difOperator L_{MN}\equiv -(X_{M}\partial_{N}-X_{N}\partial_{M})-
S \overline{\Sigma}_{MN} \partial_{S}-
\overline{S} \Sigma_{MN} \partial_{\overline{S}}.
\end{equation}
It is sometimes convenient to work with the conformal generators in $SU(2,2)$ notation
\beq
\label{eq:generators_SU(2,2)}
L_{a}{}^{b}\equiv\big[\Sigma^{MN}\big]_a{}^{b}\,L_{MN},\quad L_{i\,MN}=-\frac{1}{2}L_{a}{}^{b}\big[\Sigma_{MN}\big]_b{}^{a}.
\eeq
In this notation the conformal generators obey the commutation relations
\begin{equation}
\left[L_{a}{}^{b},L_{c}{}^{d}\right]
=2\delta_c^b\,L_{a}{}^{d}-2\delta_a^d\,L_{c}{}^{b}.
\end{equation}
We also have the following action on the primary operators
\begin{equation}
[L_a{}^b,O(X,S,\overline{S})]=\difOperator{L}_a{}^b\,O(X,S,\overline S),
\end{equation}
where $\difOperator{L}_a{}^c$ is the differential operator associated to the 6D generator $L_a{}^c$ in Hilbert space
\begin{equation}\label{eq:lorentz_generators_6D_SU(2,2)_differential}
\difOperator{L}_a{}^b\equiv
-\frac{1}{2}\,\left[\big(\mathbf X \overline{\Sigma}^M\big)_a^{\,\,b}\,\partial_{M} - \big(\Sigma^M\overline{\mathbf X}\big)_a^{\,\,b}\,\partial_{M}  \right]
+\frac{1}{2}\,\delta_a^b\,\left( S\cdot\partial_S-\overline S\cdot\partial_{\overline S} \right)
-2\,\left( S_a \partial_S^b-\overline S^b\partial_{\overline S\,a} \right).
\end{equation}

\paragraph{Embedding formalism}
In the embedding formalism the flat 4D space is identified with a particular section of the 6D light cone $X^2=0$. Namely, we take the Poincare section $X^+=1$, which then implies
\beq
	X^-=-X^\mu X_\mu.
\eeq
The 4D coordinates $x_\mu$ are identified on this section as
\beq
	x^\mu = X^\mu.
\eeq
In particular, on the Poincare section we have
\begin{equation}\label{eq:projection_coordinates}
X^M\Big|_\text{Poincare}= \{x^\mu,\,1,\,-x^2\}.
\end{equation}

Consider an operator $O^{a_1...a_l}_{b_1...b_{\bar l}}(X)$, defined on the light cone $X^2=0$, symmetric in its two sets of indices.
Following~\cite{SimmonsDuffin:2012uy}, it can be projected down to a 4D operator $\OO_{\alpha_1\ldots\alpha_\ell}^{\dot\beta_1\ldots\dot\beta_{\bar\ell}}(x)$ as
\begin{equation}\label{eq:operator_embedding}
\OO_{\alpha_1\ldots\alpha_\ell}^{\dot\beta_1\ldots\dot\beta_{\bar\ell}}(x)=%(X^{+})^{-\Delta-\frac{\ell+\bar\ell}{2}}
\mathbf{X}_{\alpha_1 a_1}\ldots
\mathbf{X}_{\alpha_\ell a_\ell}
\overline{\mathbf{X}}^{\dot{\beta}_1 b_1}\ldots
\overline{\mathbf{X}}^{\dot{\beta}_{\bar\ell} b_{\bar\ell}}
O^{a_1...a_\ell}_{b_1...b_{\bar\ell}}(X)\bigg|_\text{Poincare}.
\end{equation}
If the 6D operator satisfies the homogeneity property 
\begin{equation}
O^{a_1...a_l}_{b_1...b_{\bar l}}(\lambda\,X)=\lambda^{-\CombinedDelta_{\OO}} O^{a_1...a_l}_{b_1...b_{\bar l}}(X),
\end{equation}
where $\CombinedDelta_{\OO}$ is defined in \eqref{eq:scaling6D}, then the resulting 4D operator will transform as a primary operator of dimension $\Delta_\OO$ under conformal transformations. We call $O$ a 6D uplift of $\OO$.

Notice that the 6D uplift $O$ is not uniquely defined. Indeed as a consequence of the light cone condition in terms of the matrices in \eqref{eq:XXbar},
\begin{equation}\label{eq:light_cone_condition}
X^2=0
\;\;\implies\;\;
\hspace{1mm}_a(\mathbf{X}\overline{\mathbf{X}})^b=0
\quad\text{and}\quad \hspace{1mm}^a(\overline{\mathbf{X}}\mathbf{X})_b=0,
\end{equation}
the 6D operator is defined up to terms which vanish in \eqref{eq:operator_embedding}, leading to the following equivalence relation
\begin{equation}\label{eq:gauge_redundancy}
O^{a_1...a_\ell}_{b_1...b_{\bar\ell}}\sim O^{a_1...a_\ell}_{b_1...b_{\bar\ell}}+
\overline{\mathbf{X}}^{a_1 \, c}A^{a_2...a_\ell}_{c\,b_1...b_{\bar\ell}}+
\mathbf{X}_{b_1 c}\,B^{c \,a_1...a_\ell}_{b_2...b_{\bar\ell}}+\delta^{a_1}_{b_1}\,C^{a_2...a_\ell}_{b_2...b_{\bar\ell}}.
\end{equation}
Furthermore, in order to simplify the treatment of derivatives in the embedding space, it is convenient to  arbitrarily extend $O(X)$ away from the light cone $X^2=0$ and treat all the extensions as equivalent. This means that we can also add to $O(X)$ terms proportional to $X^2$.
Following the terminology of \cite{Costa:2011dw}, we refer to this possibility as a gauge freedom and the terms proportional to $\mathbf{X}_{ab}\;,\overline{\mathbf{X}}^{ab},\;\delta^{a}_b$ or $X^2$ will be called pure gauge terms.

It is convenient to use the index-free notation~\eqref{eq:embeddingFormulaMainText}. Contracting the 4D auxiliary spinors with \eqref{eq:operator_embedding}, we find that
\begin{equation}\label{eq:projection_rule_index}
\OO(x,s,\bar s)=O(X,S,\overline{S})\bigg|_{proj},
\end{equation}
where we introduced the formal operation $\vert_{proj}$ defined as
\beq
	\label{eq:projection_auxiliary_vectors}
X^M\proj \equiv X^M\Big|_\text{Poincare} ,\qquad
 S_a\Big|_{proj}\equiv s^{\alpha}{\mathbf{X}_{\alpha a}}\bigg|_\text{Poincare},
\qquad
\overline{S}^a\Big|_{proj}\equiv\bar{s}_{\dot{\beta}}{\overline{\mathbf{X}}^{\dot{\beta} b}}\bigg|_\text{Poincare}.
\eeq

As a consequence of the gauge freedom, the index-free 6D uplift $O(X,S,\overline S)$ is defined up to pure gauge terms proportional to $S \overline{\mathbf{X}},\; \overline S \mathbf{X},\; \overline S S$ or $X^2$. Note that they all vanish under the operation of projection~\eqref{eq:projection_rule_index} due to~\eqref{eq:light_cone_condition}
\begin{equation}\label{eq:gauge_choice}
\overline{\mathbf{X}}^{ab}S_{b}\proj=0,\quad\;
\overline{S}^b\mathbf{X}_{ba}\proj=0,\quad\;
\overline{S}^a S_a\proj=0,\quad\; X^2\proj=0,
\end{equation}

We will always work modulo the gauge terms~\eqref{eq:gauge_choice}. In practice this is taken into account by treating~\eqref{eq:gauge_choice} as explicit relations in the embedding formalism even before the projection. Note then that as a consequence of the relations ~(\ref{eq:light_cone_condition}), ~(\ref{eq:gauge_choice}), the anti-symmetric properties~(\ref{eq:coordinates}) and the relations (A.7) in appendix A of~\cite{Elkhidir:2014woa}, the following identities hold\footnote{We thank Emtinan Elkhidir for showing this simple derivation.} which we call the 6D Jacobi identities
\begin{equation}\label{eq:jacobi_identities}
S_{[a}\mathbf{X}_{bc]}=0,\quad
\overline S^{[a}\overline{\mathbf{X}}^{bc]}=0,\quad
\mathbf{X}_{[ab}\mathbf{X}_{c]d}=0,\quad
\overline{\mathbf{X}}^{[ab}\overline{\mathbf{X}}^{c]d}=0.
\end{equation}

\paragraph{Differential operators}
In section \ref{sec:Outline} we commented upon the importance of some differential operators, such as the conservation operator \eqref{eq:conservation_condition_4D}, spinning differential operators~\eqref{eq:spinning_differential_operators_A}, \eqref{eq:spinning_differential_operators_B} and the Casimir operators entering \eqref{eq:casimir_equations}. To consistently define these operators in embedding space, we require their action to be insensitive to different extensions of fields outside the light cone and the other gauge terms in~\eqref{eq:gauge_choice}. This results in the requirement\footnote{In this equation $O$ stands for the usual big-$O$ notation and not the 6D operator.}
\begin{equation}\label{eq:derivative_consistency}
D\left(\frac{\partial}{\partial X^M},\frac{\partial}{\partial S_a},\frac{\partial}{\partial \overline{S}^a},X,S,\overline{S}\right)\cdot O(X^2,S \overline{\mathbf{X}}, \overline S \mathbf{X}, \overline S S )=O(X^2,S \overline{\mathbf{X}}, \overline S \mathbf{X}, \overline S S ).
\end{equation}

To go from 6D differential operators to 4D differential operators, we need to find an explicit uplift of the 4D operators $\OO(x,s,\bar s)$ to the 6D operators $O(X,S,\overline S)$. As noted above, there are infinitely many such uplifts differing by gauge terms, but all lead to the same result for 4D differential operators if the 6D operator satisfies~\eqref{eq:derivative_consistency}. For example, we can choose the uplift
\beq
	O(X,S,\overline S) = (X^+)^{-\CombinedDelta_\OO}\OO(X^\mu/X^+,S_\alpha,\overline S_{\dot\alpha}).
\eeq
In particular, $X^-,\, S^{\dot\beta},\, \overline S^{\beta}$ derivatives of this uplift of $\OO$ vanish. By applying 6D derivatives to this expression we automatically obtain the required 4D derivatives on the right hand side. For instance, we find for the first order derivatives after the 4D projection
\begin{equation}\label{eq:derivatives_projection}
\partial/\partial X^M \proj=\left\{\partial/\partial x^\mu,-\CombinedDelta_\OO-x^\nu\partial/\partial x^\nu,0\right\},
\end{equation}
\begin{equation}\label{eq:derivatives_projection_2}
\partial/\partial S_a\proj=\left\{\partial/\partial s_{\alpha},0\right\},
\quad\quad
\partial/\partial \overline{S}^a\proj=
\left\{0,\partial/\partial \bar{s}_{\dot{\alpha}}\right\}.
\end{equation}

\paragraph{Reality properties of the basic invariants}
Using the reality properties~\eqref{eq:sigmareality} of the sigma matrices, the projection rules~\eqref{eq:projection_auxiliary_vectors} for $S$ and $\overline S$, and the reality convention for 4D auxiliary polarizations $s_\alpha=\left(\bar s_{\dot\alpha}\right)^*$, we can find the following reality properties for the basic objects hold
\beq
\left(\mathbf{X}_{ab}\right)^*=\overline{\mathbf{X}}_{\bar a\bar b},\quad
\left(\overline{\mathbf{X}}^{ab}\right)^*={\mathbf{X}}^{\bar a\bar b},\quad
(S_a)^*=i \overline S_{\bar a},\quad
(\overline S^a)^*=i S^{\bar a}.
\eeq

Due to the relations such as $Y^a W_a=Y_{\bar a} W^{\bar a}$, we have an extremely simple conjugation rule for the expressions such as $\big(\overline S_i\mathbf{X}_j\overline{\mathbf{X}}_k S_l\big)$: replace $\mathbf{X}\leftrightarrow \overline{\mathbf{X}},\, S\leftrightarrow \overline S$ and add a factor of  $i$ for each $S$ and $\overline S$.

\paragraph{Action of Space Parity}
To analyze space parity, let us denote by $P^M_N$ the 6x6 matrix which relfects the spacial components of $X^\mu$. We also denote by $\hat a$ indices transforming in the representation reflected relative to the one of $a$.\footnote{The reflected representation is the representation with the Lorentz generators $\MM^\text{refl}_{MN}$ given by $\MM^\text{refl}_{MN}=P^{M'}_{M}P^{N'}_N\MM_{M'N'}$, where $\MM$ are the original generators.} Note that the reflection of the fundamental representation is equivalent to anti-fundamental and vice versa and this equivalence should be implemented by some matrices $p^{\hat a b}$ and $p_{\hat a b}$. In terms of these matrices we then have
\algn{
	P^{M}_{N}\Sigma^N_{ab}&=\overline{\Sigma}^N_{\hat a\hat b}=p_{\hat a a'}p_{\hat b b'}\overline{\Sigma}^{M\, a'b'},\\
	P^{M}_{N}\overline{\Sigma}^{N\,ab}&=\Sigma^{N\,\hat a\hat b}=p^{\hat a a'}p^{\hat b b'}{\Sigma}^{M}_{a'b'}.
}
It is easy to check that these identities (as well as the equivalence between the representations) are achieved by choosing
\beq
	p^{\hat a b}=p^{b\hat a}=-p_{\hat a b}=-p_{b\hat a}=\begin{pmatrix}
	0 & 0 & 0 & -i\\
	0 & 0 & i & 0\\
	0 & i & 0 & 0\\
	-i & 0 & 0 & 0
	\end{pmatrix}_{ab}.
\eeq
From the above we deduce the action of parity on on $\mathbf{X}$ and $\overline{\mathbf{X}}$
\beq\label{eq:EFparity1}
	\mathbf{X}_{ab}\mapsto \overline{\mathbf{X}}_{\hat a\hat b}, \quad \overline{\mathbf{X}}^{ab}\mapsto {\mathbf{X}}^{\hat a \hat b}.
\eeq
We can also check, based on 4D projections of $S$ and $\overline S$, that 
\beq\label{eq:EFparity2}
	S_a\mapsto -\overline S_{\hat a}, \quad \overline S^a\mapsto S^{\hat a}.
\eeq
Due to the identities such as $Y^a W_a=Y_{\hat a} W^{\hat a}$, we have the following parity conjugation rule for the products like $\left(\overline S_i\mathbf{X}_j\overline{\mathbf{X}}_k S_l\right)$: replace $\mathbf{X}\leftrightarrow \overline{\mathbf{X}},\, S\leftrightarrow \overline S$ and a factor of  $-1$ for each $S$ in the original expression.

\paragraph{Action of Time Reversal}
As discussed in appendix~\ref{app:conventions4D}, see equation~\eqref{eq:time_reversal_simple_rule}, the time reversal transformation can be implemented by combining the space parity with complex conjugation. Using the above rule, $\TT$ acts simply as a multiplication by $i^{\sum_i\ell_i-\bar \ell_i}$ on each structure.

\section{Normalization of Two-point Functions and Seed CPWs}
\label{app:twopointnorm}
In this appendix our goal is to fix the normalization constants of 2-point functions~\eqref{eq:normalization_2_point_function} and the seed CPWs~\eqref{eq:normalization_seeds}.

The phase of 2-point functions is constrained by unitarity. A simple manifestation of the unitarity is the requirement that all the states in a theory have non-negative norms
\beq
	\<\Psi|\Psi\>\geq 0.
\eeq
Our strategy is to define a state whose norm is related to 2-point functions~\eqref{eq:2PointFunction} and use this relation to fix the phase~\eqref{eq:normalization_2_point_function}. In particular, we set 
\begin{equation}
|\OO(s,\bar s)\rangle
\equiv\OO(x_0,s,\bar s)\,|0\rangle,
\quad
x_0^\mu\equiv\{i\epsilon,0,0,0\},
\end{equation}
where $\epsilon>0$.
Here we are working in the standard Lorentzian quantization where the states are defined on spacelike hyperplanes. The state $|\OO(s,\bar s)\>$ can then be interpreted as a NS-quantization state in a Euclidean CFT~\cite{Rychkov:2016iqz}. Note that we have
\beq
	|\OO(s,\bar s)\>=e^{-\epsilon H}\OO(0,s,\bar s)|0\>.
\eeq
Here $H=-i P_0$ is the Hamiltonian\footnote{Recall that in our conventions $P$ is anti-Hermitian.} of the theory, and thus its spectrum is bounded from below. Therefore, we need $\epsilon>0$ in order for $|\OO(s,\bar s)\>$ to have a finite norm. To compute this norm, we first consider the conjugate state
\beq
	\<\OO(s,\bar s)|=\<0|(\OO(x_0,s,\bar s))^\dagger=\<0|\overline{\OO}(-x_0,s,\bar s),
\eeq
where we used $x_0^*=-x_0$. Then the norm is given by
\beq\label{eq:the_norm}
	\<\OO(s,\bar s)|\OO(s,\bar s)\>=\<0|\overline{\OO}(-x_0,s,\bar s)\OO(x_0,s,\bar s)|0\>.
\eeq
By using~\eqref{eq:2PointFunction} to further rewrite~\eqref{eq:the_norm}, with the invariants $x_{12}^2$, $\II^{21}$ and $\II^{12}$ taking the form
\beq
x_{12}^2=4\epsilon^2,\quad
\II^{21}=2i\epsilon\;s^\dagger s,\quad
\II^{12}=-2i\epsilon\;s^\dagger s,
\eeq
we find
\beq
	\<0|\overline{\OO}(-x_0,s,\bar s)\OO(x_0,s,\bar s)|0\>=c_{\<\overline\OO\OO\>}(2\epsilon)^{-2\Delta}(s^\dagger s)^{\ell+\bar \ell}i^{\bar\ell-\ell}\geq 0,
\eeq
where
$s^\dagger s=|s_1|^2+|s_2|^2\geq 0$. This equation fixes the phase of $c_{\<\overline\OO\OO\>}$, and we can consistently set
\beq
	c_{\<\overline\OO\OO\>}=i^{\ell-\bar\ell}.
\eeq

\paragraph{Normalization of seed CPWs}
One can find the leading OPE behavior of the seed and the dual seed conformal blocks by taking the limit $z,\bar z\rightarrow 0$, $z\sim\bar z$, of the solutions obtained in~\cite{Echeverri:2016dun}. In particular, for the seed blocks we find
\begin{equation}\label{eq:asymptotics_seed_blocks}
\lim_{z,\bar z\rightarrow 0}\,H^{(p)}_{e}=
c_{0,-p}^p\;
\frac{(-2)^{e-p}\,p!\,(p-e+1)_e}{e!\,(\ell+1)_p}\;
(z\bar z)^{\frac{\Delta+e-p/2}{2}}\;
C_{\ell-p+e}^{(p+1)}\bigg( \frac{z+\bar z}{2\,(z\bar z)^{1/2}} \bigg),
\end{equation}
and for the dual seed blocks
\begin{equation}\label{eq:asymptotics_dual_seed_blocks}
\lim_{z,\bar z\rightarrow 0}\,\overline H^{(p)}_{e}=
(-2)^p\,\bar c_{0,-p}^{p}\;
\frac{(-2)^{e-p}\,p!\,(p-e+1)_e}{e!\,(\ell+1)_p}\;
(z\bar z)^{\frac{\Delta+e-p/2}{2}}\;
C_{\ell-e}^{(p+1)}\bigg( \frac{z+\bar z}{2\,(z\bar z)^{1/2}} \bigg),
\end{equation}
where $C^{(\nu)}_j(x)$ are the Gegenbauer polynomials, which in the limit $0< z\ll\bar z\ll 1$ read as
\begin{equation}\label{eq:gegenbauers_limit}
C_{s}^{(p+1)}\bigg( \frac{z+\bar z}{2\,(z\bar z)^{1/2}} \bigg)\approx
\frac{(p+1)_s}{s!}\; z^{-\tfrac{s}{2}}\,\bar z^{\tfrac{s}{2}}.
\end{equation}
In the equations above $c_{0,-p}^p$ and $\bar c_{0,-p}^p$ are some overall normalization coefficients defined in~\cite{Echeverri:2016dun}. The purpose of this paragraph is to find the values of these coefficients appropriate for our conventions for 2- and 3-point functions.

In order to fix these coefficients, it suffices to consider the leading term in the s-channel OPE in the seed 4-point functions. We have checked that the OPE exactly reproduces the form of~\eqref{eq:asymptotics_seed_blocks} and~\eqref{eq:asymptotics_dual_seed_blocks} if one sets
\begin{equation}\label{eq:seednormalization}
c_{0,-p}^p=2^p\,\bar c_{0,-p}^p=(-1)^\ell\;i^p.
\end{equation}
Let us stress that this normalization factor is fixed by the convention~\eqref{eq:2PointFunction} and~\eqref{eq:normalization_2_point_function} for the 2-point functions, and the definitions of the seed 3-point functions. The seed 3-point tensor structures are defined as
\algn{\label{eq:seedleft}
\langle \FF_1^{(0,0)}(\point_1)\,\FF_2^{(p,0)}(\point_2)\,\OO_\Delta^{(\ell,\,\ell+p)}(\point_3)\rangle&=%\lambda_{\langle \FF_1\FF_2\OO\rangle}
[\hat \II^{32}]^p\;[\hat \JJ_{12}^3]^{\ell}\mathcal{K}_3,\\\label{eq:seedright}
\langle \overline\OO_\Delta^{(\ell+p,\,\ell)}(\point_2)\,\FF_3^{(0,0)}(\point_3)\,\FF_4^{(0,p)}(\point_4)\,\rangle&=%\lambda_{\langle \overline\OO\FF_3\FF_4\rangle}
[\hat \II^{42}]^p\;[\hat \JJ_{34}^2]^{\ell}\mathcal{K}_3,
}
and the dual seed 3-point functions are defined as
\algn{\label{eq:seedleftdual}
\langle \FF_1^{(0,0)}(\point_1)\,\FF_2^{(p,0)}(\point_2)\,\overline{\OO}_\Delta^{(\ell+p,\,\ell)}(\point_3)\rangle&=%\lambda_{\langle \FF_1\FF_2\overline\OO\rangle}
[\hat \KK^{23}_1]^p\;[\hat \JJ_{12}^3]^{\ell}\mathcal{K}_3,\\\label{eq:seedrightdual}
\langle \OO_\Delta^{(\ell,\,\ell+p)}(\point_2)\,\FF_3^{(0,0)}(\point_3)\,\FF_4^{(0,p)}(\point_4)\,\rangle&=%\lambda_{\langle \OO\FF_3\FF_4\rangle}
[\hat{\overline\KK}^{24}_3]^p\;[\hat \JJ_{34}^2]^{\ell}\mathcal{K}_3,
}
where in each equation $\mathcal{K}_3$ has to be replaced with the appropriate 3-point kinematic factor as defined in~\eqref{eq:kinematic_factor_n=3}.

Equation~\eqref{eq:seednormalization} can be derived from these three-point functions and the corresponding leading OPE terms
\algn{
	\FF_1^{(0,0)}(0)\FF_2^{(p,0)}(x_2,s_2)&=\frac{(-i)^p}{\ell!(\ell+p)!}|x_2|^{\Delta-\Delta_1-\Delta_2-\ell}(s_2\partial_s)^p(x_2^\mu \partial_s\sigma_\mu\partial_{\bar s})^\ell\overline\OO^{(\ell+p,\ell)}_{\Delta}(0,s,\bar s)+\ldots,\\
	\FF_1^{(0,0)}(0)\FF_2^{(p,0)}(x_2,s_2)&=\frac{i^p}{\ell!(\ell+p)!}|x_2|^{\Delta-\Delta_1-\Delta_2-\ell-p}(x_2^\mu s_2\sigma_\mu\partial_{\bar s})^p(x_2^\mu \partial_s\sigma_\mu\partial_{\bar s})^\ell\OO^{(\ell,\ell+p)}_{\Delta}(0,s,\bar s)+\ldots,
}
where we have defined
\beq
(\partial_s)^\alpha\equiv \frac{\partial}{\partial s_\alpha},\quad (\partial_{\bar s})^{\dot \alpha}\equiv \frac{\partial}{\partial {\bar s}_{\dot \alpha}}.
\eeq
The normalization coefficients in these OPEs can be computed by substituting the OPEs into~\eqref{eq:seedleft} and~\eqref{eq:seedleftdual} and using the two-point function~\eqref{eq:normalization_2_point_function}. The normalization coefficients for the CPWs are then obtained by using these OPEs in the seed four-point function
\beq
\<
\FF_1^{(0,0)}\,
\FF_2^{(p,0)}\,
\FF_3^{(0,0)}\,
\FF_4^{(0,p)}
\>
\eeq
and utilizing the 3-point function definitions~\eqref{eq:seedright} and~\eqref{eq:seedrightdual}. In practice, when comparing the normalization coefficients, we found it convenient to use the conformal frame~\eqref{eq:conformalframedefn_1} - \eqref{eq:conformalframedefn_4} in the limit $0<z\ll\bar z\ll 1$ and further set $\eta_2=0$ and $e=p$ for the seed CPWs or $\xi_2=0$ and $e=0$ for the dual seed CPWs.

\section{4D Form of Basic Tensor Invariants}
\label{sec:TensotInvariants}
Here we provide the form of basic tensor invariants in 4D for $n\leq 4$ point functions. They are obtained by applying the projection operation~\eqref{eq:projection_auxiliary_vectors} to the basic 6D tensor invariants constructed in section \ref{sec:constructing_tensor_structures_EF}
\begin{equation}
(\hat{\II}^{ij},\;\hat{\II}^{ij}_{kl},\;\hat{\JJ}^k_{ij},\;\hat{\KK}^{ij}_k,\;\hat{\overline{\KK}}^{ij}_k,\;\hat{\LL}^{i}_{jkl},\;\hat{\overline{\LL}}^{i}_{jkl})\equiv
(\hat{I}^{ij},\;\hat{I}^{ij}_{kl},\;J^k_{ij},\;K^{ij}_k,\;\overline{K}^{ij}_k,\;L^{i}_{jkl},\;\overline L^{i}_{jkl})\proj,
\end{equation}
where
\begin{align}\label{eq:I4D}
&\hat\II^{ij}= x_{ij}^\mu \,(\bar s_i \bar \sigma_\mu s_j),\\
&\hat\II^{ij}_{kl}=\frac{1}{2\,x_{kl}^{2}}\times\Big(
(x^2_{ik}x_{jl}^\mu-x^2_{il}x_{jk}^\mu)+(x^2_{jk}x_{il}^\mu-x^2_{jl}x_{ik}^\mu) 
-x^2_{ij}x_{kl}^\mu-x^2_{kl}x_{ij}^\mu \nonumber  \\
&\quad\quad\quad\quad\quad\quad\quad\quad\quad
-2i\epsilon^{\mu\nu\rho\sigma}x_{ik\,\nu}x_{lj\,\rho}x_{lk\,\sigma}\Big)\times (\bar s_i \bar \sigma_\mu s_j)
,\\
&\hat\JJ^{k}_{ij}= \,\frac{x_{ik}^2x_{jk}^2}{x_{ij}^{2}}\,
\times\bigg( \frac{x_{ik}^\mu}{x_{ik}^2}-\frac{x_{jk}^\mu}{x_{jk}^2} \bigg)\times\,
(\bar s_k \bar \sigma_\mu  s_k),\\
&\hat\KK^{ij}_k =\frac{1}{2}\frac{|x_{ij}|}{|x_{ik}||x_{jk}|}\,\times\,\Big( (x_{ik}^2+x_{jk}^2-x_{ij}^2)(s_is_j)-4x_{ik}^\mu x_{jk}^\nu\,(s_i\sigma_{\mu\nu}s_j)  \Big),\\
&\hat{\overline{\KK}}^{ij}_k =\frac{1}{2}\frac{|x_{ij}|}{|x_{ik}||x_{jk}|}\,\times\,\Big( (x_{ik}^2+x_{jk}^2-x_{ij}^2)(\bar s_i\bar s_j)-4x_{ik}^\mu x_{jk}^\nu\,(\bar s_i\bar\sigma_{\mu\nu}\bar s_j)  \Big),\\
&\hat\LL^{i}_{jkl} = \frac{2}{|x_{jk}||x_{kl}||x_{lj}|}\times\left(x^{2}_{ij}x_{kl}^\mu x_{il}^\nu+x^2_{ik}x_{lj}^\mu x_{ij}^\nu+ x^2_{il}x_{jk}^\mu x_{ik}^\nu\right)\times
\left(s_i\sigma_{\mu\nu}s_i\right),\\
&\hat{\overline{\LL}}^{i}_{jkl}=\frac{2}{|x_{jk}||x_{kl}||x_{lj}|}\times\left(x^{2}_{ij}x_{kl}^\mu x_{il}^\nu+x^2_{ik}x_{lj}^\mu x_{ij}^\nu+ x^2_{il}x_{jk}^\mu x_{ik}^\nu\right)\times
\left(\bar s_i \bar \sigma_{\mu\nu} \bar s_i\right).
\end{align}
We recall that $x_{ij}^\mu\equiv x_i^\mu-x_j^\mu$ and $\epsilon_{0123}=-1$ in our conventions.
From these expressions it is possible to derive the conjugation properties of the invariants. They read as follows
\begin{align}\label{eq:InvariantsConjugationA}
\left(\hat\II^{ij}\right)^*=- \hat\II^{ji}\,,&\quad\quad
\left(\hat\II^{ij}_{kl}\right)^*=-\hat\II^{ji}_{lk}\,,\quad\quad
\left(\hat\JJ^{k}_{ij}\right)^*=\hat\JJ^{k}_{ij}\,,\\\label{eq:InvariantsConjugationB}
\left(\hat\KK^{ij}_k\right)^*&=-\hat{\overline{\KK}}^{ij}_k\,,\quad\quad
\left(\hat\LL^{i}_{jkl}\right)^*=-\hat{\overline{\LL}}^{i}_{jkl}.
\end{align}
Their parity transformation can be deduced from \eqref{eq:ParityIndexFreeNotation}
\begin{align}\label{eq:InvariantsParityA}
\PP\,\hat\II^{ij}=- \hat\II^{ji}\,,&\quad\quad
\PP\,\hat\II^{ij}_{kl}=-\hat\II^{ji}_{lk}\,,\quad\quad
\PP\,\hat\JJ^{k}_{ij}=\hat\JJ^{k}_{ij}\,,\\\label{eq:InvariantsParityB}
\PP\,\hat\KK^{ij}_k&=\hat{\overline{\KK}}^{ij}_k\,,\quad\quad
\PP\,\hat\LL^{i}_{jkl}=\hat{\overline{\LL}}^{i}_{jkl}.
\end{align}
Finally, according to~\eqref{eq:time_reversal_simple_rule} one gets transformations under time reversal
\begin{align}\label{eq:InvariantsTimeReversalA}
\TT\,\hat\II^{ij}= \hat\II^{ij}\,,&\quad\quad
\TT\,\hat\II^{ij}_{kl}=\hat\II^{ij}_{kl}\,,\quad\quad
\TT\,\hat\JJ^{k}_{ij}=\hat\JJ^{k}_{ij}\,,\\\label{eq:InvariantsTimeReversalB}
\TT\,\hat\KK^{ij}_k&=-\hat{\KK}^{ij}_k\,,\quad\quad
\TT\,\hat\LL^{i}_{jkl}=
-\hat{\LL}^{i}_{jkl}.
\end{align}
The same properties follow from the discussion of $\PP$-, $\TT$-symmetries, and conjugation in appendix~\ref{sec:DetailsOfTheEmbeddingFormalism}.

\section{Covariant Bases of Three-point Tensor Structures}
\label{sec:BasisthreePointFunctions}

Let us review the construction \fref{n3ListStructures} of 3-point function tensor structures~\cite{Elkhidir:2014woa}. According to the discussion below~\eqref{eq:set_of_tensor_structures} one has
\begin{equation}\label{eq:over_complete_basis_n=3}
\hat{\mathbb{T}}_3^a
=\Big\{
\prod_{i\neq j}
\big[\hat\II^{ij}\big]^{m_{ij}}
\times
\prod_{i,\,j<k}
\big[\hat \JJ^{i}_{jk}\big]^{n_{i}}
\big[\hat \KK^{jk}_{i}\big]^{k_{i}}
\big[\hat{\overline \KK}^{jk}_{i}\big]^{\bar k_{i}}
\Big\},
\end{equation}
where the exponents satisfy the following system
\begin{align}
\ell_i &=\sum_{l\neq i} m_{li}+\sum_{l\neq i} k_{l}+ n_{i},
\\
\bar\ell_i &=\sum_{l\neq i} m_{il}+\sum_{l\neq i} \bar k_{l}+ n_{i}.
\end{align}
Let us also define the quantity
\begin{equation}
\Delta \ell \equiv \sum_{i}(\ell_i-\bar{\ell}_i).
\end{equation}
Due to relations among products of invariants, not all the structures obtained this way are independent and constraints on possible values of the exponents in~\eqref{eq:over_complete_basis_n=3} must be imposed.
Theses relations come from the Jacobi identities~\eqref{eq:jacobi_identities} by contracting them with 6D polarizations and 6D coordinate matrices in all possible ways.

The first set of relations reads
\begin{align}
\hat \KK_j^{ik}\hat{\overline{\KK}}_i^{jk}  & =-\,\hat \II^{ki}\hat \II^{jk}-\hat \II^{ji}\hat \JJ^k_{ij}\,,\label{eq:Jacobi_rel_1}\\
\hat \KK_k^{ij}\hat{\overline{\KK}}_k^{ij}  & =\,\hat \II^{ij}\hat \II^{ji}-\hat \JJ_i^{jk}\hat \JJ_j^{ik}\,.\label{eq:Jacobi_rel_2}
\end{align}
If $\Delta\ell\neq0$ we use these relations to set $\bar k_i=0$ or $k_i=0$ for $\forall \,i$ in the expression~(\ref{eq:over_complete_basis_n=3}); if $\Delta\ell=0$ we set instead $k_i=\bar k_i=0\;\;\forall \,i$.

The second set of relations reads
\begin{align}
\hat \JJ^j_{ik}\hat \KK_j^{ik}       & =\hat \II^{ji}\hat \KK_i^{jk}-\hat \II^{jk}\hat \KK_k^{ij}\,,
     \label{eq:Jacobi_rel_3}   \\
\hat \JJ^j_{ik}\hat{\overline{\KK}}_j^{ik} & =\hat \II^{ij}\hat{\overline{\KK}}_i^{kj}+\hat \II^{kj}\hat{\overline{\KK}}_k^{ij} \,.
      \label{eq:Jacobi_rel_4} 
\end{align}
This allows to set either $n_i=0$ or $k_i=0$ if $\Delta\ell>0$ and either $n_i=0$ or $\bar k_i=0$ if $\Delta\ell<0$ in~\eqref{eq:over_complete_basis_n=3}.

If $\Delta\ell=0$ it might seem that the relations~(\ref{eq:Jacobi_rel_3}) and~(\ref{eq:Jacobi_rel_4}) do not play any role, since all $K$ and $\overline K$ are removed by mean of~(\ref{eq:Jacobi_rel_1}) and~(\ref{eq:Jacobi_rel_2}). However it is not the case, by combining~(\ref{eq:Jacobi_rel_3}) and~(\ref{eq:Jacobi_rel_4}) with \eqref{eq:Jacobi_rel_1} and \eqref{eq:Jacobi_rel_2} one gets a third order relation
\be\label{eq:Jacobi_rel_5} 
\hat \JJ^1_{23}\hat \JJ^2_{13}\hat \JJ^3_{12}=  \,\big(  \hat \II^{23}\hat \II^{32}\hat \JJ^1_{23} - \hat \II^{13}\hat \II^{31}\hat \JJ^2_{13} + \hat \II^{12}\hat \II^{21}\hat \JJ^3_{12}  \big)-\,\big(\hat \II^{21}\hat \II^{13}\hat \II^{32}-\hat \II^{12}\hat \II^{31}\hat \II^{23}\big) \,.
\ee
This allows to set in~(\ref{eq:over_complete_basis_n=3}) either $n_1=0$ or $n_2=0$ or $n_3=0$ when $\Delta \ell=0$\footnote{Notice that for $\Delta \ell \neq 0$ at least one $n_i$ is always $0$ and hence \eqref{eq:Jacobi_rel_5} does not give new constraints.}. It can be verified that no other independent relations exist.

In the case when all operators are trace-less symmetric, i.e. $\ell_i=\bar \ell_i$ for each field, it is convenient to work in terms of structures manifestly even or odd under parity. Following~\cite{Echeverri:2015rwa}, the most general parity definite tensor structure reads as
\begin{equation}\label{eq:over_complete_basis_n=3_even}
\hat{\mathbb{T}}_{3}^a=
\Big\{
\big(\hat \II^{21}\hat \II^{13}\hat \II^{32}+\hat \II^{12}\hat \II^{31}\hat \II^{23}\big)^{p}\times
\prod_{i,j}\left(\hat\II^{ij} \hat\II^{ji}\right)^{m_{ij}}
\times
\prod_{i,\,j<k}\big[\hat \JJ^{i}_{jk}\big]^{n_{i}}
\Big\},
\end{equation}
where the structure is even if $p=0$ and the structure is odd if $p=1$. The form of this basis is structurally identical to the one found in~\cite{Costa:2011mg}. 
This basis has extremely simple properties under complex conjugation, parity and time reversal
\begin{equation}
\left(\hat{\mathbb{T}}_{3}^{a}\right)^*=(-1)^p\;\hat{\mathbb{T}}_{3}^{a},\quad
\PP\,\hat{\mathbb{T}}_{3}^{a}=(-1)^p\;\hat{\mathbb{T}}_{3}^{a},\quad
\TT\,\hat{\mathbb{T}}_{3}^{a}=\hat{\mathbb{T}}_{3}^{a}.
\end{equation}
This basis can be constructed using~\fref{n3ListStructuresAlternativeTS}.

\section{Casimir Differential Operators}
\label{app:casimir_differential_operators}
The Lie algebra of the 4D conformal group is a real form of the simple rank-3 algebra $\mathfrak{so}(6)$. Therefore, it has three independent Casimir operators, which can be defined using the 6D Lorentz generators~\eqref{eq:lorentz_generators} as follows 
\begin{align}
\label{eq:casimir_operator_2}
C_2 &\equiv \frac{1}{2}\,L_{MN}\;L^{NM},\\
\label{eq:casimir_operator_3}
C_3 &\equiv \frac{{1}}{24i}\,\epsilon^{MNPQRS}\;L_{MN}\;L_{PQ}\;L_{RS},\\
\label{eq:casimir_operator_4}
C_4 &\equiv \frac{1}{2}\,L_{MN}\;L^{NP}\;L_{PQ}\;L^{QM},
\end{align}
where $\epsilon^{012345}=\epsilon_{012345}=+1$.

To write out the Casimir eigenvalues for primary operators, it is convenient to introduce also the $SO(1,3)$ Casimir operators using the 4D Lorentz generator~\eqref{eq:LorentzAction}. There are two such Casimirs
\begin{equation}
c_2^+ \equiv -\frac{1}{2}L_{\mu\nu}L^{\mu\nu},\quad
c_2^- \equiv \frac{{1}}{4i}\epsilon^{\mu\nu\rho\sigma}L_{\mu\nu}L_{\rho\sigma},
\end{equation}
with the eigenvalues 
\begin{equation}
e_2^+ = \frac{1}{2}\ell(\ell+2)+\frac{1}{2}\bar\ell(\bar\ell+2),\quad
e_2^- = \frac{1}{2}\ell(\ell+2)-\frac{1}{2}\bar\ell(\bar\ell+2).
\end{equation}
The conformal Casimir eigenvalues are then given by
\algn{
E_2 &\equiv \Delta(\Delta-4)+e_2^+,\\
E_3 &\equiv \big(\Delta-2\big)\,e_2^-,\\
E_4 &\equiv \Delta^2(\Delta-4)^2+6\,\Delta(\Delta-4)
+\big(e_2^+\big)^2
-\frac{1}{2}\,\big( e_2^-\big)^2.
}
Note that $c_2^-$ is parity-odd and therefore $e_2^-$ changes the sign under $\ell\leftrightarrow\bar\ell$. The same comment applies to $C_3$ and $E_3$.

It is convenient to write the Casimir Operators in the $SU(2,2)$ language by plugging~\eqref{eq:generators_SU(2,2)} into the expression~\eqref{eq:casimir_operator_2},~\eqref{eq:casimir_operator_3} and~\eqref{eq:casimir_operator_4} 
\begin{align}
\label{eq:casimir_operator_2_Prime}
C_2 =& \frac{1}{4}\,\tr L^2,\\
\label{eq:casimir_operator_3_Prime}
C_3 =& \frac{1}{12}\,\left(\tr L^3-16\,C_2\right),\\
\label{eq:casimir_operator_4_Prime}
C_4 =&-\frac{1}{8}\,\left(\tr L^4-8\,\tr L^3-12\,C_2^2+16\,C_2\right).
\end{align}

Let us emphasize that the Casimir operators $C_n$ are the Hilbert space operators. Their differential form $\difOperator C_n$ can be obtained by replacing the Hilbert space operators $L_{MN}$ and $L_a{}^c$ with their differential representations $\difOperator L_{MN}$ and $\difOperator L_a{}^c$ given in~\eqref{eq:lorentz_generators_6D_differential} and~\eqref{eq:lorentz_generators_6D_SU(2,2)_differential} together with reverting\footnote{See the discussion below~\eqref{eq:LorentzGeneratorSpinorConjugated}.} the order of operators $L_{MN}$ and $L_a{}^c$ in equations~\eqref{eq:casimir_operator_2} - \eqref{eq:casimir_operator_4} and~\eqref{eq:casimir_operator_2_Prime} - \eqref{eq:casimir_operator_4_Prime}.

\section{Conserved Operators}
\label{app:cons}
By conserved operators we mean primary operators in short representations of the conformal group, i.e. those possessing null descendants and thus satisfying differential equations. In a unitary 4D CFT all local primary operators satisfy the unitarity bounds~\cite{Mack:1975je,Minwalla:1997ka}\footnote{An operator with $\ell=\bar\ell=0$ has an extra option $\Delta=0$. This is the identity operator.}
\begin{align}
\Delta \geq 1+\frac{\ell+\bar\ell}{2},&\;\;\ell = 0 \;\mathbf{or}\; \bar\ell = 0,\\
\Delta \geq 2+\frac{\ell+\bar\ell}{2},&\;\;\ell \neq 0 \;\mathbf{and}\; \bar\ell\neq 0,
\end{align}
and unitary null states can only appear when these bounds are saturated.

The operators of the type $\ell = 0 \;\mathbf{or}\; \bar\ell = 0$ with $\Delta=1+(\ell+\bar\ell)/2$ satisfy the free wave equation\footnote{This is not the conformally-invariant differential equation satisfied by these operators, but rather its consequence.} $\partial^2\OO_\Delta^{(\ell,\bar\ell)}=0$~\cite{Weinberg:2012cd}, which immediately implies that such operators can only come from a free subsector of the CFT. The operators of the second type, $\ell\bar\ell\neq 0, \,\Delta=2+(\ell+\bar\ell)/2$, are the conserved currents which satisfy the following operator equation\footnote{The operator $\partial$ can be applied in the conformal frame \fref{opConservation4D} or in the embedding formalism \fref{opConservationEF}.}
\begin{equation}\label{eq:conservation_condition_4D}
\partial \cdot \ops O_{\Delta}^{(\ell,\bar\ell)}(x,s,\bar s)=0,
\quad
\partial\equiv
(\epsilon\sigma^\mu)^{\alpha}_{\dot\beta}\;\partial_\mu\;
\frac{\partial^2}{\partial s^{\alpha}\;\partial\bar s_{\dot\beta}}.
\end{equation}
Of particular importance are the spin-1 currents $J^\mu$ in representation $(1,1)$, the stress tensor $T^{\mu\nu}$ in representation $(2,2)$ and the supercurrents $J_{\alpha}^{\mu}$ and $J^\mu_{\dot\alpha}$ in representations $(2,1)$ and $(1,2)$. Note that an appearance of traceless symmetric higher-spin currents is known to imply an existence of a free subsector~\cite{Maldacena:2011jn,Alba:2015upa}.

The conservation condition results in the following Ward identity for $n$-point functions
\begin{equation}\label{eq:ConservationConstraintGeneral}
\partial\cdot\langle \ldots \mathcal{O}_{\Delta}^{(\ell,\bar\ell)}(x,s,\bar s)\ldots\rangle=0+\text{contact terms},
\end{equation}
where the contact terms encode charges of operators under the symmetry generated by the conserved current $\mathcal{O}_{\Delta}^{(\ell,\bar\ell)}$. Note that since $\partial\cdot\OO_\Delta^{(\ell,\bar\ell)}$ is itself a primary operator in representation $(\ell-1,\bar\ell-1),\, \Delta=3+(\ell+\bar\ell)/2$, the left hand side of the above equation has the transformation properties of a correlation function of primary operators and thus can be expanded in a basis of appropriate tensor structures.

For 3-point functions, the Ward identities imply two kind of constraints. First, the validity of \eqref{eq:ConservationConstraintGeneral} at generic configurations of points $x_i$ implies homogeneous linear relations between the OPE coefficients entering 3-point functions. Second, the validity of \eqref{eq:ConservationConstraintGeneral} at coincident points relates some of the OPE coefficients to the charges of the other two operators in a given 3-point function (this happens only if special relations between scaling dimensions of these operators are satisfied). The solution of these constraints is of the form \eqref{eq:algebraic_constraints}, where some of $\hat\lambda$ can be related to the charges.

For 4-point functions the situation is more complicated, since \eqref{eq:ConservationConstraintGeneral} at non-coincident points leads to a system of first order differential equations for the functions $g^I_4(u,v)$ of the form
\begin{equation}\label{eq:n=4ConstraintConservation}
B^{AJ}(u,v,\partial_u,\partial_v)\; g_4^J(u,v)=0,
\end{equation}
where $A$ runs through the number of tensor structures for the correlator in the left hand side of \eqref{eq:ConservationConstraintGeneral}.
The constraints implied by these equations were analysed in~\cite{Dymarsky:2013wla}. It turns out that one can solve these equations by aribtrarily specifying a smaller number $N'_4$ of the functions $g_4^I(u,v)$ and a number of boundary conditions for the remaining $g_4^I(u,v)$.\footnote{DK thanks Anatoly Dymarsky, Jo\~ao Penedones and Alessandro Vichi for discussions on this issue.} It is generally important to take this into account when formulating an independent set of crossing symmetry equations. We refer the reader to~\cite{Dymarsky:2013wla} for details. In~\cite{Dymarsky:2013wla} the value $N_4'$ was found for 4 identical conserved spin 1 and spin 2 operators. The same values  $N_4'$ were found later by other means in~\cite{Echeverri:2015rwa} and a general counting rule was proposed in~\cite{Kravchuk:2016qvl}.

\paragraph{Conservation operator in the Embedding Formalism}
The conservation condition~(\ref{eq:conservation_condition_4D}) can be consistently reformulated in the embedding space \fref{opConservationEF} as follows
\begin{equation}\label{eq:conservation_condition_6D}
D\; O_{\Delta_{\mathcal O}}^{(\ell,\bar\ell)}\left(X,S,\overline{S}\right)=0,\quad
\Delta_{\mathcal O}=2+\frac{\ell+\bar\ell}{2}
\end{equation}
and the differential operator originally found in~\cite{Elkhidir:2014woa} is given by\footnote{We note that there is a mistake in the original paper~\cite{Elkhidir:2014woa} due to a wrong choice of the analogue of~\eqref{eq:extractingIndices6D}.}
\begin{equation}\label{eq:conservation_operator_part_1}
D\equiv
\frac{2}{\ell\,\bar{\ell}\,\left(2+\ell+\bar{\ell}\right)}
\left(X_{M}\Sigma^{MN}\partial_N\right)_a^b
\;\partial^{\,a}_{\,b},
\end{equation}
where we have defined
\begin{multline}
\partial^{\,a}_{\,b}\equiv \frac{1}{1+\ell+\bar{\ell}}\;\partial^{\,a}\partial_{\,b} =\\
\left(4+S\cdot\frac{\partial}{\partial S}+\overline{S}\cdot\frac{\partial}{\partial \overline S}\right) \frac{\partial}{\partial S_a} \frac{\partial}{\partial \overline{S}^b}-S_b\frac{\partial}{\partial S_a}\frac{\partial^{\,2}}{\partial S \cdot\partial\overline{S}}-\overline{S}^a\frac{\partial}{\partial \overline{S}^b}\frac{\partial^{\,2}}{\partial S \cdot\partial\overline{S}}.
\end{multline}
In this identity we dropped the terms which project to zero upon contraction with $\left(X_{M}\Sigma^{MN}\partial_N\right)_a^b$.

\section{Permutations Symmetries}
\label{app:permutations}
When the points in \eqref{eq:n_point_correlation_function} are space-like separated, the ordering of operators is not important up to signs coming from permutations of fermions. In particular, if some operator enters the expectation value more than once, say at points $\point_i$ and $\point_j$, the function $f_n$ enjoys the permutation symmetry
\beq	
	f_n(\ldots,\point_i,\ldots,\point_j,\ldots)=[(ij)f_n](\ldots,\point_i,\ldots,\point_j,\ldots)\equiv \pm f_n(\ldots,\point_j,\ldots,\point_i,\ldots).
\eeq
Here we used the cycle notation for permutations, for instance $(123)$ denotes $1\to 2$, $2\to 3$, $3\to 1$. In general, there may be more identical operators in the right hand side of~\eqref{eq:n_point_correlation_function} in which case $f_n$ is invariant under some subgroup of permutations $\Pi\subseteq S_n$.

The degrees of freedom in $f_n$ are described by the functions $g_n^I$ defined via~\eqref{eq:correlation_function_structure}
\begin{equation}\label{eq:correlation_function_structureAPP}
f_n(x_i,s_i,\bar s_i)=\sum_{I=1}^{N_n}g_n^I(\mathbf{u})\;\mathbb{T}_n^I(x_i,s_i,\bar s_i).
\end{equation}
One can then find the implications of the permutation symmetries directly for $g_n^I$. Note that since the exchanged operators are identical, a permutation $\pi\in\Pi$ acting on a tensor structure gives a tensor structure of the same kind, and thus we can expand it in the same basis
\beq
	\pi \mathbb{T}^I_n = \sum_J \pi^J_I(\mathbf{u})\mathbb{T}^J_n.
\eeq
This means that in general the consequence of a permutation symmetry is
\beq\label{eq:permutation_constraint}
	g_n^I(\mathbf{u}) = \sum_J \pi^I_J(\mathbf{u}) g_n^J(\pi \mathbf{u}).
\eeq

At this point we should divide all the permutations into two classes. We call the permutations which preserve the cross-rations ($\pi\mathbf{u}=\mathbf{u}$) the kinematic permutations and all the other permutations will be referred to as non-kinematic. The group of kinematic permutations $\Pi^\text{kin}_n$ is $S_n$ for $n\leq 3$ since there are no non-trivial cross-ratios in these cases. We also have $\Pi^\text{kin}_4=\mathbb{Z}_2\times \mathbb{Z}_2=\{\text{id},(12)(34),(13)(24),(14)(23)\}$ and $\Pi^\text{kin}_n$ is trivial for $n\geq 5$.

This distinction is important because for kinematic permutations the constraint~\eqref{eq:permutation_constraint} becomes a simple local linear constraint,
\beq\
	g_n^I(\mathbf{u}) = \sum_J \pi^I_J(\mathbf{u}) g_n^J(\mathbf{u}),
\eeq
which we can be solved as
\beq\label{eq:solution_to_kinematic_permutations}
	g_n^I(\mathbf{u}) = \sum_A P^I_A(\mathbf{u})\hat g^A_n(\mathbf{u}).
\eeq
In the case of 3-point functions the solution~\eqref{eq:solution_to_kinematic_permutations} has a particularly simple form~\eqref{eq:algebraic_constraints}.

Applying permutation \fref{permutePoints} and computing $\pi_J^I(\mathbf{u})$ is straightforward in the EF -- we simply need to permute the coordinates $X_i$ and the polarizations $S_i,\overline S_i$. It is somewhat trickier to figure out the permutations in the CF~\cite{Kravchuk:2016qvl}, and we describe the case $n=4$ in the remainder of this section. We also comment on how to permute non-identical operators, which is required, for example, in order to exchange $s$- and $t$-channels.

\paragraph{Semi-covariant CF Structues}
First, we describe a slight generalization of the conformal frame, which is convenient for computing the action of permutations on the CF structures. Note that the 4-point tensor structures constructed in section~\ref{sec:fourpointCFstructs} are covariant under the conformal transformations acting in $z$ plane. Indeed, it is easy to see that the structures~\eqref{eq:fourpointcfdef} transform with 2d spin $q_i+\bar q_i$ at each point. Taking into account the scaling dimensions of the operators, we see that we can assign the left- and right-moving weights 
\beq
	h_i=\frac{\Delta_i+q_i+\bar q_i}{2},\quad \bar h_i=\frac{\Delta_i-q_i-\bar q_i}{2}
\eeq
to each tensor structure. We can then easily write the value of the 4-point function represented on the conformal frame by
\beq
	f_4(0,z,1,\infty,s_i,\bar s_i)=\structgeneral g_{\{q_i,\bar q_i\}}(z,\bar z)
\eeq
in a generic configuration of the four points $z_i$ in $z$-plane as \fref{cfEvaluateInPlane}
\beq
   f_4(z_1,z_2,z_3,z_4,s_i,\bar s_i)=	\structgeneralz g_{\{q_i,\bar q_i\}}(z,\bar z),
   \label{eq:f4semicovariant}
\eeq
where
\beq
	z=\frac{(z_1-z_2)(z_3-z_4)}{(z_1-z_3)(z_2-z_4)},\quad \bar z=\frac{(\bar z_1-\bar z_2)(\bar z_3-\bar z_4)}{(\bar z_1-\bar z_3)(\bar z_2-\bar z_4)},
\eeq
and, defining $z_{ij}=z_i-z_j$,
\algn{
	\structgeneralz=\structgeneral\times& (z_{31}^{-h_1-h_2-h_3+h_4}z_{41}^{-h_1+h_2+h_3-h_4}z_{42}^{-2h_2}z_{43}^{h_1+h_2-h_3-h_4})\nn\\
	\times&(\bar z_{31}^{-\bar h_1-\bar h_2-\bar h_3+\bar h_4}\bar z_{41}^{-\bar h_1+\bar h_2+\bar h_3-\bar h_4}\bar z_{42}^{-2\bar h_2}\bar z_{43}^{\bar h_1+\bar h_2-\bar h_3-\bar h_4}).\label{eq:fourpointcovariantdef}
}
Note that the definition is chosen in such a way that the semi-covariant structure transforms with the required left and right weights and\footnote{Recall that the limit $z_4=\infty$ is defined with an extra factor $|x_4|^{2\Delta_4}$ in order to obtain a non-zero result.}
\beq
	\structgeneralconcretez{0}{z}{1}{\infty}=\structgeneral.
\eeq
In general we might need to specify the branches of the fractional powers in~\eqref{eq:fourpointcovariantdef}. The kinematic factor in this equation can be split into products of 
\beq
	(z_{ij}\bar z_{ij})^{f(\Delta_k)}\quad \text{and}\quad \left(\frac{z_{ij}}{\bar z_{ij}}\right)^{\tilde f(q_k+\bar q_k)}.
\eeq
In the region of the configuration space where all pairs of points are spacelike separated\footnote{In particular, in the whole Euclidean region.}, we have $z_{ij}\bar z_{ij}>0$, so there is no branching for the factors of the first kind. The exponent of the factors of the second kind is always half-integral, thus we only need to specify the branch of $\sqrt{\frac{z_{ij}}{\bar z_{ij}}}$ which can be chosen
\beq
	\sqrt{\frac{z_{ij}}{\bar z_{ij}}}=\sqrt{\frac{z_{ij}^2}{z_{ij}\bar z_{ij}}}=\frac{z_{ij}}{\sqrt{z_{ij}\bar z_{ij}}}.
\eeq
This is valid because it gives a smooth choice for the whole spacelike region and reduces the kinematic factor to $1$ in the standard configuration $\{z_1,z_2,z_3,z_4\}=\{0,1,z,\infty\}$.

The above discussion gives a version of the CF 4-point tensors structures which is defined for any configuration of the four points in the $z$-plane. This is sufficient for computing the action of arbitrary permutations on the tensor structures~\eqref{eq:fourpointcfdef}. Explicit formulas for permutations between identical operators can be found in~\cite{Kravchuk:2016qvl}. General permutations are implemented in \texttt{CFTs4D} package in the function \fref{permutePoints}.

\bibliographystyle{JHEP}
\bibliography{Draft}

\end{document}